%% file: quantumMarginalProblem_revised3.tex
\documentclass[aps,prx,superscriptaddress]{revtex4-2}
\usepackage{graphicx}
\usepackage{float}
\usepackage{amssymb,amsthm,amsfonts,amstext,amsmath}
\usepackage{url}
\usepackage{layouts}

\usepackage{xcolor}
\usepackage{tcolorbox}
\usepackage{enumitem}
\usepackage{booktabs}
\usepackage{bm}
\usepackage{stmaryrd}
\usepackage{tikz}
\usepackage{extarrows}
\usepackage{commath}   
\usepackage{scalerel}

\usepackage{appendix}

\usepackage{tikz-cd}
\usetikzlibrary{calc}
\usetikzlibrary{math}
\usetikzlibrary{decorations.pathreplacing,calligraphy}
\usetikzlibrary{decorations.text}
\usetikzlibrary{shapes,arrows,chains}

\usepackage[colorlinks=true,citecolor=blue,urlcolor=blue]{hyperref}
\usepackage[capitalize]{cleveref} 

\def\be{\begin{equation}}
\def\ee{\end{equation}}
\def\bea{\begin{eqnarray}}
\def\eea{\end{eqnarray}}
\def\bma{\begin{mathletters}}
\def\ema{\end{mathletters}}

\def\q0{\underline{0}}

\def\H{{\cal H}}

\def\S{{\cal S}}
\def\C{{\mathbb C}}
\def\id{{\mathbb I}}

\def\W{{\cal W}}
\def\L{{\cal L}}
\def\LL{\mathbb{L}}

\def\H{{\cal H}}
\def\epsil{{\varepsilon}}

\def\CC{\mathbb{C}}
\def\cg{B}

\def\CG{C}

\def\Rho{\Theta}

\def\tr{\mathrm{Tr}}
\def\trL{\mathrm{Tr}_{\mathrm{L}}}
\def\trR{\mathrm{Tr}_{\mathrm{R}}}

\def\trA{\mathrm{Tr}_{\mathrm{AB}}}
\def\trB{\mathrm{Tr}_{\mathrm{AD}}}
\def\trC{\mathrm{Tr}_{\mathrm{CD}}}
\def\trPrimeA{\mathrm{Tr}^\prime_{\mathrm{AB}}}
\def\trPrimeB{\mathrm{Tr}^\prime_{\mathrm{AD}}}
\def\trPrimeC{\mathrm{Tr}^\prime_{\mathrm{CD}}}

\newcommand{\rhoM}[1]{\rho^{(#1)}}
\newcommand{\omegaM}[1]{\omega^{(#1)}}

\def\W{{\cal W}}
\def\one{\leavevmode\hbox{\small1\normalsize\kern-.33em1}}

\def\bra#1{\langle#1|} 
\def\ket#1{|#1\rangle}

\def\Ind{{\cal I}}

\DeclareRobustCommand{\ckvdots}{%
	\vcenter{%
		\baselineskip=4pt 
		\kern2pt
		\hbox{.}\hbox{.}\hbox{.}
}}

\def\id{{\mathbb I}}

\include{tikzPreFabForNewSection}
 
\newcommand{\tikzPic}[1]{
	\begin{tikzpicture}[baseline=-1mm , every node/.style={scale=0.8}]
		\pic at (0,0) {#1};
 	\useasboundingbox ([shift={(0,2mm)}]current bounding box.north east) rectangle ([shift={(0,-2mm)}]current bounding box.south west);
	\end{tikzpicture}
}

\begin{document}
\title{Lower bounds on ground-state energies of local Hamiltonians through the renormalization group}

\author{Ilya Kull}
\email[Author to whom correspondence should be addressed: ]{ilya.kull@univie.ac.at}
\affiliation{University of Vienna, Faculty of Physics, Boltzmanngasse 5, A-1090 Vienna, Austria}
\affiliation{Vienna Doctoral School in Physics, Boltzmanngasse 5, A-1090 Vienna, Austria}
\author{Norbert Schuch}
\affiliation{University of Vienna, Faculty of Physics, Boltzmanngasse 5, A-1090 Vienna, Austria}
\affiliation{University of Vienna, Faculty of Mathematics, Oskar-Morgenstern-Platz 1, A-1090 Vienna, Austria}
\author{Ben Dive}
\author{Miguel Navascu\'es}
\affiliation{Institute for Quantum Optics and Quantum Information - IQOQI Vienna,\\ Austrian Academy of Sciences, Boltzmanngasse 3, A-1090 Vienna, Austria}

\begin{abstract}
Given a renormalization scheme, we show how to formulate a tractable convex relaxation of the set of feasible local density matrices of a many-body quantum system. The relaxation is obtained by introducing a hierarchy of constraints between the reduced states of ever-growing sets of lattice sites. The coarse-graining maps of the underlying renormalization procedure serve to eliminate a vast number of those constraints, such that the remaining ones can be enforced with reasonable computational means. This can be used to obtain rigorous lower bounds on the ground state energy of arbitrary local Hamiltonians, by performing a linear optimization over the resulting convex relaxation of reduced quantum states. The quality of the bounds crucially depends on the particular renormalization scheme, which must be tailored to the target Hamiltonian. We apply our method to 1D translation-invariant spin  models, obtaining energy bounds comparable to those attained by optimizing over locally translation-invariant states of $n\gtrsim 100$ spins. Beyond this demonstration, the general method can be applied to a wide range of other problems, such as spin systems in higher spatial dimensions, electronic structure problems, and various other many-body optimization problems, such as entanglement and nonlocality detection. 
\end{abstract}

\maketitle

\section{Introduction}
A central task in quantum theory is the study of quantum many-body systems.
To start, it is the key problem in condensed matter, quantum chemistry, and high
energy physics. Beyond that, the nature of the quantum correlations 
arising in such systems is a central topic in quantum information
(entanglement theory), as well as in quantum foundations, 
where the validity of physical theories is tested through the correlations
they exhibit (such as Bell experiments, or experiments testing
the limits  of quantum theory).
However, solving the quantum many-body problem is extremely
challenging, both analytically and computationally. 
Its hardness stems from the rapid growth of the number of parameters that are required to specify  a state of the system---a vector in an exponentially large Hilbert space---with the number of its constituents. 

Yet, the vast majority of these parameters is not  relevant for the study of any given system. 
At a fundamental level, this is due to \emph{locality}.
In any experiment, we can only probe a small number of  properties, such as expectation values of local   (few-body) operators, correlations, or functions thereof. 
At the same time, the laws governing physical systems---encoded in their Hamiltonians---are also written in terms of local operators.
It is thus sufficient to  deal with local  quantities in order to describe and predict the 
properties of any given system. 
This holds true whether we are studying the ground-state properties of a solid or a molecule, 
designing a measurement protocol to certify the presence of entanglement in a quantum computer, 
or characterizing the correlations that can arise in a physical theory that we aim to test.

It is thus appealing to altogether drop the description of the system in
terms of an exponentially big state vector, and instead characterize quantum many-body systems in terms of their relevant observable correlations---the
local \emph{marginals} of the full distribution.
This can be done, e.g.,  by  working directly with the set of expectation values of the relevant observables, or by formulating the problem in terms of the local reduced density matrices that encode all local expectation values \cite{MazziottiRDMbook,Nakata2,plenio}. 
Clearly, this description is significantly more efficient, as only the few quantities required to predict the properties of interest need to be captured.
However, in this description, a new difficulty arises: The fact that those marginals need to be consistent with a global quantum mechanical wavefunction imposes highly non-trivial constraints on them \cite{KlyachkoQMPreview_2006,pureQMPyu,symmQMPaloy,rdmConsist2006,rdmConsist2020,LiuNrepQMAcomp}.  
The likely simplest such constraint is that for a spin-$1/2$ system, the Pauli expectation values must satisfy
$\langle \sigma_x\rangle^2 + \langle \sigma_y\rangle^2 + \langle
\sigma_z\rangle^2\le 1$, 
but significantly more subtle constraints arise for many-body systems. 
In order to study a system based on its marginals, one therefore must be able to characterize the set of values they can---or cannot---take.

In order to get a better understanding of this phenomenon in the quantum
many-body context,  let us consider a concrete example: the antiferromagnetic Heisenberg spin-$1/2$ chain 
$H=\sum_i (\sigma_x^i\sigma_x^{i+1} + \sigma_y^i\sigma_y^{i+1} + \sigma_z^i\sigma_z^{i+1})/4$. 
Its energy (per site) is fully determined by the $2$-site reduced density matrix $\rho_2$. 
If we only impose that $\rho_2$ itself is physical (i.e.\ positive and normalized) we find that the energy is minimized by the singlet state $\lvert\Psi^-\rangle$, with value $-0.75$. 
The true ground state energy per site is however equal to  $0.25-\log(2)\approx -0.44$. The result we obtained is thus clearly unphysical. 
It requires each spin to be  in a maximally entangled singlet state with both of its neighbors simultaneously, which is well-known to be impossible by the monogamy of entanglement \cite{monogamy_entanglement}.

Remarkably, this simple analysis can still be used to obtain rigorous
estimates for the ground state energy. 
On the one hand, it is clear that when we minimized over all 2-body states in the above, we have disregarded certain constraints on $\rho_2$ which are needed to make it consistent with a global state---i.e.\ we  \emph{relaxed} the problem.
We therefore end up with an energy which is below what can be obtained from any physical wavefunction, that is, a \emph{lower bound} to the true ground state energy. 
At the same time, we can also use the solution above to construct an explicit ansatz wavefunction, by placing the energetically optimal singlet state between consecutive pairs of adjacent spins $\ket{\Psi^-}\otimes\ket{\Psi^-}\otimes\ket{\Psi^-}\otimes\ldots$    (and possibly symmetrizing with respect to translation). 
This is clearly a physical ansatz for the many-body wavefunction, and thus its energy (on average, $-.375$ per site) is above the optimal value---it yields an \emph{upper bound} to the true ground state energy.

\begin{figure}[t]
	\includegraphics[width=1\textwidth]{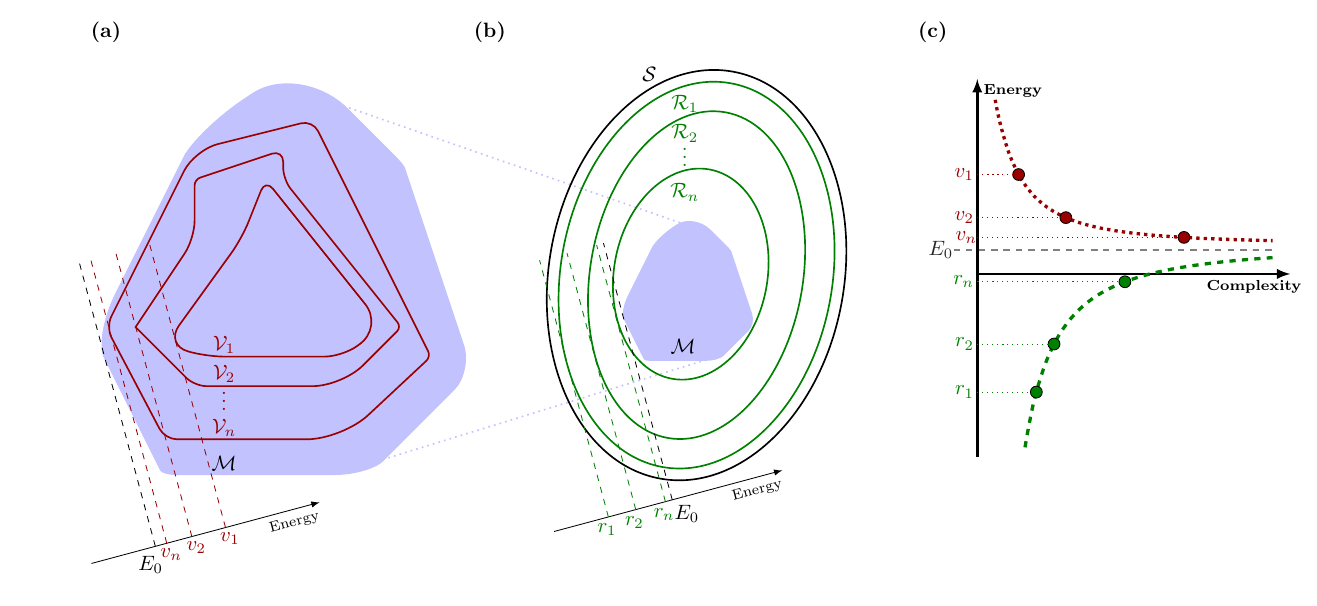}
	\caption{ The set of marginals $\mathcal{M}$ and its inner (\textbf{a}) and outer (\textbf{b}) approximations.
		The set $\mathcal{M}$ consists of those local variables (e.g.\ reduced density matrices on a fixed collection of subsystems) that are compatible with a global quantum state of the whole system, and is therefore a subset of the set of all possible local variables $\mathcal{S}$ (e.g.\ all tuples of positive semidefinite matrices of trace $1$). 
		Finding the ground-state energy $E_0$ of a local Hamiltonian amounts to minimizing a linear function over the set of marginals $\mathcal{M}$.
		This task is intractable because the set is very complex.
		Approximate solutions can be obtained by minimizing the energy over approximations of $\mathcal{M}$:
		Variational ansatz methods (\textbf{a}) minimize the energy function   over inner approximations $\mathcal{V}_k\subset\mathcal{M}$, and produce upper bounds $v_k \geq E_0$. 
		Relaxation methods (\textbf{b}) minimize the energy  over outer approximations  $\mathcal{R}_k\supset\mathcal{M}$, thereby obtaining lower bounds $r_k\leq E_0$.
		Numerical methods  often feature   hierarchies of approximating sets    that provide tighter bounds as the  level $k$ is increased (e.g.\ increasing the number of  ansatz parameters). 
		This   increases the computational complexity of the problem, and the efficiency of such methods  is then determined by the trade-off (\textbf{c}) they exhibit between the precision of the bound and   the computational complexity of the problem that has to be solved to produce it.
	} 
	\label{fig:varVsRelax}
\end{figure}

Both the upper and the lower bound can be systematically improved by
considering blocks of $n$ sites, and minimizing the energy over all
$\rho_n$. 
The resulting optimal energy gives a hierarchy of increasingly
better lower bounds to the ground state energy---known as   Anderson
bounds \cite{Anderson}---while the ansatz constructed from the optimal $\rho_n$
correspondingly gives increasingly better upper bounds. 
Clearly, this analysis is not bound to the Heisenberg model, but can be applied to any other ground state problem. 
Formally, we can understand the two approaches as either enlarging (through relaxation of constraints), or restricting (by specifying  an ansatz), the space of two-body marginals we allow in the optimization, see \cref{fig:varVsRelax} for a detailed discussion.

What is the precision of the energy obtained this way, and how does the
computational cost depend on it? 
The lower and upper bounds differ precisely by the way in which they treat the interaction term across the boundary of $n$-site regions, and thus, the resulting energies per site differ by $O(1/n)$. 
The required computation task, on the other hand, is the diagonalization of an $n$-site system, which scales exponentially in $n$. 
We thus find that the computational cost grows \emph{exponentially} in the desired accuracy---a rather unfavorable scaling. 
This raises the question whether more efficient methods to obtain both upper and lower bounds exist.

For upper bounds, any variational wavefunction gives rise to physically
allowed marginals, and thus to an upper bound for the energy. 
Variational wavefunctions are plentiful, tailored to the physics of the system at hand (such as BCS \cite{BCS}, Laughlin \cite{Laughlin}, or Gutzwiller projected wavefunctions \cite{Gutzwiller63,Gutzwiller65}).
A particularly powerful and versatile class are tensor network
wavefunctions, such as MPS, PEPS \cite{TNrevModPhys}, TTNs \cite{TTNalg}, or MERA \cite{mera}, that can  be understood as variational ansatzes arising from   various renormalization schemes \cite{Cirac_2009}.
Renormalization schemes, by construction, adapt to the given problem in the way in which they select which degrees of freedom to keep, and correspondingly, variational tensor network families come with a high degree of versatility. 
Their ability to adapt to the problem at hand, together with efficient numerical algorithms, has made them the prime tool for the study of many quantum many-body problems, most prominently in one dimension.
Importantly, the computational cost of tensor network algorithms typically  
increases at most \emph{polynomially} with the desired accuracy, 
an exponential improvement over the simple hierarchy introduced above.


For lower bounds, no comparably powerful approaches are known. 
The development of semidefinite programming \cite{VandenbergheAndBoyd}---an optimization framework that allows  positivity constraints to be imposed on matrices---opened the way to formulating relaxations that  significantly improve on the  Anderson-bounds hierarchy. 
Yet such methods are typically formulated by first deriving an exhaustive hierarchy of constraints, and then discarding all constraints that exceed a certain level of complexity (e.g.\ as   in Refs.\ \cite{Mazziotti3pos,plenio,NPA,symmetricExtensions,RDMTvariationalCorrelation}). 
Since the complexity within the hierarchy generally grows exponentially, typically all but 
the lowest-level constraints have to be dropped. 
While for some systems, even few constraints can  already produce useful lower bounds (e.g.\ \cite{plenio,MazziottiRDMbook,entanglementDetectFrerotBaccariAcin}), and cleverly selecting which constraints to keep from the different levels of the hierarchy can lead to improvements
\cite{BarthelAndHubener,TuraAndDunjko,WangRDMT2023},
those methods still exhibit the same exponential growth of the computational cost
with the desired accuracy, and therefore cannot match the performance of state-of-the-art variational methods.  
 
This situation is all the more unsatisfactory given the fundamental
importance of rigorous lower bounds. 
Firstly, of course, they are needed to complement
variationally obtained upper bounds, if one is interested in assigning
rigorous error bounds to numerically obtained estimates. 
However, the importance of lower bounds reaches down to a far more
fundamental level: 
Any attempt at falsifying a probabilistic physical theory---be it 
entanglement witnesses  rejecting a separable (i.e.\ not entangled) description \cite{ent_detect}, Bell
inequality violations ruling out a local hidden variable model \cite{bell_nonlocality}, or experiments designed to detect a breakdown of quantum theory \cite{many_body_quant_non_loc}---requires us to \emph{prove} that the  experimentally measured values lie outside of the set of values  compatible with the theory at hand. 
This is precisely what is achieved by  rigorous \emph{lower bounds} on the minimum value attainable in the compatible set: an observation of a lower value  proves incompatibility.
Given this fundamental importance of precise and rigorous lower bounds, the
exponential scaling and lack of adaptability of the existing methods for
finding such bounds is all the more pressing.

In this paper, we propose a general framework to construct efficient relaxations
of optimization problems for quantum many-body systems, such as the ground
state problem. It yields versatile families of relaxations
which can be continuously tuned and adapted to the specific scenario. Crucially, their 
computation cost 
scales polynomially with the desired accuracy, and thus overcomes
the exponential scaling of existing methods. Its
construction is guided by renormalization, which allows us to use the established knowledge  
on renormalization transformations of quantum systems to closely adapt the relaxation 
to the problem at hand. 
More generally, our framework can be seen as a comprehensive way to
efficiently and systematically relax sets of local many-body
correlations with specific global consistency conditions, and is thus applicable to a wide range of problems,
including tasks such as entanglement or non-locality detection.

The key idea of our method is as follows. We start by expanding the
compatibility constraints on the physical reduced density matrices as a
hierarchy of increasingly complex objects. However, unlike previous approaches, which 
truncate
this hierarchy at some level $n$, we first
\emph{compress} (or ``coarse-grain'') the objects in the hierarchy, such
that each of them becomes of constant (or otherwise tractable) size. This way,
we circumvent the exponential growth of complexity with the level $n$ in the
hierarchy, which allows us to reach significantly larger values of $n$. The
attainable accuracy is thus no longer limited by $n$, but rather by the
chosen compression scheme, including the degree of compression. Unlike the 
static truncation at fixed $n$, this compression can
now be adjusted continuously, and it can be tuned to the problem at hand,
guided e.g.\ by renormalization ideas. This way, the relaxation can be
adjusted to the specific scenario, and much higher accuracies can be
obtained.

We present a detailed realization of our method for the central problem of
finding the ground state energy density of a one-dimensional (1D) spin
chain. We show that, in this case, both Matrix Product States (MPS) and
Tree Tensor Networks (TTN) provide suitable coarse-graining schemes,
corresponding to Wilson's NRG and real-space renormalization, respectively
\cite{Cirac_2009}.
We test the method on a range of paradigmatic spin
chains, both critical and gapped. We find that our relaxation method
yields lower bounds which are between one and two orders of magnitude
better than what could be reached with existing approaches which truncate
the hierarchy without compression  (where for the same accuracy the exact solution of a problem involving on the
order of  $100$ spins would be required).
We find  renormalization to be  a useful  guide for adjusting the   coarse-graining scheme  to the target Hamiltonian, 
that is, that using  MPS or TTN describing  variationally optimized wavefunctions for the compression in our relaxation, performs indeed rather well. Yet, we also demonstrate that
further optimizing over those coarse-graining maps can additionally
enhance the accuracy of the method. 
Our results provide rigorous lower bounds on the energy densities 
of infinite spin chains, and significantly improve upon lower bounds 
previously obtained for such systems using reduced-density-matrix 
theory and similar approaches~\cite{plenio,BarthelAndHubener,RDMTvariationalCorrelation}.

The method presented can be straightforwardly generalized to ground state
problems in higher dimensions, different geometries, or fermionic systems, by
choosing a suitable renormalization scheme and the corresponding tensor network ansatz.
The general framework underlying the method is much broader, and can be
adapted to handle a diverse set of problems, ranging from electronic
structure problems in condensed matter and quantum chemistry all the way
to entanglement \cite{PeresSeparability,HORODECKI_separability,Gurvits,entanglement_polytopes_Walter,Sawicki}
and nonlocality detection
\cite{Tura_2014,Tura_2017,Fadel_2017,Baccari_2019,Fr_rot_2021} in
many-body systems, as well as other situations where the positivity of
intractably large matrices, subject to semidefinite constraints, has to be ensured \cite{bootstrapRandomMatrix,bootstrapMatrixQM,bootstrapQM}.
\\

The presentation of the material in the paper aims to equally bridge two
aspects: On the one hand, a comprehensive presentation of the general
method, and on the other hand, a tangible discussion of its
application to 1D spin systems. 
In order to enhance the accessibility of the manuscript to the
reader, we have thus structured it as follows:
\begin{itemize}
    \item \cref{sec:summary} 
    provides a \emph{concise summary of the
    method}: It derives the method specifically for the case of 1D spin chains with MPS as coarse-graining maps, and provides a brief summary of the central numerical findings. 
    It is \emph{self-contained and can be read as an ``executive summary''} of our work. 
    After having read \cref{sec:summary}, the reader should be able to continue directly with those sections which interest them the most.
    
    \item \cref{sec:genMethod} 
    introduces the general  framework and demonstrates its range of applicability. 
    The general formulation is presented in \cref{sec:genMethodDeriv}. It is set up such that it can be particularized to a wide range of scenarios, as well as to various underlying renormalization schemes.
    Specific realizations of the general framework then follow: 
    \begin{itemize}
        \item In \cref{sec:TTN} it is shown how a   renormalization procedure based on tree tensor networks gives rise to yet another   family of relaxations for 1D systems, distinct from the one derived in \cref{sec:summary}.
        \item The application of the method to   lattice  spin systems in higher dimensions is discussed in  \cref{sec:2D}.
        \item \cref{sec:RDMT} shows how our    approach   can be implemented within the reduced-density-matrix theory framework.
        \item \Cref{sec:quantumInfo} discusses how our approach can be applied to the problems of  non-locality and entanglement detection.
       
    \end{itemize}   
     
    \item \cref{sec:numericalStuff} 
    deals with specialized aspects pertaining to our numerical implementation, namely how to optimize coarse-grainers and how to certify solutions.
    \item \cref{sec:results} reports our numerical findings for 1D spin chains, and discusses the results and the methods used in detail.
\end{itemize}
Finally, in \cref{sec:conclusion} we present our conclusions.

\section{A case study:\  1D  translation-invariant Hamiltonians }
\label{sec:summary}
In this section, we will introduce our method and illustrate its key ideas by means of a concrete problem. 
Specifically, we will consider the paradigmatic 
problem of finding the minimum energy density of a  translation-invariant local Hamiltonian 
on a one-dimensional (1D) spin chain, also known as the 1D translation-invariant local Hamiltonian problem.
We will see that, in this scenario, a natural choice of coarse-graining maps to relax the problem will be given by translation-invariant matrix product states (MPS). Remarkably, the same MPS that have been variationally optimized to upper bound the ground state energy turn out to perform very well as coarse-graining maps for the lower bound as well. We conclude the section by providing numerical results that demonstrate the significant improvement in accuracy that can be obtained through our method.

\subsection{Setting up the problem as a hierarchy of constraints}

We start by rephrasing the ground state energy problem as a hierarchy of semidefinite constraints.
The problem that we wish to solve is finding the minimum energy density of a  translation-invariant Hamiltonian consisting of  an identical nearest-neighbors interaction term $h$, acting on every pair of consecutive  spins on an infinite chain. 
As the Hamiltonian is translation invariant, we can restrict our minimization to states that have the same symmetry. 
Because all the interaction terms are identical, the energy density can be evaluated in terms of the 2-body reduced density matrix, $\rhoM{2}$. To obtain the true ground state energy density of the infinite system we need to restrict $\rhoM{2}$ to be compatible with a global translation-invariant  (TI) state. We formally represent this condition by $\rhoM{2}\leftarrow \psi_{\text{TI}}$. 
(We do  not  elaborate on   what it means to be a state on the infinite chain  as from now on  we will only use   rather obvious properties of the reduced states of $\psi_{\text{TI}}$ on finite regions.)
With this notation the problem reads 
\begin{align}
	\label{energy}
		E_{\text{TI}}:=		\min_{\rhoM{2},\psi_{\text{TI}}}  & \tr\left( h \rhoM{2} \right) \\ \nonumber
		\mbox{s.t. } & \rhoM{2} \geq 0, \; \tr (\rhoM{2})=1 \; , \\
		& \rhoM{2} \leftarrow \psi_{\text{TI}}  \;. \nonumber
\end{align}

The existence of a global state $\psi_{\text{TI}}$ implies that all its $m$-body reduced states $\rhoM{m}$ also  exist and    are compatible with each other in the sense that $\rhoM{m-1}$ is obtained from $\rhoM{m} $ by tracing out one spin.
Since $\psi_{\text{TI}}$ is assumed to be translation invariant, tracing out the leftmost spin gives the same reduced state as tracing out the rightmost one. 
We   denote this compatibility condition as
\begin{equation}
	\label{eq:LTIconditionPRE}
	\rhoM{m-1} \leftarrow  \rhoM{m} 
	\hspace{10pt}\Leftrightarrow	\hspace{10pt}
	\rhoM{m-1} = \trL(\rhoM{m}) = \trR(\rhoM{m}) \; ,  
\end{equation}
where $\trL$ denotes the partial trace of the leftmost spin (and similarly $\trR$).
We call a state $\rhoM{m}$ with this property \emph{locally translation invariant} (LTI):
\begin{align}
	\label{eq:LTIcondition}
	 \trL(\rhoM{m}) &= \trR(\rhoM{m})  & & \text{(The LTI condition.)}
\end{align}

We now rewrite \cref{energy} by explicitly including the compatibility conditions between the reduced states $\rhoM{m}$ of $\psi_{\text{TI}}$ for $m\in\{2,\ldots,n\}$ in the constraints as follows
\begin{equation}
	\label{energyExpand}
	\setlength{\jot}{1pt}
	\begin{split}
		E_{\text{TI}}=		\min_{\{\rhoM{m}\},\psi_{\text{TI}}}  & \tr\left( h \rhoM{2} \right)\\[2pt]
		\mbox{s.t. } & \tr (\rhoM{2})=1, \; \rhoM{m} \geq 0  \mbox{ for all }m \in \{2,\ldots,n\} \; ,\\
		& \rhoM{2} \leftarrow  \rhoM{3} \leftarrow  \rhoM{4}\leftarrow \ldots \leftarrow  \rhoM{n-1} \leftarrow  \rhoM{n}   \leftarrow \psi_{\text{TI}} \; .
	\end{split}
\end{equation} 

\subsection{First relaxation: truncating the hierarchy}
Next we relax \cref{energyExpand} by dropping the condition requiring  the existence of  $\psi_{\text{TI}}$ but keeping all the other constraints. That is, we require that the reduced states $\rhoM{m}$ for $m=\{2,\ldots,n\}$ exist and are compatible in the sense of \cref{eq:LTIconditionPRE}. In particular this implies  that $\rhoM{n}$ is LTI. 
We refer to the resulting problem as the LTI problem of size $n$ and its solution as $E_{\text{LTI}}(n)$:
\begin{equation}
	\label{eq:locTIn}  
	\setlength{\jot}{1pt}
	\begin{split}
		E_{\text{LTI}}(n):=		\min_{\{\rhoM{m}\}}  & \tr\left( h \rhoM{2} \right)\\[2pt]
		\mbox{s.t. } & \tr (\rhoM{2})=1 \; , \\[2pt]
		& \rhoM{m} \geq 0 \; , \mbox{ for all }m \in \{2,\ldots,n\} \; ,\\
		& \rhoM{m-1}= \trL(\rhoM{m}) =\trR(\rhoM{m}) \; , \mbox{ for all }m \in \{3,\ldots,n\} \; .
	\end{split}
\end{equation}

By removing the constraint 
$ \rhoM{n}   \leftarrow \psi_{\text{TI}}$
 we allow more states $\{\rhoM{m}\}$ to be considered in the minimization and therefore obtain a lower bound, i.e.\ $E_{\text{TI}}\geq E_{\text{LTI}}(n)$ for all $n\geq2$. 
The resulting optimization problem, \cref{eq:locTIn}, involves matrices of finite sizes and can be solved numerically using semidefinite programming (SDP) \cite{VandenbergheAndBoyd}.
For an LTI state $\rhoM{n+1}$, \cref{eq:LTIcondition} implies that $\rhoM{n}$ is also LTI, we therefore have that the sequence $\{ E_{\text{LTI}}(n)\}_n$ is non-decreasing.  
Furthermore, it converges to the exact energy density in the limit $n\rightarrow \infty$.

The analysis so far suggests the following strategy for solving \cref{energy}:\ Solve   \cref{eq:locTIn} while keeping as many states $\rhoM{m}$ as we can fit in memory, up to some $m_{\text{max}}$.
Unfortunately, this approach does not  lead to accurate lower bounds because the resources needed to solve the resulting SDP scale exponentially with $m_{\text{max}}$.

\subsection{Second relaxation: applying coarse-graining to the constraints }
\label{sec:summary_method1}
So far we have shown that the LTI problem,  \cref{eq:locTIn}, is a relaxation of the  translation-invariant local  Hamiltonian problem, \cref{energy}. 
We also pointed out that to obtain tight lower bounds one has to solve the LTI problem for   $n\gg1$, which is not a tractable task due to the exponentially large state $\rhoM{n}$ appearing in \cref{eq:locTIn}. 
To overcome this impasse we suggest to relax the hierarchy of constraints appearing in the LTI problem, which we have formally written as 
$ \rhoM{2} \leftarrow  \rhoM{3} \leftarrow \ldots \leftarrow \rhoM{n} $,
in a smarter way than just truncating it at a low level $m_{\text{max}}\ll n$.
In particular, we would like to keep \emph{some} of the constraints arising due to the state $\rhoM{n}$ rather than discard it completely.
To achieve this we apply a renormalization procedure to the states $\rhoM{m}$.
We now explain how  one constraint of the form  $\rhoM{m-1} \leftarrow  \rhoM{m} $ can be relaxed and compressed by applying a coarse-graining map.
In the next subsection we explain how this can be iterated to relax and compress the entire LTI problem, \cref{eq:locTIn}.

To explain the following steps we   rely on graphical tensor notation, in which  the constraints $\rhoM{m-1}= \trL(\rhoM{m}) =\trR(\rhoM{m})$ appearing in \cref{eq:locTIn} are written as follows (we focus on $m=3,4,5$):
\begin{equation}
\label{eq:TNconstraints1}
\setlength{\jot}{0pt}
\begin{split}
 \tikzPic{sigma3 } &= 	 \tikzPic{TLSigma4 } \\
 \tikzPic{sigma3 } &= 	 \tikzPic{TRSigma4 } \\
 \tikzPic{sigma4 }  &= 	 \tikzPic{TLSigma5 }   \\
 \tikzPic{sigma4 } & =	 \tikzPic{TRSigma5 }  \; ,
\end{split}
\end{equation}
where the partial trace is represented by connecting the corresponding legs of the tensor. 

Now consider the first two rows in \cref{eq:TNconstraints1}. The following steps form the elementary building block of our procedure and are described in \cref{eq:TNoneStepRelaxation}.
First we apply to both sides of each equation a coarse-graining map  $W_2:\C^d\otimes\C^d\rightarrow\C^\chi$, which maps two spins of dimension $d$ into one new site of dimension $\chi$,   such that in both rows the two central spins of $\rhoM{4}$ are acted upon, leading to the  equations in the center of \cref{eq:TNoneStepRelaxation}. 
\begin{equation} 
	\label{eq:TNoneStepRelaxation}
	\begin{split}
		\begin{tikzpicture}[baseline=-1mm, scale=0.8 ,every node/.style={scale=0.8}]
			\tikzmath{ \up=0.4;  \step=0.5;   \n=2; \s=0.25;} 
			\coordinate (R) at (0,0);
			\coordinate (L) at ($ (R) - \n*(\step,0)$);
			\foreach \x in {0,...,\n}
			\draw[-] ($(L)+ (\x*\step,\up)$) -- ($(L)+ (\x*\step,-\up)$) ;
			\draw[fill=white,rounded corners=3pt] ($ (L) - (\s,\s) $) rectangle ($ (R) + (\s,\s) $) ;
			\node at ($ 0.5*(R)+0.5*(L) $) {\small$\rhoM{3}$};
		\end{tikzpicture}
		&=
		\begin{tikzpicture}[baseline=-1mm, scale=0.8 ,every node/.style={scale=0.8}]
			\tikzmath{ \up=0.4;  \step=0.5;   \n=3; \s=0.25;} 
			\coordinate (R) at (0,0);
			\coordinate (L) at ($ (R) - \n*(\step,0)$);
			\draw[rounded corners=3pt,white] ($(L)+(0,\up)$) -- ($(L)+(-\step,\up)$) --  ($(L)+(-\step,-\up)$) -- ($(L)+(0,-\up)$) --cycle;
			\foreach \x in {0,...,2}
			\draw[-] ($(L)+ (\x*\step,\up)$) -- ($(L)+ (\x*\step,-\up)$) ;
			\draw[rounded corners=3pt] ($(R)+(0,\up)$) -- ($(R)+(\step,\up)$) --  ($(R)+(\step,-\up)$) -- ($(R)+(0,-\up)$) -- cycle ;
			\draw[fill=white,rounded corners=3pt] ($ (L) - (\s,\s) $) rectangle ($ (R) + (\s,\s) $) ;
			\node at ($ 0.5*(R)+0.5*(L) $) {\small$\rhoM{4}$};
		\end{tikzpicture}
		\\[5pt]
		\begin{tikzpicture}[baseline=-1mm, scale=0.8 ,every node/.style={scale=0.8}]
			\tikzmath{ \up=0.4;  \step=0.5;   \n=2; \s=0.25;} 
			\coordinate (R) at (0,0);
			\coordinate (L) at ($ (R) - \n*(\step,0)$);
			\foreach \x in {0,...,\n}
			\draw[-] ($(L)+ (\x*\step,\up)$) -- ($(L)+ (\x*\step,-\up)$) ;
			\draw[fill=white,rounded corners=3pt] ($ (L) - (\s,\s) $) rectangle ($ (R) + (\s,\s) $) ;
			\node at ($ 0.5*(R)+0.5*(L) $) {\small$\rhoM{3}$};
		\end{tikzpicture}
		&=
		\begin{tikzpicture}[baseline=-1mm, scale=0.8 ,every node/.style={scale=0.8}]
			\tikzmath{ \up=0.4;  \step=0.5;   \n=3; \s=0.25;} 
			\coordinate (R) at (0,0);
			\coordinate (L) at ($ (R) - \n*(\step,0)$);
			\foreach \x in {1,...,\n}
			\draw[-] ($(L)+ (\x*\step,\up)$) -- ($(L)+ (\x*\step,-\up)$) ;
			\draw[rounded corners=3pt] ($(L)+(0,\up)$) -- ($(L)+(-\step,\up)$) --  ($(L)+(-\step,-\up)$) -- ($(L)+(0,-\up)$) -- cycle;
			\draw[fill=white,rounded corners=3pt] ($ (L) - (\s,\s) $) rectangle ($ (R) + (\s,\s) $) ;
			\node at ($ 0.5*(R)+0.5*(L) $) {\small$\rhoM{4}$};
		\end{tikzpicture} 
	\end{split}
	\xlongrightarrow[\text{grain}]{\text{Coarse-}}
	\begin{split}
		\begin{tikzpicture}[baseline=-1mm, scale=0.8 ,every node/.style={scale=0.8}]
			\tikzmath{ \up=0.4;  \step=0.5;   \n=2; \s=0.25;\W=0.35;} 
			\coordinate (R) at (0,0);
			\coordinate (L) at ($ (R) - \n*(\step,0)$);
			\foreach \x in {0,...,\n}
			\draw[-] ($(L)+ (\x*\step,\up)$) -- ($(L)+ (\x*\step,-\up)$) ;
			\draw[fill=white,rounded corners=3pt] ($ (L) - (\s,\s) $) rectangle ($ (R) + (\s,\s) $) ;
			\node at ($ 0.5*(R)+0.5*(L) $) {\small$\rhoM{3}$};
			\draw[line width=0.5mm] ($(L)+ (\step+\s,\up)$) -- ($(L)+ (\step+\s,\up+\s+\W)$) ;
			\draw[line width=0.5mm] ($(L)+ (\step+\s,-\up)$) -- ($(L)+ (\step+\s,-\up-\s-\W)$) ;
			\draw[fill=white,rounded corners=3pt] ($ (L)+(\step-\s,\up) $) rectangle ($ (R) + (\s,\up)+ (0,\W) $) ;
			\draw[fill=white,rounded corners=3pt] ($ (L)+(\step-\s,-\up-\W) $) rectangle ($ (R) + (\s,-\up) $) ;
			\node at ($(L)+ (\step+\s,\up+0.5*\W)$) {\tiny $W_2$}; 
			\node at ($ (L)+(\step+\s,-\up-0.5*\W) $) {\tiny $W_2^\dagger$};
		\end{tikzpicture}
		&=
		\begin{tikzpicture}[baseline=-1mm, scale=0.8 ,every node/.style={scale=0.8}]
			\tikzmath{ \up=0.5;  \step=0.5;   \n=3; \s=0.25; \W=0.35; \m=.11;} 
			\coordinate (R) at (0,0);
			\coordinate (L) at ($ (R) - \n*(\step,0)$);
			\draw[rounded corners=3pt,white] ($(L)+(0,\up)$) -- ($(L)+(-\step,\up)$) --  ($(L)+(-\step,-\up)$) -- ($(L)+(0,-\up)$) --cycle;
			\foreach \x in {0,...,2}
			\draw[-] ($(L)+ (\x*\step,\up)$) -- ($(L)+ (\x*\step,-\up)$) ;
			\draw[rounded corners=3pt] ($(R)+(0,\up)$) -- ($(R)+(\step,\up)$) --  ($(R)+(\step,-\up)$) -- ($(R)+(0,-\up)$) -- cycle ;
			\draw[fill=white,rounded corners=3pt] ($ (L) - (\s,\s) $) rectangle ($ (R) + (\s,\s) $) ;
			\node at ($ 0.5*(R)+0.5*(L) $) {\small$\rhoM{4}$};
			\draw[line width=0.5mm] ($(L)+ (\step+\s,\up)$) -- ($(L)+ (\step+\s,\up+\s+\W)$) ;
			\draw[line width=0.5mm] ($(L)+ (\step+\s,-\up)$) -- ($(L)+ (\step+\s,-\up-\s-\W)$) ;
			\draw[fill=white,rounded corners=3pt] ($ (L)+(\step-\s,\up) $) rectangle ($ (R) + (-\step+\s,\up)+ (0,\W) $) ;
			\draw[fill=white,rounded corners=3pt] ($ (L)+(\step-\s,-\up-\W) $) rectangle ($ (R) + (-\step+\s,-\up) $) ;
			\node at ($(L)+ (\step+\s,\up+0.5*\W)$) {\tiny $W_2$}; 
			\node at ($ (L)+(\step+\s,-\up-0.5*\W) $) {\tiny $W_2^\dagger$};
    \draw[rounded corners=3pt, red, dotted] 
            ($ (L) + (-\s-\m,\s+\m ) $) --
            ($ (L) + (\step-\s-\m,\s+\m ) $) -- 
            ($ (L) + (\step-\s-\m,\up+\W+\m) $) -- 
            ($ (R) + (-\step+\s+\m,\up+\W+\m) $) --
            ($ (R) + (-\step+\s+\m,\s+\m) $) --
            ($ (R) + (\s+\m,\s+\m ) $) -- 
            ($ (R) + (+\s+\m,-\s-\m ) $) --
            ($ (R) + (-\step+\s+\m,-\s-\m) $) --
            ($ (R) + (-\step+\s+\m,-\up-\W-\m) $) --
            ($ (L) + (\step-\s-\m,-\up-\W-\m) $) -- 
            ($ (L) + (\step-\s-\m,-\s-\m) $) -- 
            ($ (L) + (-\s-\m,-\s-\m ) $) 
             -- cycle;
		\end{tikzpicture} 
		\\[5pt]
		\begin{tikzpicture}[baseline=-1mm, scale=0.8 ,every node/.style={scale=0.8}]
			\tikzmath{ \up=0.4;  \step=0.5;   \n=2; \s=0.25;\W=0.35;} 
			\coordinate (R) at (0,0);
			\coordinate (L) at ($ (R) - \n*(\step,0)$);
			\foreach \x in {0,...,\n}
			\draw[-] ($(L)+ (\x*\step,\up)$) -- ($(L)+ (\x*\step,-\up)$) ;
			\draw[fill=white,rounded corners=3pt] ($ (L) - (\s,\s) $) rectangle ($ (R) + (\s,\s) $) ;
			\node at ($ 0.5*(R)+0.5*(L) $) {\small$\rhoM{3}$};
			\draw[line width=0.5mm] ($(L)+ ( +\s,\up)$) -- ($(L)+ (+\s,\up+\s+\W)$) ;
			\draw[line width=0.5mm] ($(L)+ ( +\s,-\up)$) -- ($(L)+ (+\s,-\up-\s-\W)$) ;
			\draw[fill=white,rounded corners=3pt] ($ (L)+(-\s,\up) $) rectangle ($ (R) + (-\step+\s,\up)+ (0,\W) $) ;
			\draw[fill=white,rounded corners=3pt] ($ (L)+(-\s,-\up-\W) $) rectangle ($ (R) + (-\step+\s,-\up) $) ;
			\node at ($(L)+ (\s,\up+0.5*\W)$) {\tiny $W_2$}; 
			\node at ($ (L)+(\s,-\up-0.5*\W) $) {\tiny $W_2^\dagger$};
		\end{tikzpicture}
		&=
		\begin{tikzpicture}[baseline=-1mm, scale=0.8 ,every node/.style={scale=0.8}]
			\tikzmath{ \up=0.5;  \step=0.5;   \n=3; \s=0.25;\W=0.35; \m=.11;} 
			\coordinate (R) at (0,0);
			\coordinate (L) at ($ (R) - \n*(\step,0)$);
			\foreach \x in {1,...,\n}
			\draw[-] ($(L)+ (\x*\step,\up)$) -- ($(L)+ (\x*\step,-\up)$) ;
			\draw[rounded corners=3pt] ($(L)+(0,\up)$) -- ($(L)+(-\step,\up)$) --  ($(L)+(-\step,-\up)$) -- ($(L)+(0,-\up)$) -- cycle;
			\draw[fill=white,rounded corners=3pt] ($ (L) - (\s,\s) $) rectangle ($ (R) + (\s,\s) $) ;
			\node at ($ 0.5*(R)+0.5*(L) $) {\small$\rhoM{4}$};
			\draw[line width=0.5mm] ($(L)+ (\step+\s,\up)$) -- ($(L)+ (\step+\s,\up+\s+\W)$) ;
			\draw[line width=0.5mm] ($(L)+ (\step+\s,-\up)$) -- ($(L)+ (\step+\s,-\up-\s-\W)$) ;
			\draw[fill=white,rounded corners=3pt] ($ (L)+(\step-\s,\up) $) rectangle ($ (R) + (-\step+\s,\up)+ (0,\W) $) ;
			\draw[fill=white,rounded corners=3pt] ($ (L)+(\step-\s,-\up-\W) $) rectangle ($ (R) + (-\step+\s,-\up) $) ;
			\node at ($(L)+ (\step+\s,\up+0.5*\W)$) {\tiny $W_2$}; 
			\node at ($ (L)+(\step+\s,-\up-0.5*\W) $) {\tiny $W_2^\dagger$};
    \draw[rounded corners=3pt, red, dotted] 
            ($ (L) + (-\s-\m,\s+\m ) $) --
            ($ (L) + (\step-\s-\m,\s+\m ) $) -- 
            ($ (L) + (\step-\s-\m,\up+\W+\m) $) -- 
            ($ (R) + (-\step+\s+\m,\up+\W+\m) $) --
            ($ (R) + (-\step+\s+\m,\s+\m) $) --
            ($ (R) + (\s+\m,\s+\m ) $) -- 
            ($ (R) + (+\s+\m,-\s-\m ) $) --
            ($ (R) + (-\step+\s+\m,-\s-\m) $) --
            ($ (R) + (-\step+\s+\m,-\up-\W-\m) $) --
            ($ (L) + (\step-\s-\m,-\up-\W-\m) $) -- 
            ($ (L) + (\step-\s-\m,-\s-\m) $) -- 
            ($ (L) + (-\s-\m,-\s-\m ) $) 
             -- cycle;
		\end{tikzpicture} 
	\end{split} 
	\xlongrightarrow[C_2(\rhoM{4})\hookrightarrow\omegaM{4}]{\text{Compress}}
	\begin{split}
		\begin{tikzpicture}[baseline=-1mm, scale=0.8 ,every node/.style={scale=0.8}]
			\tikzmath{ \up=0.4;  \step=0.5;   \n=2; \s=0.25;\W=0.35;} 
			\coordinate (R) at (0,0);
			\coordinate (L) at ($ (R) - \n*(\step,0)$);
			\foreach \x in {0,...,\n}
			\draw[-] ($(L)+ (\x*\step,\up)$) -- ($(L)+ (\x*\step,-\up)$) ;
			\draw[fill=white,rounded corners=3pt] ($ (L) - (\s,\s) $) rectangle ($ (R) + (\s,\s) $) ;
			\node at ($ 0.5*(R)+0.5*(L) $) {\small$\rhoM{3}$};
			\draw[line width=0.5mm] ($(L)+ (\step+\s,\up)$) -- ($(L)+ (\step+\s,\up+\s+\W)$) ;
			\draw[line width=0.5mm] ($(L)+ (\step+\s,-\up)$) -- ($(L)+ (\step+\s,-\up-\s-\W)$) ;
			\draw[fill=white,rounded corners=3pt] ($ (L)+(\step-\s,\up) $) rectangle ($ (R) + (\s,\up)+ (0,\W) $) ;
			\draw[fill=white,rounded corners=3pt] ($ (L)+(\step-\s,-\up-\W) $) rectangle ($ (R) + (\s,-\up) $) ;
			\node at ($(L)+ (\step+\s,\up+0.5*\W)$) {\tiny $W_2$}; 
			\node at ($ (L)+(\step+\s,-\up-0.5*\W) $) {\tiny $W_2^\dagger$};
		\end{tikzpicture}
		&=
		\begin{tikzpicture}[baseline=-1mm, scale=0.8 ,every node/.style={scale=0.8}]
			\tikzmath{ \up=0.5;  \step=0.5;   \n=3; \s=0.25; \m=.11;} 
			\coordinate (R) at (0,0);
			\coordinate (L) at ($ (R) - \n*(\step,0)$);
			\draw[rounded corners=3pt,white] ($(L)+(0,\up)$) -- ($(L)+(-\step,\up)$) --  ($(L)+(-\step,-\up)$) -- ($(L)+(0,-\up)$) --cycle;
			\draw[line width=0.5mm] ($(L)+(\step+\s,-1.3*\up)$) -- ($(L)+ (\step+\s,1.3*\up)$) ;
			\foreach \x in {0}
			\draw[-] ($(L)+ (\x*\step,\up)$) -- ($(L)+ (\x*\step,-\up)$) ;
			\draw[rounded corners=3pt] ($(R)+(0,\up)$) -- ($(R)+(\step,\up)$) --  ($(R)+(\step,-\up)$) -- ($(R)+(0,-\up)$) -- cycle ;
			\draw[fill=white,rounded corners=3pt] ($ (L) - (\s,\s) $) rectangle ($ (R) + (\s,\s) $) ;
           \draw[rounded corners=3pt,red, dotted] ($ (L) - (\s+\m,\s+\m) $) rectangle ($ (R) + (\s+\m,\s+\m) $) ;
			\node at ($ 0.5*(R)+0.5*(L) $) {\small$\omegaM{4}$};
		\end{tikzpicture} 
		\\[5pt]
		\begin{tikzpicture}[baseline=-1mm, scale=0.8 ,every node/.style={scale=0.8}]
			\tikzmath{ \up=0.4;  \step=0.5;   \n=2; \s=0.25;\W=0.35;} 
			\coordinate (R) at (0,0);
			\coordinate (L) at ($ (R) - \n*(\step,0)$);
			\foreach \x in {0,...,\n}
			\draw[-] ($(L)+ (\x*\step,\up)$) -- ($(L)+ (\x*\step,-\up)$) ;
			\draw[fill=white,rounded corners=3pt] ($ (L) - (\s,\s) $) rectangle ($ (R) + (\s,\s) $) ;
			\node at ($ 0.5*(R)+0.5*(L) $) {\small$\rhoM{3}$};
			\draw[line width=0.5mm] ($(L)+ ( +\s,\up)$) -- ($(L)+ (+\s,\up+\s+\W)$) ;
			\draw[line width=0.5mm] ($(L)+ ( +\s,-\up)$) -- ($(L)+ (+\s,-\up-\s-\W)$) ;
			\draw[fill=white,rounded corners=3pt] ($ (L)+(-\s,\up) $) rectangle ($ (R) + (-\step+\s,\up)+ (0,\W) $) ;
			\draw[fill=white,rounded corners=3pt] ($ (L)+(-\s,-\up-\W) $) rectangle ($ (R) + (-\step+\s,-\up) $) ;
			\node at ($(L)+ (\s,\up+0.5*\W)$) {\tiny $W_2$}; 
			\node at ($ (L)+(\s,-\up-0.5*\W) $) {\tiny $W_2^\dagger$};
		\end{tikzpicture}
		&= 
		\begin{tikzpicture}[baseline=-1mm, scale=0.8 ,every node/.style={scale=0.8}]
			\tikzmath{ \up=0.5;  \step=0.5;   \n=3; \s=0.25; \m=.11;} 
			\coordinate (R) at (0,0);
			\coordinate (L) at ($ (R) - \n*(\step,0)$);
			\draw[line width=0.5mm] ($(L)+(\step+\s,-1.3*\up)$) -- ($(L)+ (\step+\s,1.3*\up)$) ;
			\foreach \x in {\n}
			\draw[-] ($(L)+ (\x*\step,\up)$) -- ($(L)+ (\x*\step,-\up)$) ;
			\draw[rounded corners=3pt] ($(L)+(0,\up)$) -- ($(L)+(-\step,\up)$) --  ($(L)+(-\step,-\up)$) -- ($(L)+(0,-\up)$) -- cycle;
			\draw[fill=white,rounded corners=3pt] ($ (L) - (\s,\s) $) rectangle ($ (R) + (\s,\s) $) ;
            \draw[rounded corners=3pt,red, dotted] ($ (L) - (\s+\m,\s+\m) $) rectangle ($ (R) + (\s+\m,\s+\m) $) ;
            \node at ($ 0.5*(R)+0.5*(L) $) {\small$\omegaM{4}$};
   		\end{tikzpicture} 
	\end{split} 
\end{equation}
(In the diagrams $d$ is indicated by thin lines and $\chi$ by thick ones.) 
Next, we can \emph{compress} the coarse-grained relations, as described by the second arrow in  \cref{eq:TNoneStepRelaxation}: 
We notice that after the coarse-graining,  $\rhoM{3}$ is related  only to  the image of $\rhoM{4}$ under the coarse-graining map   $C_2(\rhoM{4})=(\id\otimes W_2 \otimes \id )\rhoM{4}(\id\otimes W^\dagger_2 \otimes \id )$. 
We can therefore encode the coarse-grained constraints using a smaller-dimensional variable $\omegaM{4}$. 

What we have achieved  is to formulate a relaxation of the constraints on $\rhoM{3}$ in terms of a smaller variable. In the example we just described, the size reduction was not so significant because we only coarse-grained two spins. However, by iterating the above coarse-graining and compression steps  as we will now explain, we can drastically reduce the overall size of the problem.

\subsection{Iterative coarse-graining: a renormalization procedure} 
\label{sec:summary_method2}
The next step is to iterate the above described coarse-graining and compression   in order  to coarse-grain all the states $\rhoM{m}$ up to $m=n$ for large values of $n$. 
To do so we require the coarse-graining maps   to satisfy  certain compatibility conditions that we shall now explain.
Consider the constraints involving $\rhoM{4}$ and $\rhoM{5}$ (the last two rows in \cref{eq:TNconstraints1}). 
In order to compress $\rhoM{5}$  we need to apply a 3-body coarse-graining map $W_3$. We would  like to have the following:
\begin{equation} 
	\label{eq:TNrho5constraints}
	\begin{split}
		\begin{tikzpicture}[baseline=-1mm, scale=0.8 ,every node/.style={scale=0.8}]
			\tikzmath{ \up=0.4;  \step=0.5;   \n=3; \s=0.25;\W=0.35;} 
			\coordinate (R) at (0,0);
			\coordinate (L) at ($ (R) - \n*(\step,0)$);
			\foreach \x in {0,...,\n}
			\draw[-] ($(L)+ (\x*\step,\up)$) -- ($(L)+ (\x*\step,-\up)$) ;
			\draw[fill=white,rounded corners=3pt] ($ (L) - (\s,\s) $) rectangle ($ (R) + (\s,\s) $) ;
			\node at ($ 0.5*(R)+0.5*(L) $) {\small$\rhoM{4}$};
			\draw[line width=0.5mm] ($(L)+ (\step+2*\s,\up)$) -- ($(L)+ (\step+2*\s,\up+\s+\W)$) ;
			\draw[line width=0.5mm] ($(L)+ (\step+2*\s,-\up)$) -- ($(L)+ (\step+2*\s,-\up-\s-\W)$) ;
			\draw[fill=white,rounded corners=3pt] ($ (L)+(\step-\s,\up) $) rectangle ($ (R) + (\s,\up)+ (0,\W) $) ;
			\draw[fill=white,rounded corners=3pt] ($ (L)+(\step-\s,-\up-\W) $) rectangle ($ (R) + (\s,-\up) $) ;
			\node at ($(L)+ (\step+2*\s,\up+0.5*\W)$) {\tiny $W_3$}; 
			\node at ($ (L)+(\step+2*\s,-\up-0.5*\W) $) {\tiny $W_3^\dagger$};
		\end{tikzpicture}
		&=
		\begin{tikzpicture}[baseline=-1mm, scale=0.8 ,every node/.style={scale=0.8}]
			\tikzmath{ \up=0.4;  \step=0.5;   \n=4; \s=0.25; \W=0.35;} 
			\coordinate (R) at (0,0);
			\coordinate (L) at ($ (R) - \n*(\step,0)$);
			\draw[rounded corners=3pt,white] ($(L)+(0,\up)$) -- ($(L)+(-\step,\up)$) --  ($(L)+(-\step,-\up)$) -- ($(L)+(0,-\up)$) --cycle;
			\foreach \x in {0,...,3}
			\draw[-] ($(L)+ (\x*\step,\up)$) -- ($(L)+ (\x*\step,-\up)$) ;
			\draw[rounded corners=3pt] ($(R)+(0,\up)$) -- ($(R)+(\step,\up)$) --  ($(R)+(\step,-\up)$) -- ($(R)+(0,-\up)$) -- cycle ;
			\draw[fill=white,rounded corners=3pt] ($ (L) - (\s,\s) $) rectangle ($ (R) + (\s,\s) $) ;
			\node at ($ 0.5*(R)+0.5*(L) $) {\small$\rhoM{5}$};
			\draw[line width=0.5mm] ($(L)+ (\step+2*\s,\up)$) -- ($(L)+ (\step+2*\s,\up+\s+\W)$) ;
			\draw[line width=0.5mm] ($(L)+ (\step+2*\s,-\up)$) -- ($(L)+ (\step+2*\s,-\up-\s-\W)$) ;
			\draw[fill=white,rounded corners=3pt] ($ (L)+(\step-\s,\up) $) rectangle ($ (R) + (-\step+\s,\up)+ (0,\W) $) ;
			\draw[fill=white,rounded corners=3pt] ($ (L)+(\step-\s,-\up-\W) $) rectangle ($ (R) + (-\step+\s,-\up) $) ;
			\node at ($(L)+ (\step+2*\s,\up+0.5*\W)$) {\tiny $W_3$}; 
			\node at ($ (L)+(\step+2*\s,-\up-0.5*\W) $) {\tiny $W_3^\dagger$};
		\end{tikzpicture} 
		\\[5pt]
		\begin{tikzpicture}[baseline=-1mm, scale=0.8 ,every node/.style={scale=0.8}]
			\tikzmath{ \up=0.4;  \step=0.5;   \n=3; \s=0.25; \W=0.35;} 
			\coordinate (R) at (0,0);
			\coordinate (L) at ($ (R) - \n*(\step,0)$);
			\foreach \x in {0,...,\n}
			\draw[-] ($(L)+ (\x*\step,\up)$) -- ($(L)+ (\x*\step,-\up)$) ;
			\draw[fill=white,rounded corners=3pt] ($ (L) - (\s,\s) $) rectangle ($ (R) + (\s,\s) $) ;
			\node at ($ 0.5*(R)+0.5*(L) $) {\small$\rhoM{4}$};
			\draw[line width=0.5mm] ($(L)+ ( +2*\s,\up)$) -- ($(L)+ (+2*\s,\up+\s+\W)$) ;
			\draw[line width=0.5mm] ($(L)+ ( +2*\s,-\up)$) -- ($(L)+ (+2*\s,-\up-\s-\W)$) ;
			\draw[fill=white,rounded corners=3pt] ($ (L)+(-\s,\up) $) rectangle ($ (R) + (-\step+\s,\up)+ (0,\W) $) ;
			\draw[fill=white,rounded corners=3pt] ($ (L)+(-\s,-\up-\W) $) rectangle ($ (R) + (-\step+\s,-\up) $) ;
			\node at ($(L)+ (2*\s,\up+0.5*\W)$) {\tiny $W_3$}; 
			\node at ($ (L)+(2*\s,-\up-0.5*\W) $) {\tiny $W_3^\dagger$};
		\end{tikzpicture}
		&=
		\begin{tikzpicture}[baseline=-1mm, scale=0.8 ,every node/.style={scale=0.8}]
			\tikzmath{ \up=0.4;  \step=0.5;   \n=4; \s=0.25; \W=0.35;} 
			\coordinate (R) at (0,0);
			\coordinate (L) at ($ (R) - \n*(\step,0)$);
			\foreach \x in {1,...,\n}
			\draw[-] ($(L)+ (\x*\step,\up)$) -- ($(L)+ (\x*\step,-\up)$) ;
			\draw[rounded corners=3pt] ($(L)+(0,\up)$) -- ($(L)+(-\step,\up)$) --  ($(L)+(-\step,-\up)$) -- ($(L)+(0,-\up)$) -- cycle;
			\draw[fill=white,rounded corners=3pt] ($ (L) - (\s,\s) $) rectangle ($ (R) + (\s,\s) $) ;
			\node at ($ 0.5*(R)+0.5*(L) $) {\small$\rhoM{5}$};
			\draw[line width=0.5mm] ($(L)+ (\step+2*\s,\up)$) -- ($(L)+ (\step+2*\s,\up+\s+\W)$) ;
			\draw[line width=0.5mm] ($(L)+ (\step+2*\s,-\up)$) -- ($(L)+ (\step+2*\s,-\up-\s-\W)$) ;
			\draw[fill=white,rounded corners=3pt] ($ (L)+(\step-\s,\up) $) rectangle ($ (R) + (-\step+\s,\up)+ (0,\W) $) ;
			\draw[fill=white,rounded corners=3pt] ($ (L)+(\step-\s,-\up-\W) $) rectangle ($ (R) + (-\step+\s,-\up) $) ;
			\node at ($(L)+ (\step+2*\s,\up+0.5*\W)$) {\tiny $W_3$}; 
			\node at ($ (L)+(\step+2*\s,-\up-0.5*\W) $) {\tiny $W_3^\dagger$};
		\end{tikzpicture} \; .
	\end{split} 
\end{equation}
However, we notice that we may no longer write conditions involving $\rhoM{4}$ as we have already compressed it and replaced it with $\omegaM{4}$ when we treated the first two constraints. 
In order to continue we thus need to be able to express the left-hand side of \cref{eq:TNrho5constraints} in terms of $\omegaM{4}$ which appears in   \cref{eq:TNoneStepRelaxation}. This can be done if the coarse-graining maps $W_3$ and $W_2$ are related as follows.
\begin{equation} 
	\label{eq:MPSrelation}
	\begin{tikzpicture}[baseline=5mm, scale=1 ,every node/.style={scale=1}]
		\tikzmath{ \up=0.4;  \step=0.5;   \n=3; \s=0.25;\W=0.35;} 
		\coordinate (R) at (0,0);
		\coordinate (L) at ($ (R) - \n*(\step,0)$);
		\foreach \x in {1,...,\n}
		\draw[-] ($(L)+ (\x*\step,\up)$) -- ($(L)+ (\x*\step,0.5*\up)$) ;
		\draw[line width=0.5mm] ($(L)+ (\step+2*\s,\up)$) -- ($(L)+ (\step+2*\s,\up+\s+\W)$) ;
		\draw[fill=white,rounded corners=3pt] ($ (L)+(\step-\s,\up) $) rectangle ($ (R) + (\s,\up)+ (0,\W) $) ;
		\node at ($(L)+ (\step+2*\s,\up+0.5*\W)$) {\tiny $W_3$}; 
	\end{tikzpicture}
	\hspace{6pt} = \hspace{6pt}
	\begin{tikzpicture}[baseline=5mm, scale=1 ,every node/.style={scale=1}]
		\tikzmath{ \up=0.4;  \step=0.5;   \n=2; \s=0.25;\W=0.35;} 
		\coordinate (R) at (0,0);
		\coordinate (L) at ($ (R) - \n*(\step,0)$);
		\foreach \x in {1,...,\n}
		\draw[-] ($(L)+ (\x*\step,\up)$) -- ($(L)+ (\x*\step,0.5*\up)$) ;
		\draw[line width=0.5mm] ($(L)+ (\step+\s,\up)$) -- ($(L)+ (\step+\s,\up+\s+\W)$) ;
		\draw[fill=white,rounded corners=3pt] ($ (L)+(\step-\s,\up) $) rectangle ($ (R) + (\s,\up)+ (0,\W) $) ;
		\node at ($(L)+ (\step+\s,\up+0.5*\W)$) {\tiny $W_2$}; 
		\foreach \x in {3} {
			\draw[-] ($(L)+ (\x*\step,\up+\s+\W)$) -- ($(L)+ (\x*\step,0.5*\up)$) ;
			\draw[fill=white,rounded corners=3pt]  ($(L)+ (2*\step-2*\s,\up+0.5*\s+\W)$) rectangle  ($(L)+ (\x*\step+\s,\up+0.5*\s+2*\W)$)  ;
		};
		\node at ($(L)+ (2*\step+0.5*\s,\up+0.5*\s+\W+0.5*\W)$)   {\tiny $R_2$}; 
		\draw[line width=0.5mm] ($(L)+ (2*\step+0.5*\s,\up+0.5*\s+2*\W)$) --  ($(L)+ (2*\step+0.5*\s,\up+0.5*\s+2*\W+\s)$) ;
	\end{tikzpicture}
	\hspace{6pt} = \hspace{6pt}
	\begin{tikzpicture}[baseline=5mm, scale=1 ,every node/.style={scale=1}]
		\tikzmath{ \up=0.4;  \step=0.5;   \n=2; \s=0.25;\W=0.35;} 
		\coordinate (R) at (0,0);
		\coordinate (L) at ($ (R) - \n*(\step,0)$);
		\foreach \x in {1,...,\n}
		\draw[-] ($(L)+ (\x*\step,\up)$) -- ($(L)+ (\x*\step,0.5*\up)$) ;
		\draw[line width=0.5mm] ($(L)+ (\step+\s,\up)$) -- ($(L)+ (\step+\s,\up+\s+\W)$) ;
		\draw[fill=white,rounded corners=3pt] ($ (L)+(\step-\s,\up) $) rectangle ($ (R) + (\s,\up)+ (0,\W) $) ;
		\node at ($(L)+ (\step+\s,\up+0.5*\W)$) {\tiny $W_2$}; 
		\foreach \x in {0} {
			\draw[-] ($(L)+ (\x*\step,\up+\s+\W)$) -- ($(L)+ (\x*\step,0.5*\up)$) ;
			\draw[fill=white,rounded corners=3pt]  ($(L)+ (-\s,\up+0.5*\s+\W)$) rectangle  ($(L)+ (\step+2*\s,\up+0.5*\s+2*\W)$)  ;
		};
		\node at ($(L)+ (\step-0.5*\s,\up+0.5*\s+\W+0.5*\W)$)   {\tiny $L_2$}; 
		\draw[line width=0.5mm] ($(L)+ (\step-0.5*\s,\up+0.5*\s+2*\W)$) -- ($(L)+ (\step-0.5*\s,\up+0.5*\s+2*\W+\s)$) ;
	\end{tikzpicture} \; ,
\end{equation}
i.e.\ we require the existence  of two additional coarse-graining maps, $L_2$ and $R_2$, that act on the output of $W_2$ together with  an additional spin on  the left  or right side respectively. 
Given maps that satisfy \cref{eq:MPSrelation} we can substitute $W_3$ in \cref{eq:TNrho5constraints}  with the appropriate expression, such that $W_2$ acts on the two central spins in $\rhoM{4}$, as shown in \cref{eq:TNrho5constraintsSUB}. 
Then the left-hand side of  \cref{eq:TNrho5constraints} can be expressed in terms of $\omegaM{4}$ and $L_2$ (respectively $R_2$), and the right-hand side---in terms of $\omegaM{5}$:
\begin{equation} 
	\label{eq:TNrho5constraintsSUB}
	\begin{split}
		\begin{tikzpicture}[baseline=-1mm, scale=0.8 ,every node/.style={scale=0.8}]
			\tikzmath{ \up=0.5;  \step=0.5;   \n=3; \s=0.25;\W=0.35;\m=.11;} 
			\coordinate (R) at (0,0);
			\coordinate (L) at ($ (R) - \n*(\step,0)$);
			\foreach \x in {0,...,\n}
			\draw[-] ($(L)+ (\x*\step,\up)$) -- ($(L)+ (\x*\step,-\up)$) ;
			\draw[line width=0.5mm] ($(L)+ (\step+\s,\up)$) -- ($(L)+ (\step+\s,\up+\s+\W)$) ;
			\draw[line width=0.5mm] ($(L)+ (\step+\s,-\up)$) -- ($(L)+ (\step+\s,-\up-\s-\W)$) ;
			\draw[fill=white,rounded corners=3pt] ($ (L)+(\step-\s,\up) $) rectangle ($ (L) + (2*\step+\s,\up)+ (0,\W) $) ;
			\draw[fill=white,rounded corners=3pt] ($ (L)+(\step-\s,-\up-\W) $) rectangle ($ (L) + (2*\step+\s,-\up) $) ;
			\node at ($(L)+ (\step+\s,\up+0.5*\W)$) {\tiny $W_2$}; 
			\node at ($ (L)+(\step+\s,-\up-0.5*\W) $) {\tiny $W_2^\dagger$};
			\foreach \x in {3} {
				\draw[-] ($(L)+ (\x*\step,\up+\s+\W)$) -- ($(L)+ (\x*\step,0.5*\up)$) ;
				\draw[fill=white,rounded corners=3pt]  ($(L)+ (2*\step-2*\s,\up+0.75*\s+\W)$) rectangle  ($(L)+ (\x*\step+\s,\up+0.75*\s+2*\W)$)  ;
			};
			\node at ($(L)+ (2*\step+0.5*\s,\up+0.75*\s+\W+0.5*\W)$)   {\tiny $R_2$}; 
			\draw[line width=0.5mm] ($(L)+ (2*\step+0.5*\s,\up+0.75*\s+2*\W)$) --  ($(L)+ (2*\step+0.5*\s,\up+0.75*\s+2*\W+\s)$) ;
			\foreach \x in {3} {
				\draw[-] ($(L)+ (\x*\step,-\up-\s-\W)$) -- ($(L)+ (\x*\step,-0.5*\up)$) ;
				\draw[fill=white,rounded corners=3pt]  ($(L)+ (2*\step-2*\s,-\up-0.75*\s-\W)$) 	rectangle  ($(L)+ (\x*\step+\s,-\up-0.75*\s-2*\W)$)  ;
			};
			\node at ($(L)+ (2*\step+0.5*\s,-\up-0.75*\s-\W-0.5*\W)$)   {\tiny $R_2^\dagger$}; 
			\draw[line width=0.5mm] ($(L)+ (2*\step+0.5*\s,-\up-0.75*\s-2*\W)$) --  ($(L)+(2*\step+0.5*\s,-\up-0.75*\s-2*\W-\s)$) ;
			\draw[fill=white,rounded corners=3pt] ($ (L) - (\s,\s) $) rectangle ($ (R) + (\s,\s) $) ;
			\node at ($ 0.5*(R)+0.5*(L) $) {\small$\rhoM{4}$};
	\draw[rounded corners=3pt, red, dotted] 
            ($ (L) + (-\s-\m,\s+\m ) $) --
            ($ (L) + (\step-\s-\m,\s+\m ) $) -- 
            ($ (L) + (\step-\s-\m,\up+\W+\m) $) -- 
            ($ (R) + (-\step+\s+\m,\up+\W+\m) $) --
            ($ (R) + (-\step+\s+\m,\s+\m) $) --
            ($ (R) + (\s+\m,\s+\m ) $) -- 
            ($ (R) + (+\s+\m,-\s-\m ) $) --
            ($ (R) + (-\step+\s+\m,-\s-\m) $) --
            ($ (R) + (-\step+\s+\m,-\up-\W-\m) $) --
            ($ (L) + (\step-\s-\m,-\up-\W-\m) $) -- 
            ($ (L) + (\step-\s-\m,-\s-\m) $) -- 
            ($ (L) + (-\s-\m,-\s-\m ) $) 
             -- cycle;	
        \end{tikzpicture}
		&=
		\begin{tikzpicture}[baseline=-1mm, scale=0.8 ,every node/.style={scale=0.8}]
			\tikzmath{ \up=0.5;  \step=0.5;   \n=4; \s=0.25; \W=0.35; \m=0.11;} 
			\coordinate (R) at (0,0);
			\coordinate (L) at ($ (R) - \n*(\step,0)$);
			\draw[rounded corners=3pt,white] ($(L)+(0,\up)$) -- ($(L)+(-\step,\up)$) --  ($(L)+(-\step,-\up)$) -- ($(L)+(0,-\up)$) --cycle;
            \foreach \x in {0,...,3}
			\draw[-] ($(L)+ (\x*\step,\up)$) -- ($(L)+ (\x*\step,-\up)$) ;
			\draw[rounded corners=3pt] ($(R)+(0,\up+\m)$) -- ($(R)+(\step,\up+\m)$) --  ($(R)+(\step,-\up-\m)$) -- ($(R)+(0,-\up-\m)$) -- cycle ;
            \draw[fill=white,rounded corners=3pt] ($ (L) - (\s,\s) $) rectangle ($ (R) + (\s,\s) $) ;
			\node at ($ 0.5*(R)+0.5*(L) $) {\small$\rhoM{5}$};
			\draw[line width=0.5mm] ($(L)+ (\step+2*\s,\up)$) -- ($(L)+ (\step+2*\s,\up+\s+\W)$) ;
            \draw[line width=0.5mm] ($(L)+ (\step+2*\s,-\up)$) -- ($(L)+ (\step+2*\s,-\up-\s-\W)$) ;
            \draw[fill=white,rounded corners=3pt] ($ (L)+(\step-\s,\up) $) rectangle ($ (R) + (-\step+\s,\up)+ (0,\W) $) ;
			\draw[fill=white,rounded corners=3pt] ($ (L)+(\step-\s,-\up-\W) $) rectangle ($ (R) + (-\step+\s,-\up) $) ;
			\node at ($(L)+ (\step+2*\s,\up+0.5*\W)$) {\tiny $W_3$}; 
			\node at ($ (L)+(\step+2*\s,-\up-0.5*\W) $) {\tiny $W_3^\dagger$};
            \draw[rounded corners=3pt, green!80!black, ultra thin, dashed] 
            ($ (L) + (-\s-\m,\s+\m ) $) --
            ($ (L) + (\step-\s-\m,\s+\m ) $) -- 
            ($ (L) + (\step-\s-\m,\up+\W+\m) $) -- 
            ($ (R) + (-\step+\s+\m,\up+\W+\m) $) --
            ($ (R) + (-\step+\s+\m,\s+\m) $) --
            ($ (R) + (\s+\m,\s+\m ) $) -- 
            ($ (R) + (+\s+\m,-\s-\m ) $) --
            ($ (R) + (-\step+\s+\m,-\s-\m) $) --
            ($ (R) + (-\step+\s+\m,-\up-\W-\m) $) --
            ($ (L) + (\step-\s-\m,-\up-\W-\m) $) -- 
            ($ (L) + (\step-\s-\m,-\s-\m) $) -- 
            ($ (L) + (-\s-\m,-\s-\m ) $) 
             -- cycle;
		\end{tikzpicture} 
		\\[5pt]
		\begin{tikzpicture}[baseline=-1mm, scale=0.8 ,every node/.style={scale=0.8}]
		\tikzmath{ \up=0.5;  \step=0.5;   \n=3; \s=0.25;\W=0.35; \m=.11;} 
		\coordinate (R) at (0,0);
		\coordinate (L) at ($ (R) - \n*(\step,0)$);
		\foreach \x in {0,...,\n}
		\draw[-] ($(L)+ (\x*\step,\up)$) -- ($(L)+ (\x*\step,-\up)$) ;
		\draw[line width=0.5mm] ($(L)+ (\step+\s,\up)$) -- ($(L)+ (\step+\s,\up+\s+\W)$) ;
		\draw[line width=0.5mm] ($(L)+ (\step+\s,-\up)$) -- ($(L)+ (\step+\s,-\up-\s-\W)$) ;
		\draw[fill=white,rounded corners=3pt] ($ (L)+(\step-\s,\up) $) rectangle ($ (L) + (2*\step+\s,\up)+ (0,\W) $) ;
		\draw[fill=white,rounded corners=3pt] ($ (L)+(\step-\s,-\up-\W) $) rectangle ($ (L) + (2*\step+\s,-\up) $) ;
		\node at ($(L)+ (\step+\s,\up+0.5*\W)$) {\tiny $W_2$}; 
		\node at ($ (L)+(\step+\s,-\up-0.5*\W) $) {\tiny $W_2^\dagger$};
		\foreach \x in {0} {
		\draw[-] ($(L)+ (\x*\step,\up+\s+\W)$) -- ($(L)+ (\x*\step,0.5*\up)$) ;
		\draw[fill=white,rounded corners=3pt]  ($(L)+ (-\s,\up+0.75*\s+\W)$) rectangle  ($(L)+ (\step+2*\s,\up+0.75*\s+2*\W)$)  ;
	};
	\node at ($(L)+ (\step-0.5*\s,\up+0.75*\s+\W+0.5*\W)$)   {\tiny $L_2$}; 
	\draw[line width=0.5mm] ($(L)+ (\step-0.5*\s,\up+0.75*\s+2*\W)$) -- ($(L)+ (\step-0.5*\s,\up+0.75*\s+2*\W+\s)$) ;
		\foreach \x in {0} {
		\draw[-] ($(L)+ (\x*\step,-\up-\s-\W)$) -- ($(L)+ (\x*\step,-0.5*\up)$) ;
		\draw[fill=white,rounded corners=3pt]  ($(L)+ (-\s,-\up-0.75*\s-\W)$) rectangle  ($(L)+ (\step+2*\s,-\up-0.75*\s-2*\W)$)  ;
	};
	\node at ($(L)+ (\step-0.5*\s,-\up-0.75*\s-\W-0.5*\W)$)   {\tiny $L_2^\dagger$}; 
	\draw[line width=0.5mm] ($(L)+ (\step-0.5*\s,-\up-0.75*\s-2*\W)$) -- ($(L)+ (\step-0.5*\s,-\up-0.75*\s-2*\W-\s)$) ;
		\draw[fill=white,rounded corners=3pt] ($ (L) - (\s,\s) $) rectangle ($ (R) + (\s,\s) $) ;
		\node at ($ 0.5*(R)+0.5*(L) $) {\small$\rhoM{4}$};
         \draw[rounded corners=3pt, red, dotted] 
            ($ (L) + (-\s-\m,\s+\m ) $) --
            ($ (L) + (\step-\s-\m,\s+\m ) $) -- 
            ($ (L) + (\step-\s-\m,\up+\W+\m) $) -- 
            ($ (R) + (-\step+\s+\m,\up+\W+\m) $) --
            ($ (R) + (-\step+\s+\m,\s+\m) $) --
            ($ (R) + (\s+\m,\s+\m ) $) -- 
            ($ (R) + (+\s+\m,-\s-\m ) $) --
            ($ (R) + (-\step+\s+\m,-\s-\m) $) --
            ($ (R) + (-\step+\s+\m,-\up-\W-\m) $) --
            ($ (L) + (\step-\s-\m,-\up-\W-\m) $) -- 
            ($ (L) + (\step-\s-\m,-\s-\m) $) -- 
            ($ (L) + (-\s-\m,-\s-\m ) $) 
             -- cycle;
	\end{tikzpicture}
		&=
		\begin{tikzpicture}[baseline=-1mm, scale=0.8 ,every node/.style={scale=0.8}]
			\tikzmath{ \up=0.5;  \step=0.5;   \n=4; \s=0.25; \W=0.35; \m=0.11;} 
			\coordinate (R) at (0,0);
			\coordinate (L) at ($ (R) - \n*(\step,0)$);
			\foreach \x in {1,...,\n}
			\draw[-] ($(L)+ (\x*\step,\up)$) -- ($(L)+ (\x*\step,-\up)$) ;
			\draw[rounded corners=3pt] ($(L)+(0,\up)$) -- ($(L)+(-\step,\up)$) --  ($(L)+(-\step,-\up)$) -- ($(L)+(0,-\up)$) -- cycle;
			\draw[fill=white,rounded corners=3pt] ($ (L) - (\s,\s) $) rectangle ($ (R) + (\s,\s) $) ;
			\node at ($ 0.5*(R)+0.5*(L) $) {\small$\rhoM{5}$};
			\draw[line width=0.5mm] ($(L)+ (\step+2*\s,\up)$) -- ($(L)+ (\step+2*\s,\up+\s+\W)$) ;
			\draw[line width=0.5mm] ($(L)+ (\step+2*\s,-\up)$) -- ($(L)+ (\step+2*\s,-\up-\s-\W)$) ;
			\draw[fill=white,rounded corners=3pt] ($ (L)+(\step-\s,\up) $) rectangle ($ (R) + (-\step+\s,\up)+ (0,\W) $) ;
			\draw[fill=white,rounded corners=3pt] ($ (L)+(\step-\s,-\up-\W) $) rectangle ($ (R) + (-\step+\s,-\up) $) ;
			\node at ($(L)+ (\step+2*\s,\up+0.5*\W)$) {\tiny $W_3$}; 
			\node at ($ (L)+(\step+2*\s,-\up-0.5*\W) $) {\tiny $W_3^\dagger$};
            \draw[rounded corners=3pt, green!80!black, ultra thin, dashed] 
            ($ (L) + (-\s-\m,\s+\m ) $) --
            ($ (L) + (\step-\s-\m,\s+\m ) $) -- 
            ($ (L) + (\step-\s-\m,\up+\W+\m) $) -- 
            ($ (R) + (-\step+\s+\m,\up+\W+\m) $) --
            ($ (R) + (-\step+\s+\m,\s+\m) $) --
            ($ (R) + (\s+\m,\s+\m ) $) -- 
            ($ (R) + (+\s+\m,-\s-\m ) $) --
            ($ (R) + (-\step+\s+\m,-\s-\m) $) --
            ($ (R) + (-\step+\s+\m,-\up-\W-\m) $) --
            ($ (L) + (\step-\s-\m,-\up-\W-\m) $) -- 
            ($ (L) + (\step-\s-\m,-\s-\m) $) -- 
            ($ (L) + (-\s-\m,-\s-\m ) $) 
             -- cycle;
		\end{tikzpicture} 
	\end{split} 
	\xlongrightarrow[C_3(\rhoM{5})\hookrightarrow\omegaM{5}]{C_2(\rhoM{4})\hookrightarrow\omegaM{4}}
	\begin{split}
		\begin{tikzpicture}[baseline=-1mm, scale=0.8 ,every node/.style={scale=0.8}]
			\tikzmath{ \up=0.5;  \step=0.5;   \n=3; \s=0.25;\W=0.35;\m=.11;} 
			\coordinate (R) at (0,0);
			\coordinate (L) at ($ (R) - \n*(\step,0)$);
			\draw[line width=0.5mm,white] ($(L)+ (2*\step+0.5*\s,-\up-0.5*\s-2*\W-\s)$) --  ($(L)+ (2*\step+0.5*\s,\up+0.5*\s+2*\W+\s)$) ;
			\foreach \x in {0,3}
			\draw[-] ($(L)+ (\x*\step,\up)$) -- ($(L)+ (\x*\step,-\up)$) ;
			\draw[line width=0.5mm] ($(L)+ (\step+\s,\up)$) -- ($(L)+ (\step+\s,-\up)$) ;
			\draw[fill=white,rounded corners=3pt] ($ (L) - (\s,\s) $) rectangle ($ (R) + (\s,\s) $) ;
           \draw[rounded corners=3pt,red, dotted] ($ (L) - (\s+\m,\s+\m) $) rectangle ($ (R) + (\s+\m,\s+\m) $) ;
			\node at ($ 0.5*(R)+0.5*(L) $) {\small$\omegaM{4}$};
			\draw[line width=0.5mm] ($(L)+ (2*\step+0.5*\s,\up)$) -- ($(L)+ (2*\step+0.5*\s,\up+\s+\W)$) ;
			\draw[line width=0.5mm] ($(L)+ (2*\step+0.5*\s,-\up)$) -- ($(L)+ (2*\step+0.5*\s,-\up-\s-\W)$) ;
			\draw[fill=white,rounded corners=3pt] ($ (L)+(\step,\up) $) rectangle ($ (R) + (\s,\up)+ (0,\W) $) ;
			\draw[fill=white,rounded corners=3pt] ($ (L)+(\step,-\up-\W) $) rectangle ($ (R) + (\s,-\up) $) ;
			\node at ($(L)+ (2*\step+0.5*\s,\up+0.5*\W)$) {\tiny $R_2$}; 
			\node at ($ (L)+(2*\step+0.5*\s,-\up-0.5*\W) $) {\tiny $R_2^\dagger$};
			
		\end{tikzpicture}
		&=
		\begin{tikzpicture}[baseline=-1mm, scale=0.8 ,every node/.style={scale=0.8}]
			\tikzmath{ \up=0.5;  \step=0.5;   \n=3; \s=0.25; \m=.11;} 
			\coordinate (R) at (0,0);
			\coordinate (L) at ($ (R) - \n*(\step,0)$);
			\draw[rounded corners=3pt,white] ($(L)+(0,\up)$) -- ($(L)+(-\step,\up)$) --  ($(L)+(-\step,-\up)$) -- ($(L)+(0,-\up)$) --cycle;
			\draw[line width=0.5mm] ($(L)+(\step+\s,-1.3*\up)$) -- ($(L)+ (\step+\s,1.3*\up)$) ;
			\foreach \x in {0}
			\draw[-] ($(L)+ (\x*\step,\up)$) -- ($(L)+ (\x*\step,-\up)$) ;
			\draw[rounded corners=3pt] ($(R)+(0,\up)$) -- ($(R)+(\step,\up)$) --  ($(R)+(\step,-\up)$) -- ($(R)+(0,-\up)$) -- cycle ;
			\draw[fill=white,rounded corners=3pt] ($ (L) - (\s,\s) $) rectangle ($ (R) + (\s,\s) $) ;
            \draw[rounded corners=3pt,green!80!black, ultra thin, dashed] ($ (L) - (\s+\m,\s+\m) $) rectangle ($ (R) + (\s+\m,\s+\m) $) ;
			\node at ($ 0.5*(R)+0.5*(L) $) {\small$\omegaM{5}$};
		\end{tikzpicture} 
		\\[5pt]
		\begin{tikzpicture}[baseline=-1mm, scale=0.8 ,every node/.style={scale=0.8}]
		\tikzmath{ \up=0.5;  \step=0.5;   \n=3; \s=0.25;\W=0.35;\m=.11;} 
		\coordinate (R) at (0,0);
		\coordinate (L) at ($ (R) - \n*(\step,0)$);
			\draw[line width=0.5mm,white] ($(L)+ (2*\step+0.5*\s,-\up-0.5*\s-2*\W-\s)$) --  ($(L)+ (2*\step+0.5*\s,\up+0.5*\s+2*\W+\s)$) ;
		\foreach \x in {0,3}
		\draw[-] ($(L)+ (\x*\step,\up)$) -- ($(L)+ (\x*\step,-\up)$) ;
		\draw[line width=0.5mm] ($(L)+ (\step+\s,\up)$) -- ($(L)+ (\step+\s,-\up)$) ;
		\draw[fill=white,rounded corners=3pt] ($ (L) - (\s,\s) $) rectangle ($ (R) + (\s,\s) $) ;
          \draw[rounded corners=3pt,red, dotted] ($ (L) - (\s+\m,\s+\m) $) rectangle ($ (R) + (\s+\m,\s+\m) $) ;
		\node at ($ 0.5*(R)+0.5*(L) $) {\small$\omegaM{4}$};
		\draw[line width=0.5mm] ($(L)+ (\step-0.5*\s,\up)$) -- ($(L)+ (\step-0.5*\s,\up+\s+\W)$) ;
		\draw[line width=0.5mm] ($(L)+ (\step-0.5*\s,-\up)$) -- ($(L)+ (\step-0.5*\s,-\up-\s-\W)$) ;
		\draw[fill=white,rounded corners=3pt] ($ (L)+(-\s,\up) $) rectangle ($ (L) + (2*\step,\up)+ (0,\W) $) ;
		\draw[fill=white,rounded corners=3pt] ($ (L)+(-\s,-\up) $) rectangle ($ (L) + (2*\step,-\up)+ (0,-\W) $) ;
		\node at ($(L)+ (\step-0.5*\s,\up+0.5*\W)$) {\tiny $L_2$}; 
		\node at ($ (L)+(\step-0.5*\s,-\up-0.5*\W) $) {\tiny $L_2^\dagger$};
	\end{tikzpicture}
	&=
	\begin{tikzpicture}[baseline=-1mm, scale=0.8 ,every node/.style={scale=0.8}]
		\tikzmath{ \up=0.5;  \step=0.5;   \n=3; \s=0.25; \m=.11;} 
		\coordinate (R) at (0,0);
		\coordinate (L) at ($ (R) - \n*(\step,0)$);
		\draw[rounded corners=3pt] ($(L)+(0,\up)$) -- ($(L)+(-\step,\up)$) --  ($(L)+(-\step,-\up)$) -- ($(L)+(0,-\up)$) --cycle;
		\draw[line width=0.5mm] ($(L)+(\step+\s,-1.3*\up)$) -- ($(L)+ (\step+\s,1.3*\up)$) ;
		\foreach \x in {3}
		\draw[-] ($(L)+ (\x*\step,\up)$) -- ($(L)+ (\x*\step,-\up)$) ;
		\draw[fill=white,rounded corners=3pt] ($ (L) - (\s,\s) $) rectangle ($ (R) + (\s,\s) $) ;
          \draw[rounded corners=3pt,green!80!black, ultra thin, dashed] ($ (L) - (\s+\m,\s+\m) $) rectangle ($ (R) + (\s+\m,\s+\m) $) ;
		\node at ($ 0.5*(R)+0.5*(L) $) {\small$\omegaM{5}$};
	\end{tikzpicture} 
	\end{split} 
\end{equation}

We can continue in this fashion and coarse-grain all the states up to $\rhoM{n}$ if for all $m=2,\ldots,n-3$ we have (analogously to \cref{eq:MPSrelation})
\begin{equation}
	\label{eq:MPSrelationM}
	W_{m+1}= R_{m}\circ (W_m\otimes \id) = L_{m}\circ (\id \otimes W_m) \; .
\end{equation}

\subsection{An ansatz for the coarse-graining maps: matrix product states}
We have just shown that we can coarse-grain and compress all the states $\rhoM{m}$ appearing in \cref{eq:locTIn} consistently if  the coarse-graining maps satisfy  \cref{eq:MPSrelationM}.
There is a simple way to satisfy  \cref{eq:MPSrelationM}:\ Uniform matrix-product states (MPS) are translation-invariant states encoded by a single rank-3 tensor, $A$, with dimensions $(D,d,D)$ where $d$ is the dimension of each   spin and $D$ is the so-called bond dimension (or virtual dimension). Contracting $m$ copies of the tensor along the bonds  results in a map from $m$ spins to the bonds at the edges. 
Representing the MPS tensor $A$ by a circle and   the bond indices by red lines, we define  the maps  $W^A_m$ as follows (e.g.\ for $m=3$) 
\begin{equation} 
	\label{eq:MPSiso}
	\begin{tikzpicture}[baseline=4mm, scale=1 ,every node/.style={scale=1}]
		\tikzmath{ \up=0.4;  \step=0.5;   \n=3; \s=0.25;\W=0.35;\d=0.05;} 
		\coordinate (R) at (0,0);
		\coordinate (L) at ($ (R) - \n*(\step,0)$);
		\foreach \x in {1,...,\n}
		\draw[-] ($(L)+ (\x*\step,\up)$) -- ($(L)+ (\x*\step,0.5*\up)$) ;
		\draw[red] ($(L)+ (\step+2*\s-\d,\up)$) -- ($(L)+ (\step+2*\s-\d,\up+\s+\W)$) ;
		\draw[red] ($(L)+ (\step+2*\s+\d,\up)$) -- ($(L)+ (\step+2*\s+\d,\up+\s+\W)$) ;
		\draw[fill=white,rounded corners=3pt] ($ (L)+(\step-\s,\up) $) rectangle ($ (R) + (\s,\up)+ (0,\W) $) ;
		\node at ($(L)+ (\step+2*\s,\up+0.5*\W)$) {\tiny $W^A_3$}; 
	\end{tikzpicture}
	\hspace{6pt} = \hspace{6pt}
	\begin{tikzpicture}[baseline=4mm, scale=1 ,every node/.style={scale=1}]
		\tikzmath{ \up=0.4;  \step=0.5;   \n=4;\s=0.25;} 
		\coordinate (R) at (0,0);
		\coordinate (L) at ($ (R) - \n*(\step,0)$);
		\foreach \x in {1,...,3}
		\draw[-] ($(L)+ (\x*\step,\up)$) -- ($(L)+ (\x*\step,0.5*\up)$) ;
		\draw[rounded corners=3pt, line width=0.2mm, red]  ($(L)+ (\step,\up)$) -- ($(L)+ (3*\step,\up)$);
		\draw[rounded corners=3pt, line width=0.2mm, red] ($(L)+(\step,\up)$) -- ($(L)+(\step,\up)+(-0.5*\step,0)$) -- ($(L)+ (\step,\up)+(-0.5*\step,0.5*\step)$);
		\draw[rounded corners=3pt, line width=0.2mm, red] ($(R)+(-\step,\up)$) -- ($(R)+(-\step,\up)+(0.5*\step,0)$) -- ($(R)+(-\step,\up)+(0.5*\step,0.5*\step)$);
		\foreach \x in {1,...,3}{ 
		\draw[fill] ($(L)+ (\x*\step,\up)$)  circle (.5ex);
              \node at ($(L)+ (\x*\step,1.5*\up)$) {\tiny$A$};
            }
	\end{tikzpicture} \; .
\end{equation}
In this case  the coarse-graining dimension $\chi$ equals $D^2$. 
The maps $L_m$ (and respectively $R_m$) are  then chosen to be the same for all $m$: $L_m\equiv L^A$ (and $R_m\equiv R^A$) and act by contracting an additional MPS tensor from the left (and respectively from the right) and as the identity on the other bond.

With this choice of maps, we apply the above described steps to the LTI problem, \cref{eq:locTIn}, to obtain the following  (shown up to  $m$=5):
\begin{equation}
	\label{eq:TNfullRelax5}
	\setlength{\jot}{0pt}
	\begin{split}
		\tikzPic{cgLSigma3 } &= 	 \tikzPic{TLcgSigma4 } \\
		\tikzPic{cgRSigma3 } &= 	 \tikzPic{TRcgSigma4 } \\
		\tikzPic{cgLSigma4 }  &= 	 \tikzPic{TLcgSigma5 }   \\
		\tikzPic{cgRSigma4 } & =	 \tikzPic{TRcgSigma5 } 
	\end{split}
	\hspace{10pt} 		\xlongrightarrow[C_{m-2}(\rhoM{m})\hookrightarrow\omegaM{m}]{\text{Compress}} 
	\hspace{10pt} 
	\begin{split}
		\tikzPic{cgLSigma3 } &= 	 \tikzPic{TLtilde4 } \\
		\tikzPic{cgRSigma3 } &= 	 \tikzPic{TRtilde4 } \\
		\tikzPic{cgLTilde4 }  &= 	 \tikzPic{TLtilde5 }   \\ 
		\tikzPic{cgRTilde4 } & =	 \tikzPic{TRtilde5 } \; .
	\end{split}
\end{equation}
For $m>5$ the last two constraints on the right side of \cref{eq:TNfullRelax5} are  repeated between the pairs of states $\omegaM{m-1},\omegaM{m}$ for $m=6,\ldots,n$.
The final relaxation is then obtained by replacing the constraints in \cref{eq:locTIn} starting from $m=3$ by the ones in the right hand side of \cref{eq:TNfullRelax5} and in addition  demanding  $\omegaM{m}\geq 0$ for $m=4,\ldots,n$. 
	 For a any  MPS tensor  $A$,   we obtain the following relaxation of  \cref{eq:locTIn}:
\begin{equation}
    \label{eq:locTInRelaxExpl}
    \setlength{\jot}{1pt}
    \begin{split}
        E^{\text{relax.}}_{\text{MPS}(A)}(n):=		\min_{\rhoM{3},\omegaM{m}}  & \tr\left( (h\otimes\id) \rhoM{3} \right)\\[2pt]
        \mbox{s.t. } & \tr (\rhoM{3})=1 \; , \\[2pt]
        & \rhoM{3} \geq 0 \; , \omegaM{m} \geq 0 \; ,  \mbox{ for all }m \in \{4,\ldots,n\} \;, \\
        & \trL (\rhoM{3}) = \trR(\rhoM{3}) \; , \\
        & \mathbb{W}_2^A\otimes\id(\rhoM{3})= \trL(\omegaM{4}) \; , \\
        & \id\otimes \mathbb{W}^A_2(\rhoM{3})  =\trR(\omegaM{4}) \; ,\\
        & \mathbb{L}^A\otimes\id(\omegaM{m})=  \trL(\omegaM{m+1}) \; , \mbox{ for all }m \in \{4,\ldots,n-1\}  \; ,\\
        & \id\otimes \mathbb{R}^A(\omegaM{m})=  \trR(\omegaM{m+1}) \; , \mbox{ for all }m \in \{4,\ldots,n-1\}  \; ,
    \end{split}
\end{equation}
where   $\mathbb{W}_2^A(\bullet):= W_2^A(\bullet)(W_2^A)^\dagger$ and where  $W_2^A$ is given by \cref{eq:MPSiso}, and similarly  $\mathbb{R}^A$ and $\mathbb{L}^A$ act as $ R^A(\bullet)(R^A)^\dagger$ and $L^A(\bullet)(L^A)^\dagger$ respectively (see the last two rows on the  right had side of \cref{eq:TNfullRelax5}).  
Note that   every state $\omegaM{m}$ has dimensions $d^2D^2\times d^2D^2$, where $D$ is the bond dimension of the MPS,  and thus the memory scaling of the relaxation is   $O(n d^4D^4)$.

The error of the lower bound obtained with the relaxed constraints \cref{eq:TNfullRelax5} depends on the MPS tensor $A$ used to perform the coarse-graining. 
In the resulting SDP, the entries of the MPS tensor appear as  hyperparameters which can be optimized in order to tighten the lower bound.
The bond dimension $D$ determines the dimension of the subspace to which the original constraints are  restricted due to  the action of the coarse-graining maps. Increasing $D$ allows to keep more constraints.

Viewing MPS as describing coarse-graining maps is by no means a new insight. This idea lies at the heart of the density matrix renormalization group algorithm \cite{dmrg,dmrgInAgeOfMPS} where the MPS produced at each iteration serves to coarse-grain the Hamiltonian in the next one. 
This connection suggests that an MPS approximation to the ground state, when used in our scheme, might give rise to a tight relaxation for the same Hamiltonian. 

Following this reasoning, in our implementation  we used the tensor describing the MPS ground-state approximation obtained from a variational algorithm based on  a uniform-MPS ansatz---the VUMPS algorithm \cite{vumps}. 
The results which we will now present demonstrate that this heuristic choice performed very well. (For further analysis of the     choice of  MPS tensor for the relaxation  refer to \cref{sec:vumpsVSsdp}   below.)

\subsection{Numerical results}
\label{sec:resultsPREV}
Let us now demonstrate the power of the method through the numerical study of some selected spin models. The models presented here serve to demonstrate the general performance of the method within the framework of this introductory section; results on a wider range of models, together with a more detailed discussion of the numerical findings and an analysis of the performance of the method, can be found in \cref{sec:results}.

In the following, we present results obtained for two paradigmatic spin $1/2$ models:\ the critical transverse-field Ising (TFI) model, and the isotropic antiferromagnetic Heisenberg chain.
For each model, we solved the relaxation   
in \cref{eq:locTInRelaxExpl} with the coarse-graining maps constructed from a variationally optimized MPS, $A_D^\star$,  of bond  dimensions $D$, for various values of $n$ and $D$. 
The obtained energies are denoted by $E^{\mathrm{relax.}}(n,D)$ and are equal to  $ E^{\mathrm{relax.}}_{\mathrm{MPS}(A^\star_D)}(n)$ of \cref{eq:locTInRelaxExpl}.
We then determined their deviations  from the exact ground state energy density $E_{\mathrm{TI}}$:  
$\Delta{E}^{\mathrm{relax.}}_{}(n,D):=E_{\text{TI}}-{E}^{\text{relax.}}(n,D)\ge0$.

For comparison, we also computed the LTI energy density $E_{\text{LTI}}{(n)}$ of the exact hierarchy truncated at the $n$-th level, by solving \cref{eq:locTIn} for attainable values of $n$, and the corresponding deviation 
$\Delta{E}_{\text{LTI}}{(n)}:=E_{\text{TI}}-E_{\text{LTI}}{(n)}$ from the exact value. 
Since our method is a relaxation of the LTI problem \cref{eq:locTIn}, it holds that 
$\Delta{E}_{\text{LTI}}{(n)}\leq \Delta{E}^{\text{relax.}}(n,D)$ for all $D$.

\begin{figure}[t]
	\includegraphics[trim={0 15cm 0 0},clip,width=0.95\textwidth]{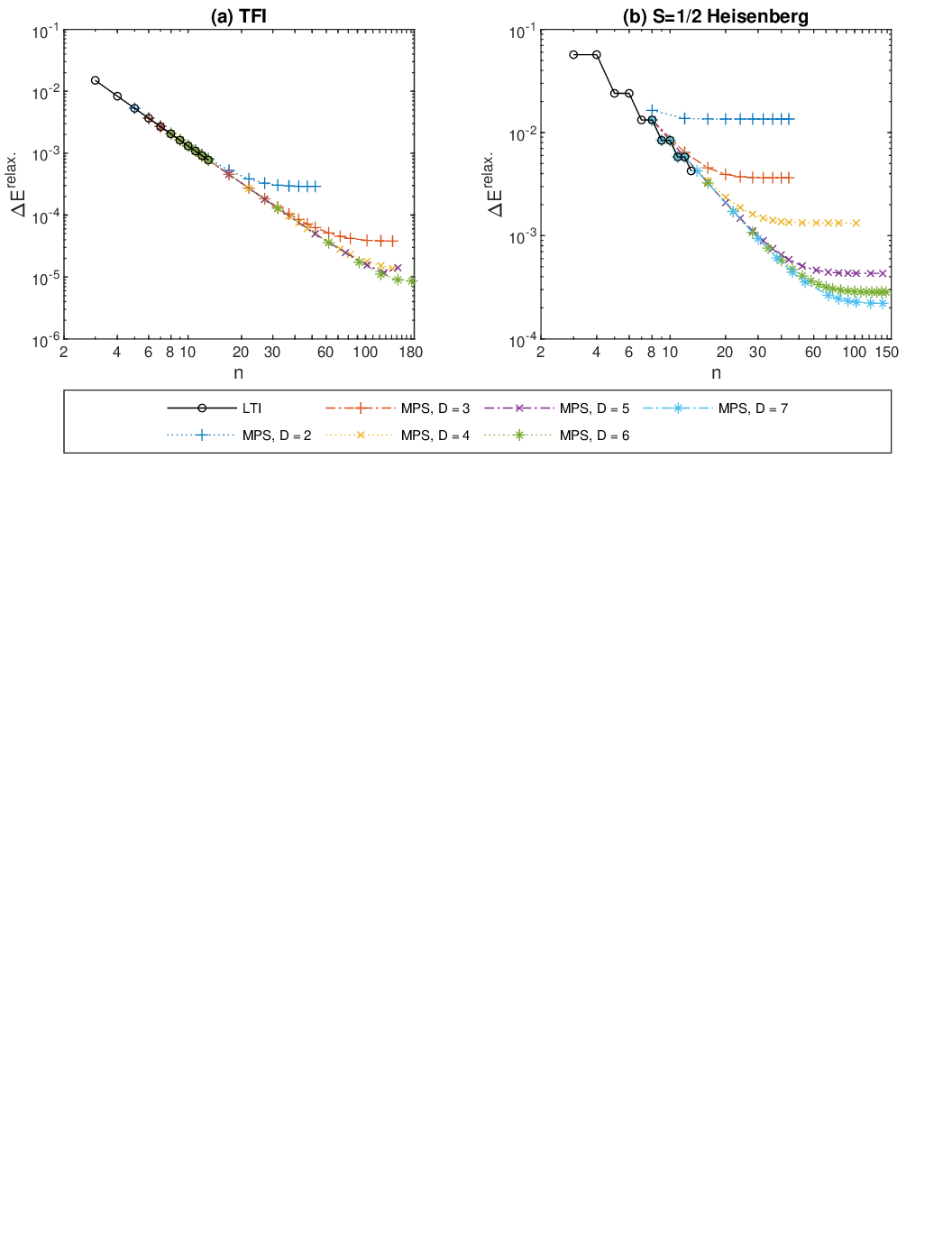}
	\caption{
		Distance $\Delta E^\text{relax.}$ of the lower bounds to the ground state energy density,
        obtained by solving the MPS-based relaxation outlined in \cref{sec:summary}, to the true value, 
        for (a) the transverse field Ising model at the critical point, and (b) the spin $1/2$ Heisenberg antiferromagnet. 
        The differently colored dashed and dash-dotted lines show results obtained using MPS with different bond dimensions $D$ (leading to better relaxations). The black circles give the energy obtained by solving the exact hierarchy up to $n$ sites, whose computational cost grows exponentially in $n$; it displays an algebraic scaling. The MPS relaxations approximate its behavior increasingly well as $D$ increases, at a fraction of the computation cost (which scales algebraically in the effective $n$), thus allowing to reach accuracies of between one and two orders of magnitude higher.
	} 
	\label{fig:resultsPreviw}
\end{figure}

\Cref{fig:resultsPreviw} shows $\Delta{E}^{\text{relax.}}(n,D)$ for different values of $D$ (colored dashed and dash-dotted lines), as well as $\Delta{E}_{\text{LTI}}{(n)}$ (black line with circles) as a function of $n$ for the TFI model  (panel \textbf{a}) and the Heisenberg model  (panel \textbf{b}). 
We observe that the accuracy obtained from the exact hierarchy truncated at
level $n$, $\Delta{E}_{\text{LTI}}$, displays an algebraic decay $n^{-\alpha}$ with $\alpha\approx 2$; however, one is limited to rather small $n$ and thus low accuracies. The curves obtained from our relaxation initially follow the same curve as $\Delta E_{\text{LTI}}$ and in fact extrapolate it to larger $n$, and as $n$ is increased eventually saturate at some $\Delta E^\text{relax.}(n=\infty,D)$.  It is immediately evident from the plots that the accuracy obtained from our relaxation is between one and two orders of magnitude higher than that obtained from the LTI hierarchy. From the plots, we can estimate the effective $n\equiv n_\text{eff.}(D)$ required to reach the same accuracy from the exact LTI hierarchy (by linear extrapolation of the latter in the log-log plot); we find 
$n_{\text{eff.}}(6)\simeq 120$ for the critical TFI model, and $n_{\text{eff.}}(7)\simeq 60$ for the Heisenberg antiferromagnet, both significantly beyond what can be achieved without the relaxation.

The same features were observed in all other models on which we tested our method; in fact, for gapped models, the observed convergence was even more rapid.  For an overview of all results and a detailed discussion of the various findings and numerical methods, we refer the reader to \cref{sec:results}.

\section{The general method}
\label{sec:genMethod}
In \cref{sec:summary}   we dealt with the ground state problem of a 1D spin chain. We have shown how the constraints imposed on a local reduced density matrix by a global  state can be organized in a hierarchical fashion due to the many-body structure of the problem. The hierarchy involves constraints arising from reduced states of increasing groups of spins. By applying a renormalization procedure to this hierarchy we were able to compress the  ground state energy problem. Choosing the renormalization procedure according to the given target Hamiltonian allowed us to efficiently compute tight lower bounds on its ground state energy.

The many-body structure of the constraints, 
on which our renormalization approach hinged, is  however not a property special to 1D spin systems. 
In this section we present the structure of our method in the most general terms and provide  blueprints for its implementation in a wide range of scenarios. 
As we will see, the same ideas that we implemented in the 1D translation-invariant setting can be applied to lattice systems in any dimension, or with other connectivity graphs. They can be applied to fermionic or bosonic particles, and even to systems that do not follow the laws of quantum mechanics. Furthermore, the renormalization procedure we followed in \cref{sec:summary}, by which we coarse-grained the spin chain one spin at a time, is also just one possible choice: other renormalization flows fit within our general framework as well, and each such flow gives rise to a different hierarchy of relaxations.  

We first present the general formulation  of the method in \cref{sec:genMethodDeriv}. We generalize the procedure we followed in \cref{sec:summary} while making minimal assumptions on the  structure or even kind of system. 
Yet, the purpose of \cref{sec:genMethodDeriv} is   not to  merely provide an abstract reformulation of what was already presented above, rather, it is to highlight the essential ingredients and the conditions that should be verified for setting up similar renormalization-based relaxations in various settings. 
Later, in the rest of this section, 
we will follow the prescription outlined in \cref{sec:genMethodDeriv} while  applying our general method to different settings.

In \cref{sec:TTN} we will demonstrate how a block-spin renormalization flow gives rise to yet another relaxation method for 1D TI systems. 
In  \cref{sec:2D} we discuss how our method can be applied to higher dimensional spin-lattice-systems and analyze which renormalization flows are suitable for this purpose. 
In \cref{sec:RDMT} we introduce the Reduced-Density-Matrix-Theory (RDMT) relaxation framework, whose scope of application includes any quantum mechanical system with local interactions, and show how to implement our method within this framework.
Finally, in \cref{sec:quantumInfo} we explain how our method can be used to boost the performance of detection and certification methods in quantum information theory.

\subsection{A general correspondence between renormalization flows and hierarchies of relaxations}
\label{sec:genMethodDeriv}

Let us abstract what we did in   \cref{sec:summary}. Our goal there was to minimize an objective function, namely, the energy density of a 1D TI quantum system. This energy density was a linear function of $\rho^{(2)}$, the $2$-particle reduced density matrix of the 1D TI spin chain.
Correspondingly, the starting point of a general formulation of the technique described in   \cref{sec:summary} is a linear function $H$ of a small number of parameters of an otherwise large (or infinite) physical system. 

Depending on the problem and the representation we choose to work in, those parameters could be the entries of a reduced density matrix (as in \cref{sec:summary}),  or multiple such reduced density matrices, as in the case when $H$ is not translation invariant;  they could be several expectation values which appear in $H$, e.g. the terms $\langle \sigma_x\otimes\sigma_x \rangle$, $\langle \sigma_y\otimes\sigma_y \rangle$, and $\langle \sigma_z\otimes\sigma_z \rangle$ which appear in the Heisenberg $S=1/2$ Hamiltonian; or they could be some other representation of local states. 
Whatever the case may be, we denote these local parameters by $\Rho_{0}$.

Like in the 1D example, our goal is to minimize $H(\Rho_0)$ over all local data $\Rho_0$ compatible with some global state $\Psi$; we represent this constraint formally by $\Rho_0\leftarrow\Psi$ and say that $\Psi$ is a \textit{realization} of $\Rho_0$. 
The problem we wish to solve is thus
\begin{align}
\min_{\Psi\in \mathcal{K}} &H(\Rho_0)\nonumber\\
\mbox{such that }&\Rho_0\leftarrow\Psi,
\label{problem_general}
\end{align}
where the optimization is over the set of all `physical states' $\Psi\in\mathcal{K}$, a notion that is also problem specific:  $\mathcal{K}$ could be the entire  global Hilbert space;\ a   symmetric subspace thereof (as in \cref{sec:summary} where we restricted the optimization to translation invariant states);\ a convex set of states with a property of interest, such the set of separable (i.e.\ not entangled) states;\ or the state space of a certain so-called general probabilistic theory (not necessarily  quantum theory, see \cref{sec:quantumInfo} below).

In our 1D example, we observed that, if $\rho^{(2)}$ admits a TI quantum realization $\Psi$, then there exists a sequence of quantum states of increasing size $(\rho^{(m)})_m$ such that $\rho^{(m-1)} =\tr_m(\rho^{(m)})$. Each such state $\rho^{(m)}$ was not an arbitrary matrix; on the contrary, it had to belong to the set ${\cal S}^{\text{LTI}}_m$ of positive semidefinite and locally translation invariant  matrices (see \cref{eq:LTIcondition}) of trace one.
Note that the proposition $\rho^{(m)}\in {\cal S}^{\text{LTI}}_m$ can be verified efficiently  (in time polynomial in the size of $\rho^{(m)}$). We therefore say that  ${\cal S}^{\text{LTI}}_m$ is a tractable set.

In a general setting, we similarly assume that the constraint ``$\Rho_0\leftarrow\Psi$, for some physical $\Psi$'' can be broken down into a (finite or infinite) hierarchy of conditions that hold between objects of increasing size. More specifically, we assume that there exists a sequence of (tractable) convex sets $({\cal S}_m)_{m=1}^n$ (where we allow $n=\infty$), and a sequence of marginalization maps $(\tr_{m})_{m=1}^n$, i.e.\ maps that discard information, with the following property: for all $\Rho_0$ admitting a physical realization, there exists a sequence $(\Rho_m)_{m=1}^n$, with $\Rho_m\in \S_m$, for $m\geq 1$, such that
\begin{equation}
	\label{eq:shortConstrChain}
	\Rho_0 \xlongleftarrow[]{\tr_{1}}
	\Rho_1 \xlongleftarrow[]{\tr_{2}}
	\Rho_2 \xlongleftarrow[]{\tr_{3}} 
	\ldots \xlongleftarrow[]{\tr_{n}}
	\Rho_n.
\end{equation}
We might further assume that this hierarchy of conditions is complete, i.e. that \cref{eq:shortConstrChain} is both a necessary and sufficient condition for  $\Rho_0$ having a  physical realization (this was the case in \cref{sec:summary}). This assumption is, however, unnecessary for what comes next.

Since the size of $\Rho_m$ might scale badly with $m$ (in the 1D example, it scaled exponentially with $m$), it is impractical to optimize $H(\Rho_0)$ via the hierarchy of conditions (\ref{eq:shortConstrChain}), as was proposed in   \cref{energyExpand}. In the 1D example, we escaped from this predicament by introducing a coarse-graining transformation $\mathbb{W}_{m}(\bullet):=W_m(\bullet) W_m^\dagger$ that, acting on all the qubits of $\rho^{(m+2)}$ but the first and the last, mapped this $m$-body density matrix to the four-partite (and low dimensional) matrix $\omega^{(m+2)}$. 
Our choice of the coarse-graining map $\mathbb{W}_m$ ensured that $\omegaM{m+2}$ is also a state (up to normalization).

Following the same idea, for a general setting we require a 
sequence of coarse-graining transformations $(C_m)_{m=1}^n$, which map each `state' $\Rho_m\in \S_m$ to some $\Omega_m\in \S'_m$, 
where $\S_m'$ is the set of `states' of the coarse-grained system. 
We choose the sets $\S_m'$ such that their  dimension grows at most polynomially  with $m$. 
The maps $C_m$ are thus to be understood as physical  coarse-graining operations.
This notion of course depends on the definition of what are valid states in any given scenario. 
As we will see, for many scenarios of interest, the set of physical operations is easy to characterize.

Our next aim is to find relations between the coarse-grained variables $\Omega_m$, that 
capture some of the relations  between the variables $\Rho_m$  (\cref{eq:shortConstrChain}). 
More precisely, in order to get a relaxation of the original problem,  we need to find  relations  that follow from (and are therefore weaker than) \cref{eq:shortConstrChain}.
This can be achieved by imposing certain compatibility conditions between the maps $C_m$ and the marginalization maps $\tr_m$. 
In order to  formulate  those conditions in the general setting we look back at \cref{eq:TNrho5constraintsSUB} and analyse more closely the interplay between coarse-graining and marginalization maps there.

First, we realized that we need to express $\mathbb{W}_3$ as the composition of $\id\otimes\mathbb{W}_2$ (resp.\ $\mathbb{W}_2\otimes \id$) with the coarse-graining map $L_2$ (resp.\ $R_2$) (see \cref{eq:MPSrelation} and similarly for all $m\geq 2$ in \cref{eq:MPSrelationM}). In the general case, we therefore require that the maps $(C_m)_m$ form a \textit{renormalization flow}, i.e.\ that each $C_{m+1}$ implements one coarse-graining step more than its predecessor $C_m$. 
Second, looking at what happens to $\rhoM{5}$ in \cref{eq:TNrho5constraintsSUB}, 
 we note that in order to substitute it with  $\omega^{(5)}$ we used the fact that the partial trace maps commute with $(\id\otimes\mathbb{W}_3\otimes\id)$. 
Overall, the relation we used in \cref{eq:TNrho5constraintsSUB} can be formally written as:
\begin{equation}
\label{eq:maps_cond_mps_vec}
\left(\begin{array}{c} \id\otimes R_2\\ L_2\otimes\id\end{array}\right)\circ (\id\otimes\mathbb{W}_2\otimes\id)\circ \tr_{R/L}=
\left(\begin{array}{c}  \tr^\prime_R\\ \tr^\prime_L \end{array}\right)\circ (\id\otimes\mathbb{W}_3\otimes\id) ,
\end{equation}
where $\tr_{R/L}$ means $\tr_{R}$ or $\tr_{L}$ as appropriate:\ $\tr_{R}$  was used in the first row of \cref{eq:TNrho5constraintsSUB} and $\tr_{L}$ in the second row, and where we introduced the $\tr^\prime$ notation to emphasize that $\tr$  and $\tr^\prime$ act on systems of different dimensions.

By analogy, we require in the general case that, for $m\geq 1$, there exists a coarse-graining map $B_m$ and a marginalization map $\tr^\prime_m$ such that:
\begin{equation}
B_m\circ C_m\circ\tr_{m+1}=\tr'_{m+1}\circ C_{m+1} \ ,
\label{map_conds_general}
\end{equation}
where the maps in \cref{map_conds_general} are in 1-to-1 correspondence with those in \cref{eq:maps_cond_mps_vec}.
 
Given maps $C_m,B_m,\tr_m$, and $\tr^\prime_m$ that satisfy \cref{map_conds_general} we can perform the analogue of the procedure described in \cref{sec:summary_method2}:  namely, we apply each side of \cref{map_conds_general} to $\Rho_{m+1}$ and invoke the relation $\Rho_m = \tr_{m+1}\Rho_{m+1}$ on the LHS. 
This leaves us with
\begin{equation}
\label{eq:Pre_relaxedConstrGen}
B_m(C_{m}(\Rho_{m}))
= \tr^\prime_{m+1}(C_{m+1}(\Rho_{m+1})).
\end{equation}
Then by substituting all $C_m(\Rho_m)$ by the  compressed states  $\Omega_m$ in the above, we obtain  the desired relation:
\begin{equation}
\label{eq:relaxedConstrGen}
B_m(\Omega_m)=\tr_{m+1}'(\Omega_{m+1}).
\end{equation}


 
We can  now put everything together. Suppose that  \cref{eq:shortConstrChain}  defines a relaxation of the set of physical local parameters $\Rho_0$ and that we find a renormalization flow with maps $(C_m,B_m)_m$ and marginalization maps $(\tr^\prime_m)_m$ satisfying \cref{map_conds_general}. Then, if we start with $C_1:=\id$ (and thus  $\Omega_{1}=\Rho_{1}$), the following optimization problem is a relaxation of the original minimization task (\ref{problem_general}):
\begin{equation}
	\label{eq:primalGeneral} 
	\begin{split}
	\min_{\Rho_{0},\Omega_{m}} & H(\Rho_0), \\
 		\mbox{s.t. } 
 		& \Rho_0\in  \S_0, \\
		& \Rho_0 =  \tr^{\prime}_{1}(\Omega_{1}),\\
		& \cg_m(\Omega_{m})  =  \tr^{\prime}_{m+1}(\Omega_{m+1}),\;  \forall m\in \{1,\ldots ,n-1\}, \\
		& \Omega_m \in \S'_{m}, \; \forall m\in\{1,\ldots,n\} \; .
	\end{split}
\end{equation}
Since $B_m,\tr'_m$ are linear maps and the sets $\S_0,\S_1',\S'_2,...$ are convex, we have that the problem variables $\Rho_{0},\{\Omega_{m}\}_m$ in (\ref{eq:primalGeneral}) are optimized over a convex region. In addition, the objective function $H$ is linear (and therefore convex) on the variables $\Rho_0$. The relaxation (\ref{eq:primalGeneral}) is thus a convex optimization problem \cite{bertsekas2009convex} and, due to the tractability of the sets $\S_0,\S_1',\S'_2,...$, we can use general algorithms from convex optimization theory to tackle it. If, like in our 1D example in Section \ref{sec:summary}, the sets $\S_0,\S_1',\S'_2,...$ happen to be defined through affine and positive semidefinite constraints, then eq. (\ref{eq:primalGeneral}) defines a semidefinite program \cite{VandenbergheAndBoyd}.

Note that the coarse-graining maps $C_m$ appear in \cref{eq:primalGeneral} as hyperparameters. The compatibility  relation in \cref{map_conds_general},   which was the key to arriving at the  relaxation \cref{eq:primalGeneral}, typically does not determine the   maps $C_m$ completely.
Far from it, it leaves us with many    free parameters that  we can vary in order to tighten the lower bound given by  \cref{eq:primalGeneral}.
In the example in \cref{sec:summary} we saw that any uniform MPS gives rise to suitable coarse-graining maps. 
In the following \cref{sec:TTN,sec:2D,sec:RDMT} we will show further examples of how tensor networks provide us with renormalization flows compatible with \cref{map_conds_general}.

Equipped with this general prescription, we will now use it to implement further examples of renormalization-based relaxations. 


\subsection{Block spin renormalization: relaxations using tree tensor networks}
\label{sec:TTN}

In this section we give another example 
of a renormalization flow that gives rise to a relaxation hierarchy, in line with the general correspondence which we described in   \cref{sec:genMethodDeriv}. 
We consider a block-spin renormalization procedure and  apply it  to the  same 1D LTI problem that was treated in \cref{sec:summary}. 
In \cref{sec:results} we will present the full benchmarking results of both relaxation variants---the MPS-based one from \cref{sec:summary} as well as the block-spin (or tree-tensor-network) based relaxation which we will now develop. 
The relaxation based on block-spin renormalization is presented here for the 1D case, but admits straightforward generalizations to higher spatial dimensions. We will discuss this point further in \cref{sec:2D}.

We first consider the exact $n$-body LTI constraint that was our starting point in \cref{sec:summary}: $\rhoM{2}\leftarrow\rhoM{n}$, but this time we start from $n=2^N$ and break the constraint  down into a different sequence   of constraints, one which relates states that double in size at each step:  
\begin{equation}
    \rhoM{2} \leftarrow  \rhoM{4} \leftarrow  \rhoM{8}\leftarrow \ldots \leftarrow  \rhoM{2^{N-1}} \leftarrow  \rhoM{2^N} \; .
\end{equation}
If we were to keep the full LTI symmetry of the state $\rhoM{2^N}$ we would have to impose the equivalence of  all of the ways in which $\rhoM{2^{N-1}}$ can be obtained from  $\rhoM{2^{N}}$    (there are $2^{N-1}+1$ of them). 
However, since we are going  to iteratively map each pair of spins into one blocked spin, we will not be able to keep the LTI property fully: After blocking two spins   we can only consider  translation by two sites, after blocking twice - by four, and so on.

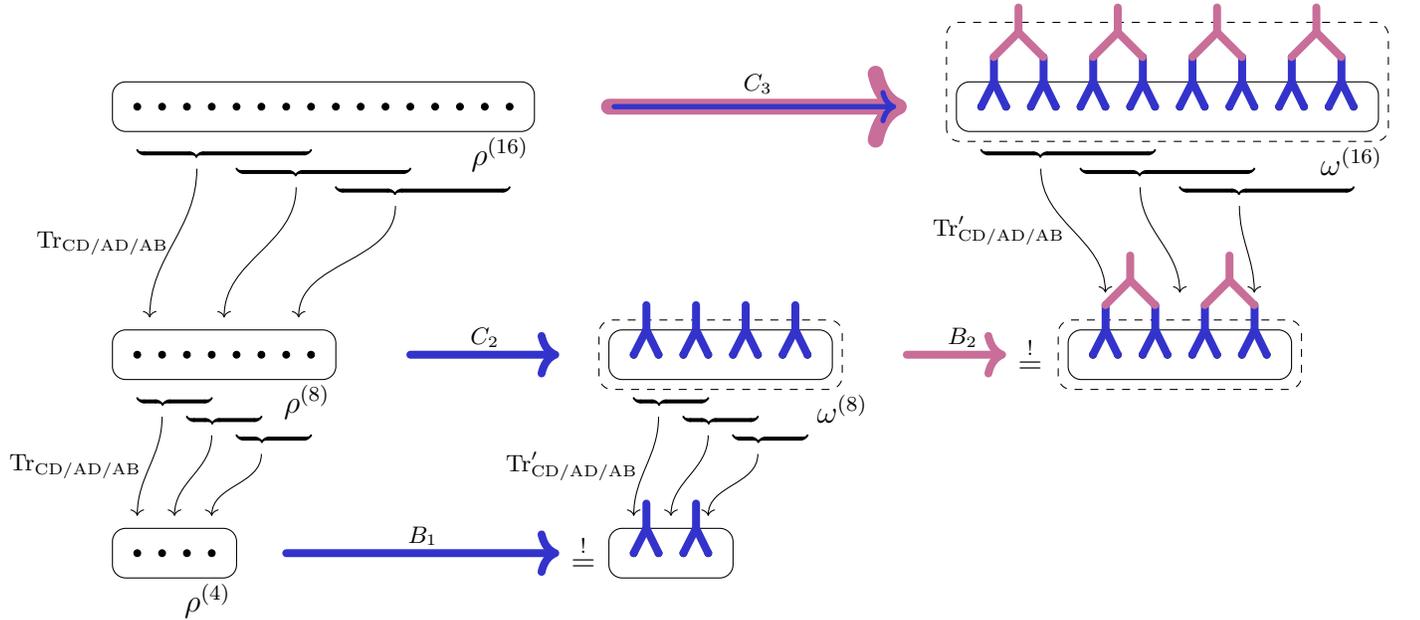
\begin{figure}[b]	
	\begin{tikzpicture}[baseline=-1mm, scale=.33, every node/.style={scale=1}]
	\tikzmath{ \d=1; \g=.75; \spaceL=20;\spaceR=14; \spaceE=4.5; \env=0.4;} 
	\coordinate (L4) at (0,0);
	\coordinate (L8) at (0,8);
	\coordinate (L16) at (0,18);
	\draw[fill=white,rounded corners=5pt]  (L4)  rectangle  ($ (L4) + (5*\d,2*\d) $);
	\node[anchor=north east,inner sep=2pt]  (rho4) at  ($(L4) + (5*\d,0 )$) {\large$\rhoM{4}$};
	\draw[fill=white,rounded corners=5pt]  (L8)  rectangle ($ (L8) + (9*\d,2*\d) $) ;
	\node[anchor=north east,inner sep=2pt]  (rho8) at  ($(L8) + (9*\d,0 )$) {\large$\rhoM{8}$};
	\draw[fill=white,rounded corners=5pt]  (L16)  rectangle ($ (L16) + (17*\d,2*\d) $) ;
	\node[anchor=north east,inner sep=2pt]  (rho16) at  ($(L16) + (17*\d,0 )$) {\large$\rhoM{16}$};
	\foreach \x in {1,...,4}
	\draw[fill] ($ (L4) + (\x*\d,\d) $)  circle (1ex); 
	\foreach \x in {1,...,8}
	\draw[fill] ($ (L8) + (\x*\d,\d) $)  circle (1ex); 
	\foreach \x in {1,...,16}
	\draw[fill] ($ (L16) + (\x*\d,\d) $)  circle (1ex); 
	\foreach \x in {0,...,2} {
		\draw [decorate,decoration = {calligraphic brace,aspect=0.66},line width=1.5pt]   
		($(L8) +(4*\d,-\g)+ (\x*2*\d,-\x*\g) $) --  ($(L8)+ (\d,-\g)+ (\x*2*\d,-\x*\g)$);
		\draw [black,->]   
		($(L8)+ (2*\d,-2*\g)+ (\x*2*\d,-\x*\g)$) to[out=-90,in=90] ($(L8)+ (\d,-\g)+ (\x*1.5*\d,-\g) - (0,4)$) ;
	}
	\foreach \x in {0}
	\path   
	($(L8)+ (2*\d,-2*\g)+ (\x*2*\d,-\x*\g)$) -- node [midway,left]{$\mathrm{Tr}_{\mathrm{CD/AD/AB}}$} ($(L8)+ (\d,-\g)+ (\x*1.5*\d,-\g) - (0,4)$) ;
	\foreach \x in {0,...,2} {
		\draw [decorate,decoration = {calligraphic brace,aspect=0.66},line width=1.5pt]  
		($(L16) +(8*\d,-\g)+ (\x*4*\d,-\x*\g) $) --  ($(L16)+ (\d,-\g)+ (\x*4*\d,-\x*\g)$);	
		\draw [black,->]    
		($(L16)+ (3.4*\d,-2*\g)+ (\x*4*\d,-\x*\g)$) to[out=-90,in=90]  ($(L16)+ (1.5*\d,-\g)+ (\x*3*\d,-\g)  - (0,6)$);	
	}
	\foreach \x in {0} 
	\path  
	($(L16)+ (3.7*\d,-2*\g)+ (\x*4*\d,-\x*\g)$) -- node [midway,left]{$\mathrm{Tr}_{\mathrm{CD/AD/AB}}$}  ($(L16)+ (1.5*\d,-\g)+ (\x*3*\d,-\g)  - (0,6)$);	
	\coordinate (R4r) at ($ (L4) + (\spaceL,0) $);
	\coordinate (R8l) at ($ (L8) + (\spaceL,0) $) ;
	\coordinate (R8r) at ($ (L8) + (\spaceL,0) + (\spaceR,0) +(\spaceE,0) $);
	\coordinate (R16) at ($ (L16) + (\spaceL,0)+ (\spaceR,0) $) ;
	\draw[fill=white,rounded corners=5pt]   (R4r)  rectangle ($ (R4r) + (5*\d,2*\d) $) ;
	\draw[fill=white,rounded corners=5pt]   (R8l)  rectangle ($ (R8l) + (9*\d,2*\d) $) ;
	\draw[rounded corners=5pt,dashed]   ($(R8l)+ (-\env,-\env) $)  rectangle ($ (R8l) + (9*\d,2*\d) +(\env,\env) $) ;
	\node[anchor=north ,inner sep=2pt]  (omega8) at  ($ (R8l) + (9*\d,0) +(\env,-\env) $) {\large$\omegaM{8}$};
	\draw[fill=white,rounded corners=5pt]   (R8r)  rectangle ($ (R8r) + (9*\d,2*\d) $) ;
	\draw[rounded corners=5pt,dashed]   ($(R8r)+ (-\env,-\env) $)  rectangle ($ (R8r) + (9*\d,2*\d) +(\env,\env) $) ;
	\draw[fill=white,rounded corners=5pt]   (R16)  rectangle ($ (R16) + (17*\d,2*\d) $) ;
	\draw[rounded corners=5pt,dashed]   ($(R16)+ (-\env,-\env) $)  rectangle ($ (R16) + (17*\d,2*\d) +(\env,2*\d+\env) $) ;
	\node[anchor=north east ,inner sep=2pt]  (omega16) at  ($ (R16) + (17*\d,0) +(\env,-\env) $) {\large$\omegaM{16}$};
	\draw[color=blue!60!gray,line width=3pt,->,line cap=round] ($ (L4) + (7*\d,\d) $) -- ($ (R4r) + (-2*\d,\d) $) 
	node [anchor=west,color=black] {\large$\overset{!}{=}$}; 
	\draw[decoration={text along path,text={{$\cg_1$}{}},text align={center},raise=4pt},decorate]  ($ (L4) + (7*\d,\d) $) -- ($ (R4r) + (-2*\d,\d) $) ; 
	\draw[color=blue!60!gray,line width=3pt,->,line cap=round] ($ (L8) + (12*\d,\d) $) -- ($ (R8l) + (-2*\d,\d) $) ; 
	\draw[decoration={text along path,text={{$\CG_2$}{}},text align={center},raise=4pt},decorate]  ($ (L8) + (12*\d,\d) $) -- ($ (R8l) + (-2*\d,\d) $) ;
	\draw[color=magenta!60!gray,line width=3pt,->,line cap=round] ($ (R8l) + (12*\d,\d) $) -- ($ (R8r) + (-2.5*\d,\d) $)
	node [anchor=west,color=black] {\large$\overset{!}{=}$} ;  
	\draw[decoration={text along path,text={{$\cg_2$}{}},text align={center},raise=4pt},decorate]  ($ (R8l) + (12*\d,\d) $) -- ($ (R8r) + (-2*\d,\d) $) ;
	\draw[double arrow=6pt colored by magenta!60!gray and blue!60!gray] ($ (L16) + (20*\d,\d) $) -- ($ (R16) + (-2*\d,\d) $) ;
	\draw[decoration={text along path,text={{$\CG_3$}{}},text align={center},raise=6pt},decorate]  ($ (L16) + (20*\d,\d) $) -- ($ (R16) + (-2*\d,\d) $) ; 
	
	\foreach \x in {1,...,4}
	\draw[fill] ($ (R4r) + (\x*\d,\d) $)  circle (1ex); 
	\foreach \x in {1,...,8} {
		\draw[fill] ($ (R8l) + (\x*\d,\d) $)  circle (1ex); 
		\draw[fill] ($ (R8r) + (\x*\d,\d) $)  circle (1ex); };
	\foreach \x in {1,...,16}
	\draw[fill] ($ (R16) + (\x*\d,\d) $)  circle (1ex); 
	\foreach \x in {0,...,2} {
		\draw 
		[decorate,decoration = {calligraphic brace,aspect=0.66},line width=1.5pt]  
		($(R8l) +(4*\d,-\g)+ (\x*2*\d,-\x*\g) $) --  ($(R8l)+ (\d,-\g)+ (\x*2*\d,-\x*\g)$);
		\draw [black,->]   
		($(R8l)+ (2*\d,-2*\g)+ (\x*2*\d,-\x*\g)$) to[out=-90,in=90] ($(R8l)+ (\d,-\g)+ (\x*1.5*\d,-\g) - (0,4)$) ;
	}
	\foreach \x in {0} 
	\path  
	($(R8l)+ (2*\d,-2*\g)+ (\x*2*\d,-\x*\g)$) to[out=-90,in=90] node [midway,left]{$\mathrm{Tr}^\prime_{\mathrm{CD/AD/AB}}$} ($(R8l)+ (\d,-\g)+ (\x*1.5*\d,-\g) - (0,4)$) ;
	\foreach \x in {0,...,2}{
		\draw [decorate,decoration = {calligraphic brace,aspect=0.66},line width=1.5pt]   
		($(R16) +(8*\d,-\g)+ (\x*4*\d,-\x*\g) $) --  ($(R16)+ (\d,-\g)+ (\x*4*\d,-\x*\g)$);	
		\draw [black,->]    
		($(R16)+ (3.4*\d,-2*\g)+ (\x*4*\d,-\x*\g)$) to[out=-90,in=90]  ($(R16)+ (1.5*\d,-\g)+ (\x*3*\d,-\g) + (\spaceE,-5)$);	
	}
	\foreach \x in {0}
	\path  
	($(R16)+ (3.4*\d,-2*\g)+ (\x*4*\d,-\x*\g)$) to[out=-90,in=90] node [midway,left]{$\mathrm{Tr}^\prime_{\mathrm{CD/AD/AB}}$}  ($(R16)+ (1.5*\d,-\g)+ (\x*3*\d,-\g) + (\spaceE,-5)$);	
	\foreach \x in {1,3} {
		\draw[color=blue!60!gray,line width=3pt,line cap=round] ($ (R4r) + (\x*\d,\d) $) -- ($ (R4r) + (\x*\d,\d) + (0.5\d,\d) $) ; 
		\draw[color=blue!60!gray,line width=3pt,line cap=round] ($ (R4r) + (\x*\d,\d) + (0.5\d,\d) $) -- ($ (R4r) + (\x*\d+\d,\d) $) ; 
		\draw[color=blue!60!gray,line width=3pt,line cap=round]  ($ (R4r) + (\x*\d,\d) + (0.5\d,\d) $) -- ($ (R4r) + (\x*\d,\d) + (0.5\d,2*\d) $) ; 
	};
	\foreach \x in {1,3,5,7} {
		\draw[color=blue!60!gray,line width=3pt,line cap=round] ($ (R8l) + (\x*\d,\d) $) -- ($ (R8l) + (\x*\d,\d) + (0.5\d,\d) $) ; 
		\draw[color=blue!60!gray,line width=3pt,line cap=round] ($ (R8l) + (\x*\d,\d) + (0.5\d,\d) $) -- ($ (R8l) + (\x*\d+\d,\d) $) ; 
		\draw[color=blue!60!gray,line width=3pt,line cap=round]  ($ (R8l) + (\x*\d,\d) + (0.5\d,\d) $) -- ($ (R8l) + (\x*\d,\d) + (0.5\d,2*\d) $) ; 
	};
	\foreach \x in {1,3,5,7} {
		\draw[color=blue!60!gray,line width=3pt,line cap=round] ($ (R8r) + (\x*\d,\d) $) -- ($ (R8r) + (\x*\d,\d) + (0.5\d,\d) $) ; 
		\draw[color=blue!60!gray,line width=3pt,line cap=round] ($ (R8r) + (\x*\d,\d) + (0.5\d,\d) $) -- ($ (R8r) + (\x*\d+\d,\d) $) ; 
		\draw[color=blue!60!gray,line width=3pt,line cap=round]  ($ (R8r) + (\x*\d,\d) + (0.5\d,\d) $) -- ($ (R8r) + (\x*\d,\d) + (0.5\d,2*\d) $) ; 
	};
	\foreach \x in {1,5} {
		\draw[color=magenta!60!gray,line width=3pt,line cap=round] ($ (R8r) + (\x*\d,\d)+(0.5*\d,2*\d) $) -- ($ (R8r) + (\x*\d,\d) + (1.5*\d,3*\d) $) ; 
		\draw[color=magenta!60!gray,line width=3pt,line cap=round] ($ (R8r) + (\x*\d,\d) + (1.5*\d,3*\d) $) --($ (R8r) + (\x*\d,\d)+(2.5*\d,2*\d) $)  ; 
		\draw[color=magenta!60!gray,line width=3pt,line cap=round]  ($ (R8r) + (\x*\d,\d) + (1.5*\d,3*\d) $)--  ($ (R8r) + (\x*\d,\d) + (1.5*\d,4*\d) $) ; 
	};
	\foreach \x in {1,3,5,7,9,11,13,15} {
		\draw[color=blue!60!gray,line width=3pt,line cap=round] ($ (R16) + (\x*\d,\d) $) -- ($ (R16) + (\x*\d,\d) + (0.5\d,\d) $) ; 
		\draw[color=blue!60!gray,line width=3pt,line cap=round] ($ (R16) + (\x*\d,\d) + (0.5\d,\d) $) -- ($ (R16) + (\x*\d+\d,\d) $) ; 
		\draw[color=blue!60!gray,line width=3pt,line cap=round]  ($ (R16) + (\x*\d,\d) + (0.5\d,\d) $) -- ($ (R16) + (\x*\d,\d) + (0.5\d,2*\d) $) ; 
	};
	\foreach \x in {1,5,9,13} {
		\draw[color=magenta!60!gray,line width=3pt,line cap=round] ($ (R16) + (\x*\d,\d)+(0.5*\d,2*\d) $) -- ($ (R16) + (\x*\d,\d) + (1.5*\d,3*\d) $) ; 
		\draw[color=magenta!60!gray,line width=3pt,line cap=round] ($ (R16) + (\x*\d,\d) + (1.5*\d,3*\d) $) --($ (R16) + (\x*\d,\d)+(2.5*\d,2*\d) $)  ; 
		\draw[color=magenta!60!gray,line width=3pt,line cap=round]  ($ (R16) + (\x*\d,\d) + (1.5*\d,3*\d) $)--  ($ (R16) + (\x*\d,\d) + (1.5*\d,4*\d) $) ; 
	};
	\end{tikzpicture}
	
    \caption{Graphical representation of the tree-tensor-network-based relaxation scheme  for $n=16$. 
    The left side of the figure shows the constraints   \cref{eq:treeLTI} before coarse-graining is applied:\ Each state $\rhoM{2^k}$ is obtained from the state above it in three different ways by tracing out the appropriate spins, as indicated by the curved black  arrows.
    On the right side, the coarse-grained 4-partite states $ \omegaM{2^k}$ are represented by the dashed rectangles. 
    Each $ \omegaM{2^k}$  is obtained from   $\rhoM{2^k}$ by applying the   $\CG_k$ maps that consist of layers of  2-to-1 coarse-graining maps indicated by inverted `Y' shapes,   see \cref{eq:treeCGlayers}. The maps $\cg_{k-1}$ then implement an additional coarse-graining step on $\omegaM{2^k}$.   The relaxed constraints between the states $\omegaM{2^k}$,  \cref{eq:treeRelaxConstr}, are indicated by $\overset{!}{=}$.}
    \label{fig:treeRelax}
\end{figure}

Our starting point for this scheme is therefore a \textit{relaxation} of the $2^N$-body LTI problem where we impose only the constraints that are compatible with the block coarse-graining, namely, we require that for all $k=2,\ldots,N-1$ the following three ways of obtaining $\rhoM{2^{k-1}}$ from $\rhoM{2^k}$ are equivalent (see also the left side of  \cref{fig:treeRelax}):
\begin{equation}
    \label{eq:treeLTI}
    \rhoM{2^{k-1}} = 
        \trA(\rhoM{2^k} ) =
        \trB(\rhoM{2^k} ) =
        \trC(\rhoM{2^k}) 
        \; ,  
\end{equation}
where  $\trA$, and $\trC$ trace out half of the spins on the left ($1,\ldots, 2^{k-1}$), and on the right ($2^{k-1}+1,\ldots, 2^{k}$) respectively, and $\trB$ traces out everything but  the  $2^{k-1}$ spins in the middle.

 
Next we coarse-grain each state $\rhoM{2^k}$ by applying $k-2$ layers  of block coarse-graining maps 
$ V^{(k-2)}_{8\rightarrow 4}\circ \ldots \circ  V^{(1)}_{2^k\rightarrow2^{k-1}}$,
where at each layer the systems size is  halved (as indicated by the subscript ${2^k\rightarrow2^{k-1}}$) until we end up with a 4-body state.
Each layer is composed as follows
\begin{equation}
\label{eq:treeCGlayers}
    V^{(l)}_{2^k\rightarrow2^{k-1}} = \bigotimes_{j=1}^{2^{k-1}} W^{(l)}\;,
\end{equation}
with $W^{(l)}$ a completely positive map that maps two (possibly already blocked) spins  into  one blocked spin, i.e.\ $W^{(l)}$ maps states on $\CC^{D_{l-1}} \otimes  \CC^{D_{l-1}}$ to states on $ \CC^{D_{l}} $, where $D_0=d$ is the dimension of the physical  spins and  the   dimension of the blocked spins after $l$ layers of coarse-graining is $D_l$.   
We denote by $\omegaM{2^k}$ the resulting states, which are all 4-partite states with local dimensions $D_{k-2}$.
\begin{equation}
\label{eq:treeOmegaDef}
    \omegaM{2^k} =
     V^{(k-2)}_{8\rightarrow 4}\circ \ldots \circ  V^{(1)}_{2^k\rightarrow2^{k-1}} 
     \left( \rhoM{2^k} \right)\;.
\end{equation}
Each state $\omegaM{2^k}$ satisfies the LTI property  $\trL (\omegaM{2^k})=\trR (\omegaM{2^k})$ (as in \cref{eq:LTIcondition}), which captures the invariance of  $\rhoM{2^k}$  with respect to translations by $2^{k-2}$ spins.

To obtain a relaxation in terms of $(\omegaM{2^k})_k$ we follow the prescription in \cref{sec:genMethodDeriv}, where the key step was  making sure that  \cref{map_conds_general} was satisfied. We now turn to this equation.   
Having chosen the coarse-graining maps $\CG_m$ as we have done in \cref{eq:treeOmegaDef}, and the marginalization maps $\tr$ as in \cref{eq:treeLTI}, we see that \cref{map_conds_general} in our setting is to be understood as follows:
taking the partial trace of $\omegaM{2^k}$ with respect to, say, the leftmost two blocked spins should give the same result as first tracing out the $2^{k-1}$ leftmost spins in $\rhoM{2^k}$  and then coarse-graining the resulting state until  only two blocked spins remain. 
The equivalence between the two different orders of operations (first partial trace and then coarse-grain or vice versa) can be ensured by choosing the maps  $W^{(l)}$ composing each layer $V^{(l)}$ to be  completely positive \textit{and trace preserving} (CPTP) rather than just completely positive. 
In this setting \cref{map_conds_general} thus reads:
\begin{equation}
\label{eq:maps_cond_tree}
    \left( W^{(k-2)} \otimes W^{(k-2)} \right)
    \circ 
    \left( V^{(k-3)}_{8\rightarrow 4}\circ \ldots \circ  V^{(1)}_{2^{k-1}\rightarrow2^{k-2}} \right)
    \circ
    \left( \begin{array}{c}
        \trA \\ \trB \\ \trC
    \end{array}\right)
    =
    \left( \begin{array}{c}
        \trPrimeA \\ \trPrimeB \\ \trPrimeC
    \end{array}\right)
    \circ 
     \left( V^{(k-2)}_{8\rightarrow 4}\circ \ldots \circ  V^{(1)}_{2^k\rightarrow2^{k-1}} \right)\; ,
\end{equation}
where  $\trPrimeA, \trPrimeB$ and  $\trPrimeC$ act on  the 4-body state $\omegaM{2^k}$ and trace out spins $(1,2),(1,4)$ and $(3,4)$  respectively. 
The constraints on $\omegaM{2^k}$  are then obtained according to \cref{eq:relaxedConstrGen} and read:
\begin{equation}
\label{eq:treeRelaxConstr}
     \left( W^{(k-2)} \otimes W^{(k-2)} \right) \omegaM{2^{k-1}}
     =\trPrimeA ( \omegaM{2^{k}})
       \ , 
 \end{equation}
 where due to the LTI property of $\omegaM{2^{k}}$ it does not matter whether we use   $\trPrimeA$ or  $\trPrimeB $ or $\trPrimeC$.

\Cref{fig:treeRelax} shows the initial states $\rhoM{2^k}$ for $k=2,3,4$ and the constraints between them (\cref{eq:treeLTI}). The figure further shows how the constraints between the coarse-grained states $\omegaM{2^k}$  are obtained through \cref{eq:maps_cond_tree}. 
We thus arrive at the following relaxation 
\begin{align}
    \label{eq:TTNRelaxExpl}
        E^{\text{relax.}}_{\mathrm{TTN}(W)}(2^N):=		
        \min_{\rhoM{4},\omegaM{2^{k}}}  & 
        \tr\left( (h\otimes\id) \rhoM{4} \right)
        \\ \nonumber
        \mbox{s.t. } & \tr (\rhoM{4})=1 \ , 
        \\ \nonumber
        & \rhoM{4} \geq 0 \ ,\; 
        \omegaM{2^k} \geq 0 \ , \;  \mbox{for all }m \in \{3,\ldots,N\} \ , 
        \\ \nonumber
        & \trL (\rhoM{4}) 
        = 
        \trR(\rhoM{4}) \ ,\;  \trL (\omegaM{2^k} ) 
        = 
        \trR(\omegaM{2^k} ) \  ,\; \mbox{ for all }m \in \{3,\ldots,N\} \ , 
        \\ \nonumber
        & \left( W^{(k-2)} \otimes W^{(k-2)} \right) \omegaM{2^{k-1}} 
        =
        \trPrimeA (\omegaM{2^{k}}) 
         \ , \mbox{for all }m \in \{3,\ldots,N\} \ , 
\end{align}
which depends on the choice of coarse-graining   maps $W^{(l)}$. 
In our numerical implementation we solved the relaxation \cref{eq:TTNRelaxExpl} starting with coarse-graining maps constructed from  variationally optimized tree-tensor-network states  obtained with the algorithm in Refs.~\cite{TTNalg,TTNimpl}. We then further optimized the maps following the procedure described in \cref{sec:optCGmaps}.
The results of the implementation of this method are presented in \cref{sec:results}. 
In \cref{sec:CPTPmaps}  we explain how we constructed the initial coarse-graining maps for this scheme (which have to be CPTP maps) from   variational tree-tensor-network states.


\subsection{Spin systems in higher spatial dimensions }
\label{sec:2D}
Our applications of the method in the 1D setting suggest straightforward generalizations to higher dimensions. 
Consider first the case of a square 2D lattice.
The most obvious generalization of our 1D implementation in \cref{sec:summary}  is to use  projected entangled-pair states (PEPS)---the 2D analogue of MPS---to do the coarse-graining. 
This approach would however quickly run into the following problem:\ As we coarse-grain larger and larger regions $R$ by contracting more PEPS tensors, the coarse-graining dimension grows as $D^{|\partial R|}$, where $|\partial R|$ is the length of the boundary of $R$. 
This means that after applying the coarse-graining procedure we would still end up with a problem that grows  exponentially in size as the region $R$ is increased.
Moreover the smallest region for which coarse-graining with PEPS would reduce the dimension, i.e.\ for which $d^{|R|}>D^{|\partial R|}$, is itself rather large (for both $D$ and $d$ equal to 2 one would need to start from a $5\times5$ square) making the problem intractable for standard SDP software.

The tree-tensor-network variant of the method does not suffer from this drawback. 
The scheme we presented for the 1D case can be directly applied in the setting of a square 2D lattice if in \cref{eq:treeLTI} we replace the states supported on chains of spins $\{1,\ldots,2^k\}$ with ones supported on rectangles 
$\{(i,j)\mid i=1,\ldots,2^k,j=1,\ldots,2^{k^\prime} \}$.  
Coarse-graining maps can then be constructed from 2D TTN states \cite{TTNalg} and applied to patches of  different sizes and geometries.
Higher dimensional tree coarse-graining schemes can similarly be implemented.

\begin{figure}[!h]
    \includegraphics[width=0.45\textwidth]{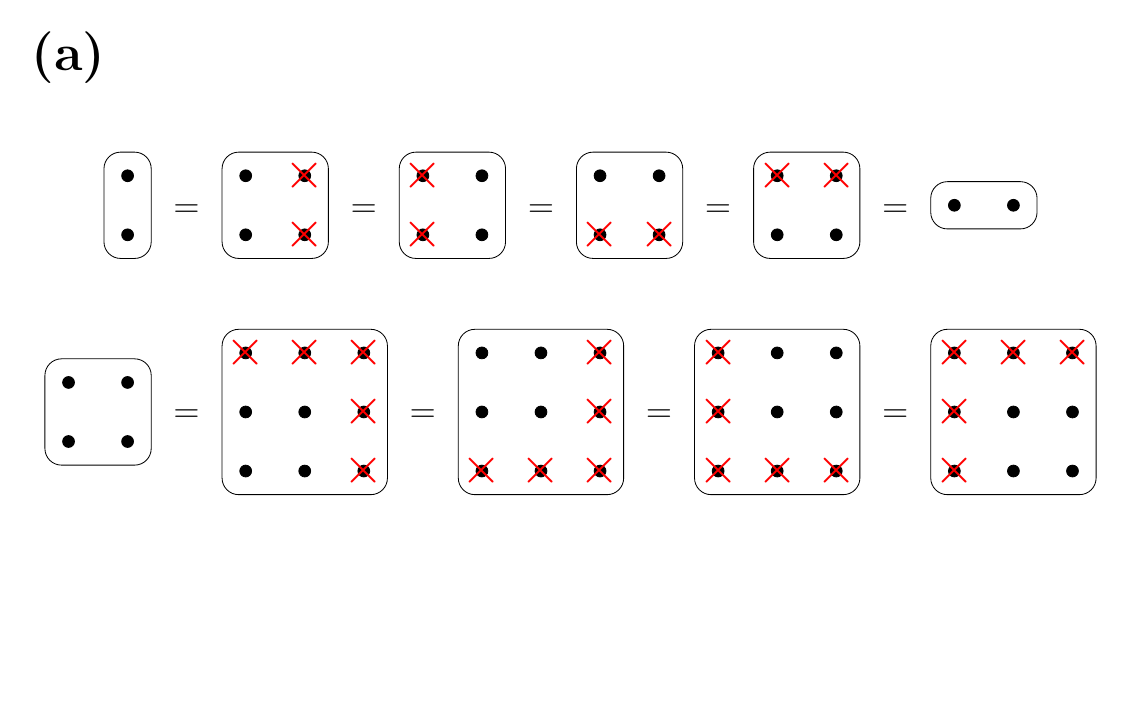}
    \hfill
    \includegraphics[width=0.45\textwidth]{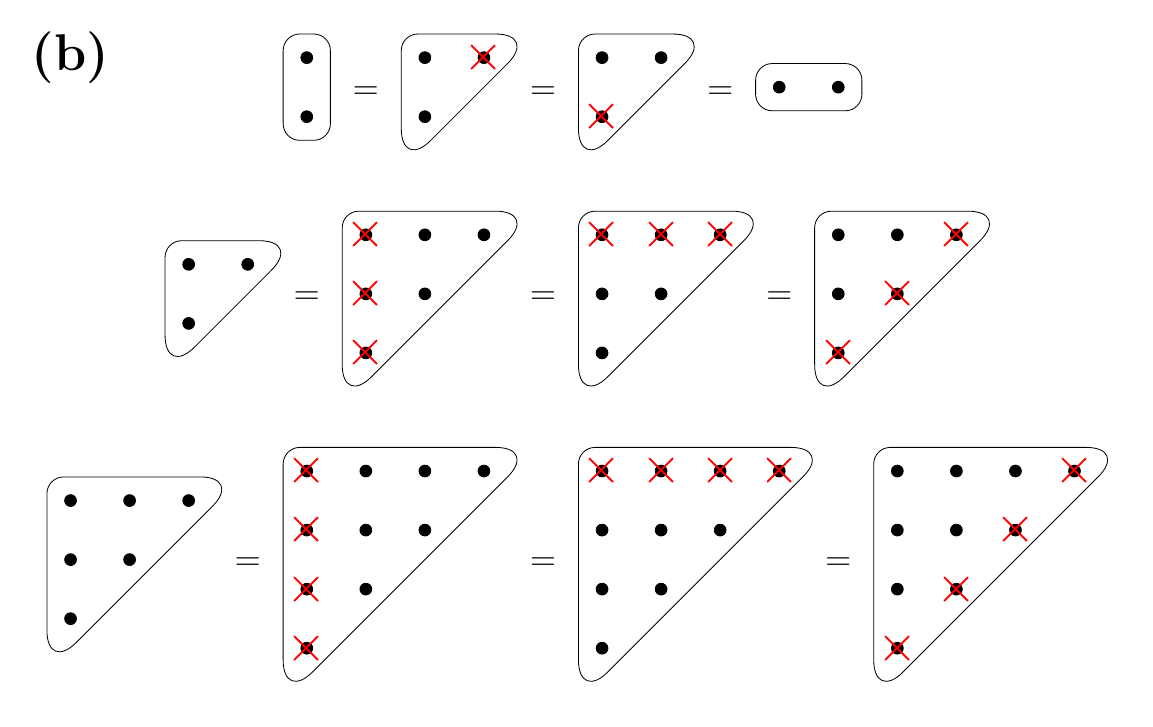}
    \caption
        {The constraints defining the 2D LTI hierarchies for square-shaped (\textbf{a}) and  triangular (\textbf{b}) regions. 
        Each shape depicts a state on a 2D region containing the spins that are  indicated by   black circles. 
        Each equality sign between two shapes defines a constraint demanding that the application of partial traces on  the spins  marked by red {\textcolor{red}{$\times$}}s on both sides results in the same state.
        Symmetry with respect to rotations of the lattice by $90^{\circ}$ is assumed. Higher levels in each hierarchy are formulated similarly.
        Both formulation lead to the same scaling of $\Delta E_{\text{LTI}}^{\text{2D}}(n)$ with $n$, as shown in \cref{fig:2DLTIresults} in \cref{sec:2D_details}.}
    \label{fig:2DLTI}
\end{figure}

Another relaxation approach that is applicable to higher dimensional systems originates in quantum chemistry and goes by the name of \textit{Reduced Density Matrix Theory} (RDMT). This approach has  so far been the most successful in treating 2D lattice systems \cite{BarthelAndHubener,WangRDMT2023}.
We will introduce this approach in \cref{sec:RDMT} and dedicate that  section to explaining  how our renormalization approach can be applied there. 
Lattice systems in 2D would be an appropriate setting to test whether this leads to an advantage over current approaches. 
 
Implementing these ideas goes beyond the scope of this paper as its  focus is on introducing the renormalization approach and  benchmarking it in 1D as a proof of concept. 
However, to get an idea of what to expect when dealing with 2D systems, we provide in \cref{sec:2D_details} some preliminary results concerning the 2D LTI problem. 
There we show that, similarly to the 1D case, the 2D LTI relaxation exhibits a clear scaling of the accuracy of the lower bound, $\Delta E_{\text{LTI}}^{\text{2D}}(n)$, with the system size $n$. 
This can be used as a benchmark for relaxation methods in 2D: one could infer from it the effective system size corresponding to a bound with a given accuracy,  similarly to our estimate of $n_{\text{eff.}}$ from the scaling of the LTI energy in 1D in  \cref{sec:resultsPREV}. 

We further demonstrate that both ways of encoding the  2D LTI constraints presented in \cref{fig:2DLTI}, using either square regions or triangular ones,   lead  to the same scaling of accuracy with system size. 
We then show that by applying one coarse-graining step to the 2D LTI constraints we can recover the energy corresponding to the next level in the hierarchy already with  coarse-graining dimension as low as $D=3$. This leads to precision that  would  have been unattainable without coarse-graining, and is comparable to state-of-the-art results \cite{BarthelAndHubener,WangRDMT2023}. 
Finally, we discuss further possible coarse-graining schemes applicable in 2D.

%

\subsection{Reduced-density-matrix theory}
\label{sec:RDMT}

Reduced-density-matrix theory (RDMT) is a relaxation method that is best known for its  application  in quantum chemistry, where it has been used  to accurately compute the shape of the  potential-energy curve of molecules  \cite{MazziottiRDMbook,Nakata2,RDMT,MazziottiUncertainty,MazziottiSDPscaling,RDMT3,PhysRevLett.108.263002,doi:10.1021/acs.jctc.6b00190} and see Ref.~\cite{PhysRevLett.130.153001} for further references.
The same method has also been applied to lattice systems, and has  appeared in the literature  under several  different  names 
\cite{plenio,BarthelAndHubener,RDMTvariationalCorrelation,WangRDMT2023,RDMThubbardMazziotti,RDMThubbard,bootstrapManyBody,RDMT_classicalSpins_Baccari,bootstrapLawrence}. 
RDMT also underlies recently proposed so-called bootstrap methods that can constrain spectral properties beyond the ground state energy \cite{bootstrapQM,bootstrapGapNancarrow}.
Viewed more  generally, RDMT can be seen as the non-commutative generalization of the Lasserre---or sum-of-squares---hierarchy from polynomial optimization \cite{lasserre2001global,nonCommLassereNPA,NPA}, and therefore underlies  approximation algorithms in quantum complexity theory \cite{gharibian_et_al:LIPIcs.APPROX-RANDOM.2019.31,King2023improved}.

\subsubsection{RDMT in a nutshell}

The basic idea of this approach is based on exactly the same  reasoning that led us to the formulation of \cref{eq:locTIn} in \cref{sec:summary}, namely: to constrain the 2-body reduced density matrix---which determines the energy of the system---by requiring its compatibility with larger and larger physical states. 
The implementation of this idea within the RDMT approach is however  more general than \cref{eq:locTIn}, where physical states were simply $n$-body reduced density matrices, and hinges  on a slightly  more abstract  representation of quantum states. 

Mathematically, a  quantum  state  is simply a function that assigns expectation values to observables $g$ (that is, $g\mapsto\mathrm{tr}[\rho g]$). More precisely, a state is  a  linear functional $L:\mathcal{A}\rightarrow \CC$ on the algebra of observables of the system $\mathcal{A}$. 
In addition, any physical state must be a \textit{positive} functional, i.e.\ it should  assign  non-negative numbers to observables  of the form $g^\dagger g$, where $g\in \mathcal{A}$ is any global observable.

Demanding global  positivity  is of course an intractable task when we are dealing with a many-body system. 
In \cref{sec:summary} this required optimizing over global wavefunctions    $\Psi$ 
(equivalently, global functionals $L_{\Psi}(g):=\bra{\Psi} g \ket{\Psi}$).
We then relaxed this  global positivity constraint by requiring  the existence of a smaller, and thus tractable, positive object:\ an $n$-body state $\rho^{(n)}\geq0$. 
In other words, we demanded that   there is a functional defined on the algebra of local $n$-body  observables   $L^{(n)}:\mathcal{A}_n\rightarrow \CC$ satisfying a weaker---local rather than global---positivity condition:
for all $g_n\in \mathcal{A}_n\subset\mathcal{A}$ we have  $L^{(n)}(g_n^\dagger g_n) := \tr(g_n^\dagger g_n \rho^{(n)})\geq 0$.

RDMT generalizes this idea as follows.
Let $O$ be a set of operators and let $M=\{m_i\}_{i=0}^s$ be a finite set of monomials (products) of  operators from $O$ (with $m_0=\id$).
Instead of considering a global functional  $L$ acting on the entire algebra of observables, 
we can now define $L$  only on the subspace 
$\mbox{span}(M^\dagger M):= \mbox{span}\{m_i^\dagger m_j| {i,j=0,\ldots,s}\}$, 
and demand its positivity there, i.e.\ that  $L(g^\dagger g)\geq 0$ for every $g\in \mbox{span}(M)$.  


If the Hamiltonian we are studying, $H$,  is in $\mbox{span}(M^\dagger M)$, then the energy of the system can be evaluated in $L$ and is given by $L(H)$. 
We can then formulate a relaxation of the ground state energy problem in terms of $L$ with  the relaxed positivity condition  $L(g^\dagger g)\geq 0 \, \forall g\in \mbox{span}(M)$.
This turns out to be an SDP as the condition that $L$ is positive  can be expressed by the positivity of $\Gamma(L)$, the so-called \textit{moment matrix} of $L$:
\begin{align}
\label{eq:momMat}
    &\Gamma(L)_{ij}:=L(m^\dagger_i m_j) \ , \; \text{for all } i,j = 0,\ldots,s \ . \\
    &\Gamma(L) \geq 0  \Leftrightarrow L(g^\dagger g)\geq 0 \; \text{for all } g\in \mbox{span}(M) .
\end{align}
The normalization condition for  the state reads $L(\id)=1$.
The linearity of $L$ means that it is defined by its values on a basis of $\mbox{span}(M^\dagger M)$, i.e.\ in terms of $\dim\mbox{span}(M^\dagger M)\leq(s+1)^2$ parameters $\bm{x}_i:=L(X_i)$, where $\{X_i\}$ is a fixed chosen basis of $\mbox{span}(M^\dagger M)$ with $X_0=\id$. 
In terms of the $\bm{x}_i$s,  the moment matrix of $L$ is given by $\Gamma(L)=\sum_l \bm{x}_i A_i$ 
where $\{[A_l]_{ij}\}_{l}$ are the coefficients of  $m_i^\dagger m_j$ in the basis $\{X_l\}$:    $m^\dagger_i m_j=\sum_l [A_l]_{ij} X_l$.
 
The  following SDP is thus a relaxation of the  ground state problem:
\begin{align}
\label{eq:momentRelaxation}
\min_{\bm{x}}\, & L(H)=\sum_i \bm{h}_i\bm{x}_i\\
\mbox{such that }& L(\id)= \bm{x}_0=1,\nonumber\\
&\Gamma(L) = \sum_i \bm{x}_i A_i \geq 0 \ ,  \nonumber
\end{align}
where the size of the matrix $\Gamma(L)$ is $(s+1)\times (s+1)$ ($s+1$ is the number of elements in $M$). 

As an example, consider a  system composed of $n$ spin-$1/2$ particles and let $H$ be  any 2-local Hamiltonian 
\begin{equation}
\label{eq:2bodyH}
    H=  \sum_{i>j}  h^{(i,j)}  \ ,
\end{equation}
 where each term $h^{(i,j)}$ acts on spins $i$ and $j$. 
If we choose the set  $M$ to include  all the single-site Pauli operators   
$M=\{m_0=\id\}\cup\{m_{(j,a)}=\sigma^{(j)}_a\}_{a=x,y,z}^{j=1,.\ldots,n}$, 
where $\sigma^{(j)}_x, \sigma^{(j)}_y$ and $\sigma^{(j)}_z$  are the Pauli x, y, and z operators  acting on the $j$th spin, we have that  $H\in\mbox{span}(M^\dagger M)$ and  we can thus obtain a lower bound on its ground-state energy by solving \cref{eq:momentRelaxation}. 
Enlarging the set $M$ to include higher degree monomials, i.e.\   products of $k\geq2$ Pauli operators,  $\prod_{i=1}^k\sigma^{(j_i)}_{a_i}$, will improve the lower bound; however, this would  cause the size of $\Gamma$ to grow exponentially with $k$.

We thus see that the RDMT approach is   limited by the same problem of exponential scaling which we highlighted  in \cref{sec:summary}. 
For spatially local Hamiltonians this  can be mitigated to some extent, as   realized in Refs.\ \cite{BarthelAndHubener,WangRDMT2023}. 
There  the  operators in the set $M$ are chosen according to a their locality: products of $k$  Pauli operators  are included only if all the spins on which they act are within a region of size $r_k$, with $r_k$ quickly decreasing to zero  with  $k$. 
Within this approach one could also optimize the choice of the regions  depending on the given Hamiltonian (in analogy to the ideas proposed in  Ref.\ \cite{TuraAndDunjko} for state ensembles). 

However, this is still a rather rigid  framework, as the choice one has is between a discrete set of parameters. 
Furthermore, as  we have demonstrated in \cref{sec:summary},  relaxation schemes based only on  locality (e.g.: the relaxation in \cref{eq:locTIn}, which is only defined by a length scale $n$), 
include a vast number of constraints that are  irrelevant for a given target Hamiltonian, and could thus be further  relaxed without compromising the precision of the resulting energy. 

Given the generality and scope of applicability of RDMT, any method to boost its performance would be a highly welcome development. We will now outline how our renormalization  procedure can be implemented within the RDMT framework, enhancing  it with the continuous degree of tunability provided by the coarse-graining maps which compose the renormalization flow. 
We leave the implementation of the proposed scheme for future work, as we expect that extensive numerical investigations will be needed to properly assess its performance on challenging many-body problems.

\subsubsection{Extending   Reduced-Density-Matrix theory through renormalization}

The relaxations obtained through the  RDMT approach  possess a hierarchical structure and thus fit naturally  within the general  framework outlined in  \cref{sec:genMethod}.
Analogous to \cref{eq:shortConstrChain}, the global positivity constraint $L(g^\dagger g)\geq 0 \ \forall g\in \mathcal{A}$ can be broken down into a sequence of  constraints, defined on an increasing sequence of sets of operators $M_k\subset M_{k+1}$. 
For each set in the sequence we define a functional $L^{(k)}$ on $\mbox{span}(M_k^\dagger M_k)$ and demand its positivity $L^{(k)}(g^\dagger g)\geq 0 \ \forall g\in \mbox{span}(M_k)$.
As the sets $M_k$ form an increasing sequence, the domain on which  each $L^{(k+1)}$ is defined includes  that of its predecessor $L^{(k)}$.
Since all of the functionals  $L^{(k)}$ are supposed to refer to the same state, we demand the consistency condition
\begin{equation}
\tr_{k+1}(L^{(k+1)}):=\left.L^{(k+1)}\right|_{\mbox{span}(M_k^\dagger M_k)  }= L^{(k)} .
\end{equation}
 
We thus arrive at the analogue of \cref{eq:shortConstrChain}, the starting point for the application of our renormalization approach:
\begin{equation}
    \label{eq:shortConstrChainMoments}
    L^{(0)} \xlongleftarrow[]{\tr_{1}}
    L^{(1)} \xlongleftarrow[]{\tr_{2}}
    L^{(2)} \xlongleftarrow[]{\tr_{3}} 
    \ldots \xlongleftarrow[]{\tr_{n}}
    L^{(n)} \ .
\end{equation}

As we have  demonstrated  in the previous sections, for a given system there could be several possible ways of choosing such a sequence depending on the renormalization procedure we seek to implement. 
Our  general method in \cref{sec:genMethodDeriv} hinged on choosing the  sequence \cref{eq:shortConstrChain} and the    renormalization flow in a compatible way, as expressed in \cref{map_conds_general}.
We will now give an example satisfying this compatibility relation, and derive---following the prescription in \cref{sec:genMethodDeriv}---a compressed relaxation of the ground state problem  within the RDMT framework.

For this demonstration  we  consider   a system  of $n$ qubits with a   Hamiltonian allowing for interactions between every  pair of qubits (\cref{eq:2bodyH}). 
For this setting we propose the following renormalization procedure.
We start from spins 1 and 2 and coarse-grain them into a system labeled $Q_2$, next we coarse-grain $Q_2$ and spin 3 into $Q_3$ and so on. 
If we denote those coarse-graining maps by $B_{(Q_{k-1},k)\to Q_k}$ (acting only on systems $Q_{k-1}$ and $k$, and mapping them to $Q_k$), the 
maps $C_j$ we choose are
\begin{equation}
\label{eq:cgMapsMoment}
C_k=B_{(Q_{k-1},k)\to Q_{k}}\circ\ldots\circ B_{(Q_3,4)\to Q_4}\circ B_{(Q_2,3)\to Q_3}\circ B_{(1,2)\to Q_2}\ .
\end{equation}

In order for the sequence \cref{eq:shortConstrChainMoments} to be compatible with this coarse-graining, we choose the sets $M_k$ to consist of  a full Pauli basis  for every group of   $k+1$ spins which contains spins 1 through $k$, plus one further spin: 
$\{1,\ldots,k,l\}_{l>k}$, i.e.\
\begin{equation}
    M_k=\left\{  \left. \sigma^{(l)}_{a} \prod_{j=1}^k \sigma^{(j)}_{b_j} \right|l>k, \; 
    a,b_j=0,x,y,z \right\}  \ , 
\end{equation}
where $\{\sigma^{(j)}_{a}\}_{a=x,y,z}$ are the Pauli matrices acting on spin $j$ and $\sigma^{(j)}_0=\id$.

For large $k$, the specification of $L^{(k)}$ requires  too many parameters. 
To relax the SDP optimization over $L^{(k)}$
we need to define their  compressed counterparts, which we denote as $\Lambda^{(k)}:=C_{k-1}(L^{(k)})$, where $C_{k-1}$ is our coarse-grainig map specified in \cref{eq:cgMapsMoment}  (with $C_1:=\id$).
To see what the  compressed state functional $\Lambda^{(k)}$ should be, consider the action of the map $C_{k-1}$ on the physical spins:  
After the map $C_{k-1}$ is applied, the spins $1,\ldots,k-1$ have been compressed into the system $Q_{k-1}$. 
As  $\Lambda^{(k)}$ should represent a state on the coarse-grained system, we define it  on   a space of compressed operators
$\mbox{span}((M_k')^\dagger M'_k)$, where
\begin{equation}
    M^\prime_k=\left\{  \left. \widetilde{\sigma}^{(Q_{k-1})}_{q} \sigma^{(k)}_{a} \sigma^{(l)}_{b} \right|l>k, \; 
    a,b=0,x,y,z , \; q=1,\ldots,\dim(Q_{k-1})^2 \right\}  \ , 
\end{equation}
and where $(\widetilde{\sigma}^{(Q_{k})}_q)_q$ is a basis of operators of the system $Q_{k}$.
For any operator of the form 
$R^{(Q_{k-1})}\otimes S^{(k)}\otimes
    \sigma^{(l)\dagger}_{b}  \otimes\sigma^{(l')}_{b'}$  
    on the  system $({Q_{k-1}},k,l,l')$, the action of $\Lambda^{(k)}$ is given in terms of  of $L^{(k)}$ and $C_{k-1}$ as
\begin{equation}
    \Lambda^{(k)}\left(
    R^{(Q_{k-1})} S^{(k)}
    \sigma^{(l)\dagger}_{b} \sigma^{(l')}_{b'} 
    \right)
    :=
    L^{(k)}\left(
    C_{k-1}^*\left(R^{(Q_{k-1})}\right)
     S^{(k)}
    \sigma^{(l)\dagger}_{b}  \sigma^{(l')}_{b'} 
    \right) \ ,
\end{equation}
where the map ${C}_{k-1}^*$ denotes the adjoint of $C_{k-1}$:
If $C_{k-1}$ maps states of the spins $(1,\ldots,k-1)$ to  states of the system   $Q^{k-1}$ through $C_{k-1}(\cdot)=\sum_{i}W_i (\cdot) W_i^\dagger$, then  $C_{k-1}^*$ maps \emph{operators} on  $Q^{k-1}$  to operators on the spins   $(1,\ldots,k-1)$ through 
$C_{k-1}^*(\cdot)=\sum_{i}W^\dagger_i (\cdot) W_i$.
Note that 
$ C_{k-1}^*\left(R^{(Q_{k-1})}\right) S^{(k)}\sigma^{(l)\dagger}_{b} \sigma^{(l')}_{b'}$ 
can be evaluated in $L^{(k)}$  because $M_k$ includes a full basis of operators for the spins $(1,\ldots,k)$.

Defining $Q_1:=1$, we note that $B_{(Q_{k-1},k)\to Q_k}(\Lambda^{(k)})$ is a functional on the set of monomials $(M''_k)^\dagger M''_k$, with 
\begin{equation}
    M''_k=\left\{  \left. \widetilde{\sigma}^{(Q_{k})}_{q}  \sigma^{(l)}_{a} \right|l>k, \; 
    a=0,x,y,z, \; q=1,\ldots,\dim(Q_{k})^2 \right\}  \ .
\end{equation}
Noticing that $M_k''\subset M'_{k+1}$, we arrive at the analog of \cref{map_conds_general}:
\begin{equation}
B_{(Q_{k-1},m)\to Q_k}\circ C_{k-1}\circ \tr_{k+1}= \tr_{k+1}'\circ C_{k},
\label{particular}
\end{equation}
with $\tr_{k+1}'$ defined as restricting $\Lambda^{(k+1)}$ to $\mathrm{span}\left((M''_k)^\dagger M''_k\right)$:
\begin{equation}
\tr_{k+1}'(\Lambda^{(k+1)}):=\Lambda^{(k+1)}\left|_{\mathrm{span}\left((M''_k)^\dagger M''_k\right)}\right. \ .
\end{equation}

Our proposed relaxation for the Hamiltonian minimization problem \cref{eq:momentRelaxation} is therefore:
\begin{align}
\label{eq:momentRelaxationCG}
\min_{(\Lambda^{(k)})_k} & \Lambda^{(2)}(H) \\
\mbox{such that }& \Lambda^{(2)}(\id) =1,\nonumber\\
&\Gamma(\Lambda^{(k)})   \geq 0, \; \text{for all } k=2,\ldots,n \ ,  \nonumber\\
&B_{(Q_{k-1},k)\to Q_k}(\Lambda^{(k)})=\tr_{k+1}'(\Lambda^{(k+1)}) \ , \text{for all }k=2,...,n-1. \nonumber
\end{align}
Similarly to $L$ in \cref{eq:momentRelaxation},  each $\Lambda^{(k)}$ is determined by a vector of parameters $\bm{x}^{(k)}$ specifying the  values of $\Lambda^{(k)}$ on a basis of $\mbox{span}((M'_k)^\dagger M'_k)$. 
The  moment matrices $\Gamma(\Lambda^{(k)})$ are  also linear functions of those parameters $\bm{x}^{(k)}$:  $\Gamma(\Lambda^{(k)}) = \sum_i \bm{x}^{(k)}_i A_i^{(k)}$ as in \cref{eq:momentRelaxation}.
Finally as both $B_{(Q_{k-1},k)\to Q_k}$ and $\tr_{k+1}'$ act as linear transformations on 
$\Lambda^{(k)}$ and $\Lambda^{(k+1)}$, we can represent them as matrices  $\bm{B}_k$ and $\bm{T}_{k+1}$ acting on $\bm{x}^{(k)}$ and $\bm{x}^{(k+1)}$, respectively.
The relaxation \cref{eq:momentRelaxationCG} can then be expressed explicitly in terms of the vectors $\bm{x}^{(k)}$.

%
%
\subsection{Quantum information theory: entanglement and nonlocality detection}
\label{sec:quantumInfo}

The general method described in \cref{sec:genMethodDeriv} also leads to sound relaxations when the variables $\Rho_m,\Omega_m$ and the transformations $\tr_m, \CG_m$ are interpreted not as representations of many-body quantum states and transformations thereof, but as representations of the many-body states of  generalized probabilistic theories (GPTs) \cite{HardyGPTs,DakicGPTs,BarnumGPTs,ChiribellaGPTs} and maps transforming such states. 
As observed in \cite{connectors}, key problems in quantum information theory, such as entanglement and nonlocality detection in many-body systems, can be interpreted as determining the existence of a state in a GPT.

The formalism of GPTs  was introduced to describe and reason about physical theories of all conceivable sorts:\ quantum, classical or else. A GPT is specified by a list of physical system types, together with composition rules that detail the type one must use to describe composite systems.
In quantum mechanics, system types are specified by the system's Hilbert space dimension, and the tensor composition rule tells us that the type of a composite system consisting of a subsystem with dimension $d_1$ and a subsystem with dimension $d_2$ is $d_1d_2$. Each system type $T$ comes with a \emph{state space} ${\cal S}_T$, namely, a convex set of vectors, each of which describes a possible state of system $T$. The composition rule must also specify how to derive the marginals of a composite system, given a description of its overall state, as well as how to represent independent system preparations in a joint system description. Transformations in a GPT correspond to linear maps that, acting on part of the state of a bipartite system, always return a valid bipartite state. That is, the linear map $M$, transforming systems of type $T$ into systems of type $T'$, is physical if, for all system types $U$, and all states $s_{UT}\in {\cal S}_{UT}$, $(\id_U\otimes M_T)(s_{UT})\in {\cal S}_{UT'}$. In \cite{connectors} such transformations are called \emph{connectors}.
In the case of quantum theory this is just the complete positivity condition.

Consider for example  the \emph{entanglement marginal problem} \cite{ent_marginal}, i.e., deciding if the observed state ensemble 
$\{\rho_I\}_{I\in\Ind}$ 
of a many-body system corresponds to the marginals of a fully separable quantum state. 
This is   equivalent to deciding if $\{\rho_I\}_{I\in\Ind}$ are the marginals of a global state in the GPT {\bf SEP} defined in \cite{connectors}, whose state spaces are, precisely, the sets of fully separable quantum states.
More generally, the \emph{GPT marginal problem} would ask if the ensemble of GPT states $\Rho_0=\{\rho_I\}_{I\in\Ind}$ arises as marginalizations of a global GPT state $\Psi\in \mathcal{K}$, where $\mathcal{K}$ is the cone of global states of the GPT. 
The dual of this problem can be called the \emph{GPT local Hamiltonian problem}, and it can be cast in exactly the  same form as \cref{eq:primalGeneral}:

\begin{align}
\min & H(\{\rho_{I}\}_I),\nonumber\\
\mbox{s.t. } & \{\rho_{I}\}_{I\in\Ind}, \mbox{ admit a global GPT state realization}.
\label{GPTenergy}
\end{align}
Finding a tight lower bound to \cref{GPTenergy} then allows to use the Hamiltonian $H$ as a witness for the presence of the  property of interest, e.g.: entanglement.
(In the standard marginal problem setting, the detected  property is \emph{incompatibility} with a global state:\ An energy value lower than the ground state energy  indicates that the state ensemble with that energy is incompatible with a global state.)

Clearly, \cref{eq:primalGeneral} corresponds to a lower bound on the GPT local Hamiltonian problem, \cref{GPTenergy}, if we regard $\Rho_{0},\Omega_m$ as ensembles of states of a GPT; $ \tr_{(\cdot)}$ and $\tr'_{(\cdot)}$ as the marginalization maps provided by the GPT's composition rule; $\cg_m$ as  connectors; and the sets $\S_0,\S'_m$ as the corresponding  GPT state spaces.
 

\section{Optimizing the coarse-graining maps and certifying numerical solutions}
\label{sec:numericalStuff}

In this section, we discuss two further numerical aspects of the optimization algorithms. First, we show that it is possible to include an optimization over the coarse-graining maps into the algorithm. This optimization takes itself the form of an SDP; thus, the entire optimization can be carried out as a sequence of alternating SDPs. Second, we discuss how the use of the dual SDP allows us to certify numerical solutions, yielding fully rigorous lower bounds.

\subsection{Dual SDP}

Both methods rely on the dual SDP of the relaxation in \cref{eq:primalGeneral}, which we shall now introduce. 

Recall that in  \cref{eq:primalGeneral} the  variables  are restricted to the sets of physical states:\  $\Rho_0\in\S_0$ and ${\Omega}_{m}\in \S^{\prime}_{m}$. In the cases we are dealing with, those sets are given by the intersection of a cone of positive semidefinite matrices with a subspace defined by a linear constraint (e.g.\ the LTI condition), where for  $\Rho_0$ we also have  an additional normalization constraint. 
In the dual SDP we will need to deal with the dual  cones ${\S^{\prime}}^*_m$. 
Let $\S= \left\{ \Omega\geq 0 | \LL(\Omega)=0 \right\}$ be  the intersection of the cone $ \Omega\geq0$ with the subspace defined in terms of a linear operator $\LL$ as $\LL(\Omega) = 0$.
Its dual cone, i.e.\ the set of Hermitian  matrices $Z$ satisfying $\tr(Z\Omega)\geq0$ for all $\Omega\in\S$,  is given by 
$\S^* =\left\{ Y+ \LL^*(X) | Y \geq 0, X=X^\dagger \right\}$, where $\LL^*$ is the adjoint of $\LL$ and $X$ is any Hermitian matrix. 
A constraint of the form $Z\in \S^*$   thus can be written as a positivity constraint: $Z-\LL^*(X) \geq 0$ for some $X$.
We denote by $\lambda_{m}$ the Lagrange multiplier of the  $m$th constraint in \cref{eq:primalGeneral}.
We further write the normalization constraint explicitly as 
$\tr(\Rho_0)=\mu$, and denote its Lagrange multiplier by $\epsil$. With this notation the dual of \cref{eq:primalGeneral} is
 
\begin{equation}
	\label{dualGeneral}	
	\begin{split}
		\max_{\epsil, \lambda_{m},X_m} & \epsil \\
		\mbox{s.t. } & H + \lambda_0 -\frac{\epsil}{\mu} - \LL^*_0(X_0)
		\geq 0 \\
		& \cg_m^*(\lambda_m) - {\tr^{\prime *}_m}(\lambda_{m-1}) - \LL^*_m(X_m) \geq 0,  \; \forall m= {1,\ldots,n-1},\\
		  & -\tr^{\prime *}_n(\lambda_{n-1}) - \LL^*_{n}(X_{n}) 	\geq 0  \\
            & X_m=X_m^\dagger\   \; \forall m= {1,\ldots,n},
	\end{split}
\end{equation}
where $(\cdot)^*$ denotes the adjoint of an operator.
The solution of \cref{dualGeneral} is always a lower bound to the solution of \cref{eq:primalGeneral} and the two are equal when strong duality holds \cite{VandenbergheAndBoyd}. 
 
\subsection{Optimizing over the coarse-graining maps}
\label{sec:optCGmaps}
In both of our implementations we used tensor networks as an ansatz for the coarse-graining maps, and we have provided heuristic arguments for constructing them from a variational  ground-state approximation when formulating our relaxation.
Our numerical results show that while there is merit to this heuristic, the results can be improved by optimizing the coarse-graining maps. In the MPS-based method (\cref{eq:locTInRelaxExpl}, \cref{sec:summary}) this was not necessary as the results obtained using the MPS ground-state approximation  were very good. 
In the  tree-tensor-network-based variant (\cref{eq:TTNRelaxExpl}, \cref{sec:TTN}) there was more room for improvement.
We therefore  implemented the following procedure to optimize the coarse-graining maps used in the tree-tensor-network-based variant.

The form of the dual  problem  \cref{dualGeneral} suggests a simple scheme to optimize over the maps $(\cg_m)_m$. Namely, first solve the problem \cref{dualGeneral} with the initial maps, hence obtaining the solution $\epsil^{(1)},(\lambda^{(1)}_m)_m$. Next, choose $j\in \{1,\ldots,n-1\}$, fix the variable $\lambda_{j}=\lambda^{(1)}_j$ and solve \cref{dualGeneral} optimizing over $\epsil,\cg_j,(X_m)_m$ and   $(\lambda_{m})_{m\neq j}$ with   additional positive semidefinite and linear constraints on the Choi representation  of $\cg_j$ to ensure it is completely positive and trace preserving.
This provides a solution $\epsil^{(2)},(\lambda^{(2)}_m)_m, \cg^{(2)}_j$ with objective value at least as high as the previous one, i.e., $\epsil^{(2)}\geq \epsil^{(1)}$. Next we can either fix  $\cg_j=\cg^{(2)}_j$ and optimize again over all variables $(\lambda_m)_m$ and $\epsil$ or choose another index $j'$ and optimize $\cg_{j'}$ fixing $\lambda_{j'}$. By conducting coordinate descent in this fashion, it is possible to   improve the lower bound obtained with the initial coarse-graining maps. 

We implemented this procedure in  the tree-tensor-network-based example. In this setting there is an additional snag as the maps $\cg_m$ are always of the form $W^{(l)}\otimes W^{(l)}$ (recall \cref{eq:treeRelaxConstr}).   
This presents a problem since $W^{(l)}\otimes W^{(l)}$ is not linear in $W^{(l)}$. 
To overcome this we parametrize $\cg_m$ in terms of  two maps $ W_1^{(l)}$ and $ W_2^{(l)}$ as follows
\begin{equation}
	\label{eq:cgMod}
	\widetilde{\cg}_{l+1}(W_1^{(l)},W_2^{(l)}) = \frac{1}{2} \left(W_1^{(l)} \otimes W_2^{(l)} + W_2^{(l)} \otimes W_1^{(l)}\right) \; .
\end{equation}
This  parametrization  is linear in both $ W_1^{(l)}$ and $ W_2^{(l)}$. We can then optimize over one map while holding the other one  fixed and alternate between the two.
This more general form of the maps $(\cg_m)_m$  corresponds to a modified choice of the layers $V^{(l)}_{2^k\rightarrow 2^{k-1}}$ comprising  the block-coarse-graining procedure  in  \cref{eq:treeRelaxConstr}: 
\begin{equation}
\label{eq:treeCGlayersMOD}
    \widetilde{V}^{(l)}_{2^k\rightarrow2^{k-1}} = \frac{1}{2}\bigotimes_{j=1}^{2^{k-2}} (W_1^{(l)}\otimes W_2^{(l)}) + \frac{1}{2}\bigotimes_{j=1}^{2^{k-2}} (W_2^{(l)}\otimes W_1^{(l)})\;,
\end{equation}
which is still a valid relaxation of the LTI problem:  each such layer preserves the LTI symmetry corresponding to its length scale and  the maps $\widetilde{V}^{(l)}_{2^k\rightarrow2^{k-1}}$ defined in \cref{eq:treeCGlayersMOD} and $\widetilde{\cg}_{l+1}(W_1^{(l)},W_2^{(l)})$ defined in   \cref{eq:cgMod} satisfy the required relation \cref{eq:maps_cond_tree}. 

We note that as an alternative to the procedure described above, one could also pursue other methods to optimize over the coarse-graining maps, such as gradient-decent methods obtained by differentiating the SDP solution with respect to the  entries of the tensors as described in  Refs.\ \cite{autoDiffSDP,rychkov}. 
 
\subsection{Rigorous bounds from  finite solver precision}
\label{sec:solverPrecision}
In theory a solution to the relaxation in \cref{eq:primalGeneral} provides a rigorous lower bound to the problem in \cref{problem_general}. However, when   performing the numerical optimization the solution will always be accurate up to some finite precision. Furthermore, for large-scale problems some SDP solvers exhibit slow convergence and one might wish to obtain a rigorous result from the last computed iteration and not wait for full convergence. According to basic SDP theory any feasible point of the dual of an SDP gives a lower bound to the primal's optimal value \cite{VandenbergheAndBoyd}.
To obtain a rigorous bound from a numerically obtained solution  it is sufficient to take the dual variable produced by the solver and modify it to make it strictly feasible. 

In our scheme this can be done by using the structure of  the dual problem \cref{dualGeneral}. Starting from a candidate solution $(\epsil,\{\lambda_m\},\{X_m\})$ produced by the solver, which is not known to be feasible, we can check if it satisfies the constraints in \cref{dualGeneral} by exact diagonalization of the matrices involved (each one of them  is typically of modest size). If it is not feasible, we can correct it as follows:\ 
Starting from the last constraint, we replace $\lambda_{n-1}\mapsto \widetilde{\lambda}_{n-1}= \lambda_{n-1} -e_{n-1}\id$, where  $e_{n-1}$ is the minimal  eigenvalue of the left hand side of the last constraint,
$- {\tr^{\prime *}_n}(\lambda_{n-1}) - \LL^*_{n}(X_{n})$. 
After this substitution the last constraint is satisfied because the adjoint of the partial trace  ${\tr^\prime_n}$   simply inserts tensor products with identity terms 
(e.g.\ for a bipartite system $AB$: $\tr_A^*(X_B) = \id_A\otimes X_B $).

We then proceed to compute   $e_{n-2}$, the minimum eigenvalue of the second to last constraint after the substitution with $\widetilde{\lambda}_{n-1}$, and replace $\lambda_{n-2}\mapsto \widetilde{\lambda}_{n-2}= \lambda_{n-2} -e_{n-2}\id$. 
We proceed in this way until in the last step we add the minimum eigenvalue of 
$ H + \widetilde{\lambda}_0 -\frac{\epsil}{\mu} - \LL^*_0(X_0)$ to $\epsil$, the objective energy function, thereby satisfying all constraints with a modified set of variables and  a possibly lower bound on the ground state energy.

Note that after, the substitution $\lambda_{m}\mapsto \widetilde{\lambda}_{m}= \lambda_{m} -e_{m}\id$, a term $e_m\cg_m^*(\id)$ is added to the previous constraint which can inflate the next required correction $e_{m-1}$.
However, if the maps $\cg_m$  satisfy $\cg^*_{m}(\id)\leq \id$, i.e.\ if $\cg_{m}$ are chosen to be trace non-increasing, then the errors of all the individual corrections  at most add up in the final step where the energy $\epsil$ is corrected. 
In our MPS-based implementation this condition on  $\cg_{m}$  can be satisfied by transforming  the MPS into e.g.\ the left  gauge (see Ref.\ \cite{vumps} for details) before constructing the coarse-graining  maps. In the tree-tensor-network-based implementation this condition is always satisfied as the maps used there are trace preserving. In theory, gauge transformations of the MPS should not change the value of the relaxation \cref{eq:locTInRelaxExpl} as the constraints before and after changing the gauge are related by an invertible map. In practice, however, the numerical result can be affected by the choice of gauge if the map implementing the gauge transformation is poorly conditioned.  
   
Finally note that depending on the specific scenario there might be other ways to modify the dual solution to make it strictly feasible. In our examples shifting  the $(\lambda_m)_m$ variables by a constant is the straightforward choice since  the $(X_m)_m$ variables correspond to the LTI constraint, meaning they appear in the equations as $\LL_m^*(X_m)=\id\otimes X_m - X_m\otimes \id$ and shifting them by a constant does not affect the constraints. 
 
\section{Benchmarking  the Method on 1D Hamiltonians}
\label{sec:results}

In this section, we present numerical results obtained with our relaxation method on a broad range of models, where we have implemented the two variants of the method: the MPS-based variant presented in \cref{sec:summary}, and the TTN-based variant presented in \cref{sec:TTN}. In addition to the results themselves, we also provide a detailed analysis of the performance of the method.

\subsection{Studied models}

We have studied the following 1D spin models: The critical Ising model,  the Heisenberg antiferromagnet, the XX model, the XXZ model in the symmetry broken Ising phase (the last three all being instances of the XXZ chain), a critical spin-$1/2$ $J_1$-$J_2$ model, and the spin-$1$ Haldane chain.
The precise Hamiltonians used are 
\begin{align}
	\label{eq:Hamiltonians}
	H_{\text{TFI}}(h_z) &= -\sum_i S^X_iS^X_{i+1} - \frac{h_Z}{2}\sum_i S^Z_i  && \text{Transverse field Ising model (TFI) } \nonumber\\
	H_{\text{XXZ}}^S(\Delta) &= \sum_i S^X_iS^X_{i+1} + S^Y_iS^Y_{i+1} + \Delta S^Z_iS^Z_{i+1} && \text{XXZ spin }S \text{ chain} \\
	H_{\text{J$_1$-J$_2$}}(J_1,J_2) &= \sum_i J_1 \vec{S}_i\cdot \vec{S}_{i+1} +
	J_2\vec{S}_i\cdot \vec{S}_{i+2} && \text{J}_1\text{-J}_2 \text{ Heisenberg spin } 1/2 \text{ chain} \nonumber
\end{align}
(with $\vec{S}_i\cdot \vec{S}_j:=S^X_iS^X_{j} + S^Y_iS^Y_{j} + S^Z_iS^Z_{j}$, and $S^X_i,S^Y_i,S^Z_i$ the spin operators acting on site $i$ of the chain, with eigenvalues $\pm 1/2$ for spin $1/2$).
An overview of of all the models and the corresponding parameters is given in 
\cref{table:models}. The short names used for the models in the following (especially in \cref{fig:results}) are typeset in boldface in the 
table.

\begin{table}[h] 
\centering
	\begin{tabular}{l l  c r}
		\hline
		 & {\bfseries Model Name} & {\bfseries Hamiltonian} & \hspace{5pt}{\bfseries Gapped/Critical}\\
		\hline
		(\textbf{a}) & Critical \textbf{TFI} & $H_{\text{TFI}}(1)$ & Critical \\
		(\textbf{b}) & Isotropic antiferromagnetic $\mathbf{S=1/2}$ \textbf{Heisenberg} & $H^{1/2}_{\text{XXZ}}(1)$  & Critical\\
		(\textbf{c}) & $S=1/2$ symmetry broken   $\mathbf{XXZ(\Delta=2)}$ & $H^{1/2}_{\text{XXZ}}(2)$  & Gapped\\
		(\textbf{d}) & $S=1/2$ $\mathbf{XX}$ model & $H^{1/2}_{\text{XXZ}}(0)$  & Critical\\
		(\textbf{e}) & Isotropic antiferromagnetic $\mathbf{S=1}$ \textbf{Heisenberg} & $H^1_{\text{XXZ}}(1)$  & Gapped\\
		(\textbf{f}) & $S=1/2$ $\mathbf{J_1\text{-}J_2}$ \textbf{Heisenberg} & $H_{\text{J$_1$-J$_2$}}(4.15,1)$  & Critical\\
	\end{tabular}
\caption{Models on which we benchmarked the method. The Hamiltonians of the models are specified in \cref{eq:Hamiltonians}. In the main text and in the figures we refer to each model by the letter (\textbf a--\textbf f) and abbreviated name appearing in boldface in the table. See Ref.~\cite{XXZphase} for the phase diagram of the $S=1/2$ XXZ Hamiltonian (models \textbf{b, c} and \textbf{d}), and Ref.~\cite{J1J2phase} for that of model \textbf{f}.}
\label{table:models}
\end{table}

\subsection{Numerical results}
For all of the above models we solved  the MPS-based relaxation, \cref{eq:locTInRelaxExpl}, with   coarse-graining maps   constructed from  variationally optimized MPS tensors.
For models $\mathbf{a \text{ through }d}$ we solved the  tree-tensor-network-based relaxation, \cref{eq:TTNRelaxExpl},  with maps first constructed from variationally optimized tree tensors,   and then further optimized as  described in \cref{sec:optCGmaps}.
For each model, we solved the relaxations for various values of $n$ and bond dimensions $D$. We denote  the obtained 
energy densities by 
${E}^{\text{relax.}}_{\text{MPS}/\text{TTN}}(n,D)$:
\begin{itemize}
    \item
    For the MPS-based method, 
    ${E}^{\text{relax.}}_{\text{MPS}}(n,D):= {E}^{\text{relax.}}_{\text{MPS}(A^\star_D)}(n)$ from \cref{eq:locTInRelaxExpl}, where $A^\star_D$ is the variationally  optimized   MPS with bond dimension $D$ obtained with the VUMPS algorithm \cite{vumps}.
    \item
    For the TTN-based method,   
    ${E}^{\text{relax.}}_{\text{TTN}}(n,D):={E}^{\text{relax.}}_{\text{TTN}(W^{\mathrm{opt.}}_D)}(n)$ from \cref{eq:TTNRelaxExpl}, where $(W^{\mathrm{opt.}(l)}_D)_l$ are the optimized maps resulting from applying the procedure described in \cref{sec:optCGmaps}   to initial maps constructed  from a variational TTN with dimension $D$, which was obtained with the TTN algorithm in  Refs.~\cite{TTNalg,TTNimpl}.
\end{itemize}
For both variants the solutions  have been certified as described in \cref{sec:solverPrecision}.
 
For comparison, we have also computed the LTI energy density $E_{\text{LTI}}{(n)}$ of the exact SDP hierarchy truncated at the $n$-th level, \cref{eq:locTIn}.  Since 
both variants of our method are relaxations of the LTI problem, it holds that 
${E}^{\text{relax.}}_{\text{MPS/TTN}}(n,D)\le E_{\text{LTI}}{(n)}$. 


For certain systems, the optimization with VUMPS can fail to converge for specific bond dimensions $D$ for a single-site unit cell (see \cite{vumps}). In some cases, this could be remedied by a sublattice rotation [specifically, a $\pi$ rotation around the $z$-axis on every second spin which transforms $H^{1/2}_{\text{XXZ}}(\Delta)$ to $-H^{1/2}_{\text{XXZ}}(-\Delta)$]. Yet, VUMPS would not converge for all values of $D$, leading to missing data points ($D=5,7$ in model \textbf{c} and $D=2,3$ in models \textbf{d} and \textbf{f}).

\begin{figure}[p]
	\includegraphics[trim={0 0.3cm 0 0.15cm},clip,width=0.93\textwidth]{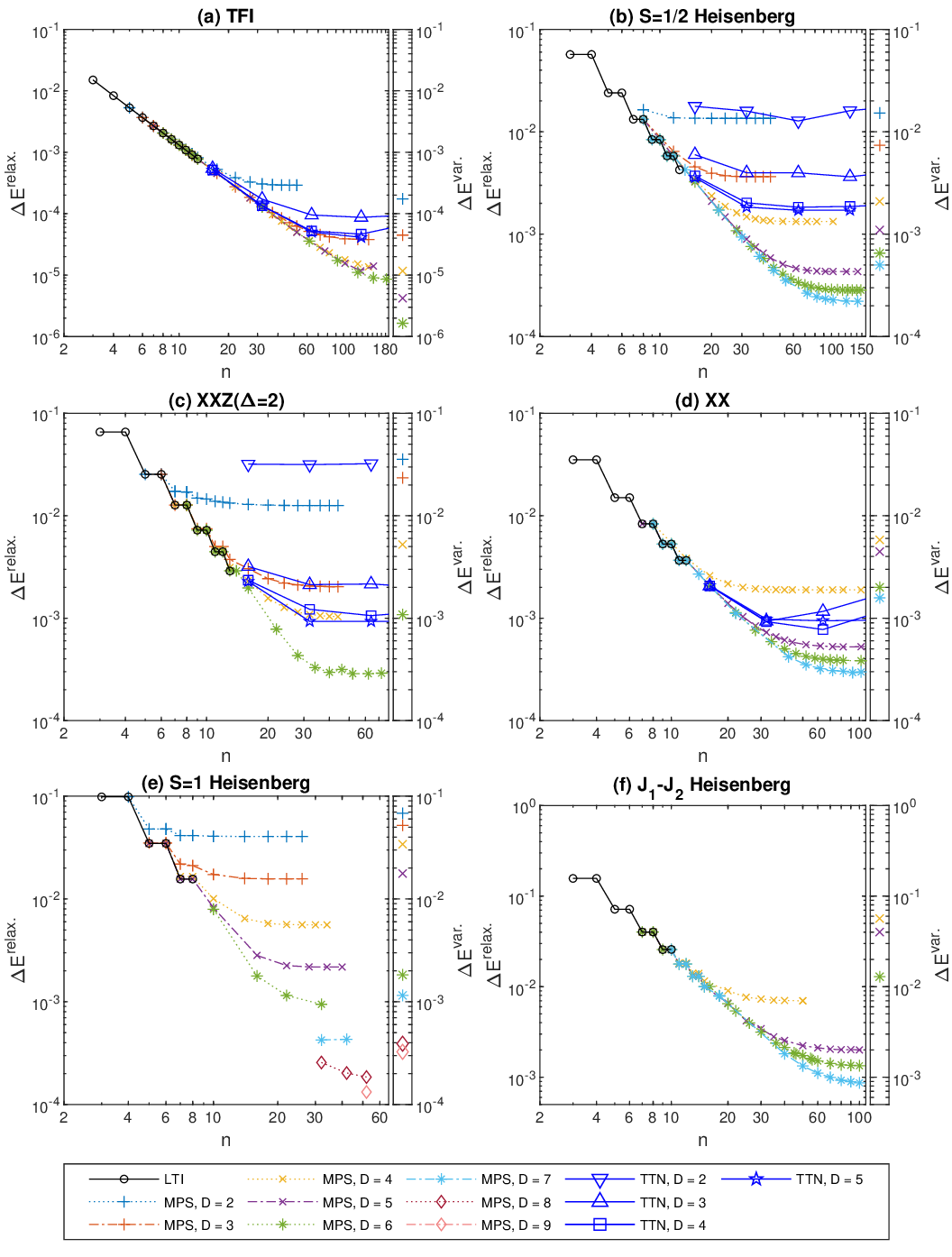}
	\caption{
		The error of the lower bounds to the ground-state energies of the models listed in \cref{table:models}	  	
		obtained by solving the  MPS-based and tree-tensor-network (TTN) based relaxations,  \cref{eq:locTInRelaxExpl,eq:TTNRelaxExpl}, of the   translation-invariant (TI)  local Hamiltonian problem \cref{energy}. 
		Both methods are relaxations of the LTI problem, \cref{eq:locTIn}.
		The error of the LTI lower bound,
		$\Delta{E}_{\text{LTI}}(n):=E_{\text{TI}}-E_{\text{LTI}}(n)$, is plotted by the solid black line with circles.
		The dotted and dash-dotted lines plot $\Delta{E}^{\text{relax.}}_{\text{MPS}}(n,D):=E_{\text{TI}}-{E}^{\text{relax.}}_{\text{MPS}}(n,D)$
		obtained from the MPS-based implementation, and the solid blue lines 
		correspond to the TTN-based variant and plot $\Delta{E}^{\text{relax.}}_{\text{TTN}}(n,D)$, both of which are shown for various   sizes $n$  and various bond  dimensions  $D$.
		In the TTN-based method the coarse-graining maps were optimized numerically as described in \cref{sec:optCGmaps}.
		The bar to the right of each panel shows the precision of the variational \emph{upper bound}, $\Delta{E}^{\text{var.}}$,  corresponding to the MPS used in  the MPS-based relaxation. 
	} 
	\label{fig:results}
\end{figure}

\cref{fig:results} shows the distance of the obtained lower bounds from the actual ground state energy density $E_{\text{TI}}$, denoted by 
$\Delta{E}^{\text{relax.}}_{\text{MPS/TTN}}(n,D):=E_{\text{TI}}-E^\text{relax}_{\text{MPS/TTN}}(n,D)$ and 
$\Delta E_\text{LTI}(n) = E_\text{TI}-E_{\text{LTI}}(n)$, respectively, for all six models. Where known, we used the 
exact ground state energy density for $E_{\text{TI}}$ (models \textbf{a}, \textbf{b}, \textbf{c}, \textbf{d} and \textbf{e}),  and otherwise 
(model \textbf{f})
a high-accuracy estimate obtained using VUMPS with a much larger MPS bond dimension $200$; note that in the latter case,  the deviation of our relaxations from the true energy density is strictly \emph{smaller} than the error $\Delta E$ shown in \cref{fig:results} (due to the variational nature of the obtained energy density). In addition, for comparison 
we also provide the
distance  $\Delta{E}^{\text{var.}}(D)$ of the energy of the variationally optimized MPS which we employed for the coarse-graining, i.e.\ the precision of the corresponding \emph{upper bound};  it is shown in the column to the right of each panel. 
(Note that the MPS data for models \textbf{(a)} and \textbf{(b)} has already been shown in \cref{fig:resultsPreviw}.)

The first feature observed in all plots is the linear decay (on a log-log-scale) of the LTI energy (black line with circles), $\Delta E_\text{LTI}(n)\propto n^{-\alpha}$, with an exponent $\alpha\approx 2$ throughout all models. The MPS curves
$\Delta{E}_{\text{MPS}}^{\text{relax.}}(n,D)$ for different $D$ (colored dotted and dash-dotted lines) initially follow the (extrapolated) LTI line, and eventually flatten out and converge to a constant value 
$\Delta{E}_{\text{MPS}}^{\text{relax.}}(n=\infty,D)$ 
as $n$ gets larger.  As $D$ is increased, the LTI slope is followed for longer, and  the attainable energy is improved.

A closer inspection of this behavior for the gapped models \textbf{c} and \textbf{e} reveals that the 
curves  obtained from our relaxation in fact fall below the linear extrapolation of the LTI line (whether for even or odd $n$);
the same behavior was observed for the TFI model in its gapped phase (not shown). 
Since $\Delta E^\text{relax.}_\text{MPS}(n,D)\ge \Delta E_\text{LTI}(n)$, 
this shows that the LTI energy actually converges super-algebraically (or at least with a larger slope) in gapped models.
This improved convergence rate could not have been inferred from the LTI results themselves, as they are limited to small $n$, but was only possible using the data obtained from our relaxation.

In all the critical models (all but  $\mathbf{c}$ and $\mathbf{e}$), the   line traced by the tightest result obtained with our relaxation for each $n$ coincides with the linear extrapolation of the LTI results in the log-log-plot. (The relaxation results were computed for even values of $n$, and the slope obtained   from a linear fit to those points lies between the even-$n$ and odd-$n$ LTI slopes.) 
From this extrapolation, we can determine the effective LTI problem size, $n_\text{eff.}(D)$, required to obtain the same accuracy as the one achievable with a given $D$
$\Delta{E}_{\text{MPS}}^{\text{relax.}}(n=\infty,D)$. 
We find that for the critical TFI model (\textbf{a}) $n_{\text{eff.}}(6)\simeq 120$, while for the Heisenberg model (\textbf{b}) $n_{\text{eff.}}(7)\simeq 60$, and for models \textbf{d} and \textbf{f} $n_{\text{eff.}}(7)\simeq 40$. In all cases, this lies significantly above the $n$ reachable for the LTI problem, whose complexity grows exponentially in $n$. More generally, the effective $n$  exhibits an algebraic scaling 
$n_{\text{eff.}}(D)\propto D^\kappa$,
with a slope $\kappa$ consistent with the well-known correlation length vs.\ bond dimension scaling for MPS,
$\kappa=6/(c(\sqrt{12/c}+1))$
(with $c$ the central charge) \cite{Tagliacozzo1,Tagliacozzo2,Tagliacozzo3,Pollmann1}.  Together with the $n^{-2}$ scaling of the LTI accuracy, this gives a scaling prediction $\Delta E^\text{relax.}_\text{MPS} \propto D^{-2\kappa}$ attainable with the MPS-based relaxation for a given $D$.

A remarkable feature is that the distance of the lower bounds from the true energy $\Delta{E}_{\text{MPS}}^{\text{relax.}}(n=\infty,D)$ is generally comparable to the upper bounds obtained from the variationally optimized MPS $\Delta{E}^{\text{var.}}(D)$, and in some cases even better by up to half an order. This is rather surprising, given that the MPS has been explicitly optimized to give the optimal upper bound, while we have simply used the same MPS for the coarse-graining for the lower bound, without optimizing over the MPS in any way to obtain optimal lower bounds. This seems to hint that a good renormalization-based variational ansatz also forms a good basis for a coarse-graining relaxation, a connection that deserves further study. 
Motivated by this, we have carried out additional numerical studies on the 
correlation between the precision of the lower and upper bounds, on which we report in 
\cref{sec:vumpsVSsdp} below.

In addition, \cref{fig:results} reports data obtained using the tree-tensor-network-based method for different $D$ (blue lines).
The tree-tensor-network-based variant did not perform as well as the MPS-based one; nevertheless, 
it   improved  upon the  precision attainable by solving the LTI problem. 
That this variant performed less well than the MPS-based one  was partly to be expected because  
the TTN-based method starts from   the constraints in \cref{eq:treeLTI}, which   are  already a   relaxation of the LTI problem as was explained in \cref{sec:TTN}.  
In this variant,  we have  included the optimization  over the coarse-graining maps described in \cref{sec:optCGmaps}; 
the data reported in \cref{fig:results} is the optimized data.
Through this optimization, we were able to decrease the error of the lower bound by factors ranging from 2 to 6. (E.g., in the TFI model with $D=3$  and $n=64$ the error was $\Delta{E}^{\text{relax.}}_{\text{TTN}}(64,3)=0.0006$ prior to optimizing, and $0.000096$ afterwards.)  
Let us note that the way we construct the initial CPTP maps from the tree tensors for this relaxation 
is ad hoc and could certainly be improved. Alternatively, one could pursue other methods to optimize over the coarse-graining maps, such as 
gradient-decent methods described at the end of \cref{sec:optCGmaps}.

A feature which merits discussion are the even-odd steps displayed both by the LTI energy and its approximation through the MPS relaxation for all models but the Ising model. All those models have antiferromagnetic couplings, suggesting that the origin of those steps could be traced back to magnetic frustration occurring in XXZ rings 
\cite{XXZodd,XXZoddFrust,XXZoddRingFrust}. 
To understand this, note that the LTI problem on $n$ sites can be obtained from the ground state 
problem on any periodic chain of size $m\ge n$ by omitting some constraints: The LTI energy density $E_\text{LTI}(n)$ is thus upper bounded by the smallest ground state energy density for any ring of size $m\ge n$, $\min_{m\ge n} E_\text{PBC}(m)$. 
For the models \textbf{(b)} to \textbf{(f)}, the ground state energy density of the periodic chain shows an oscillatory behavior, with the odd-$m$ energy density being higher due to frustration.  Thus, the upper bound 
$\min_{m\ge n} E_\text{PBC}(m)$ shows precisely the same step-like behavior as observed for $E_\text{LTI}(n)$ in \cref{fig:results}. It is thus plausible that the step-like behavior of $E_\text{LTI}(n)$ has the same origin. Remarkably, we find that for 
models $\mathbf{b}$ through $\mathbf{f}$, as well as in all spin-$1/2$ XXZ models with $\Delta\in[-2,2]$, 
$E_\text{LTI}(n)=\min_{m\ge n} E_\text{PBC}(m)$ (up to solver precision), i.e., the 
the solution of the LTI problem  
for even $n$ is in fact \emph{equal} to $E_\text{PBC}(n)$. This is certainly remarkable, as the LTI problem is a relaxation of the PBC problem, and deserves further study. Let us note that this equality, however, breaks down for more general models (e.g.\ in the XYZ family), and thus must be rooted in some special properties of the models considered.

\begin{figure}[b]
	\centering
	\includegraphics[width=0.7\textwidth]{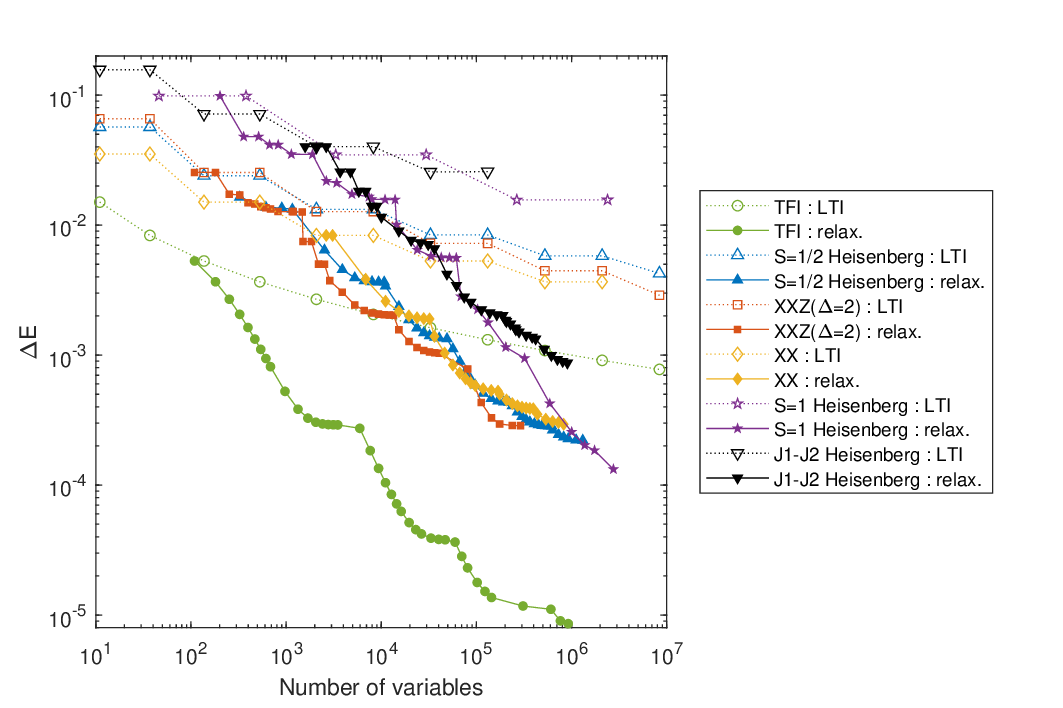}
	\caption{
		The precision of the tightest lower bound obtained for a given memory budget (the number of  variables in the problem) by solving the MPS-based relaxation    \cref{eq:locTInRelaxExpl} (solid lines),   as well as  the LTI problem \cref{eq:locTIn} (dotted lines) for the models listed in \cref{table:models}.  The LTI lines follow a logarithmic curves whereas the  MPS-relaxation data follow linear trends. The slopes of the lines were found to be between $-0.59$ and $-0.74$ depending on the model.  
	}
	\label{fig:numVarsScaling}
\end{figure}

\subsection{Efficiency analysis }
\label{sec:effAnalys}

Let us now assess the performance of the method, that is, 
the way in which the resources required to solve the problem scale with the precision of the lower bound.
To this end, we compare in \cref{fig:numVarsScaling} the 
improvement in precision as a function of the number of variables in the problem, $N_{\text{vars.}}$, 
for the MPS-based relaxation [\cref{eq:locTInRelaxExpl}] and the exact LTI problem [\cref{eq:locTIn}], 
using the entire data set obtained for all the models (see \cref{table:models}) on which we tested our method. 
In \cref{fig:numVarsScaling}, 
the dotted lines represent the LTI problem and the solid lines our MPS-based relaxation,
for each of the models \textbf{a} through \textbf{f} (shown in different colors).
We observe that LTI lines become increasingly flat; indeed, from the exponential scaling 
$N_{\text{vars.}}(\text{LTI(n)})\propto d^{2n}$, together with the observed $\Delta E_{\text{LTI}}(n)\propto n^{-\alpha}$, 
we expect it to scale as $-\log(\,\cdot\,)$ in the log-log-plot. Our relaxation results, on the other hand, seems to follow a clear linear trend in the plot.
This demonstrates that our method results in algebraic convergence of the energy as a function of the size of the SDP. The observed rates of convergence for the different models scale as $N^{-\beta}_{\text{vars.}}$ with $\beta\in[0.59,0.74]$, depending on the model;
this is in line with the different exponents for $n_\text{eff}(D)\propto D^\kappa$ and the $n^{-2}$ scaling of the energy accuracy.

\subsection{Correlation between upper and lower bounds}
\label{sec:vumpsVSsdp}

\begin{figure}[b]
	\centering
	\includegraphics[width=0.99\textwidth]{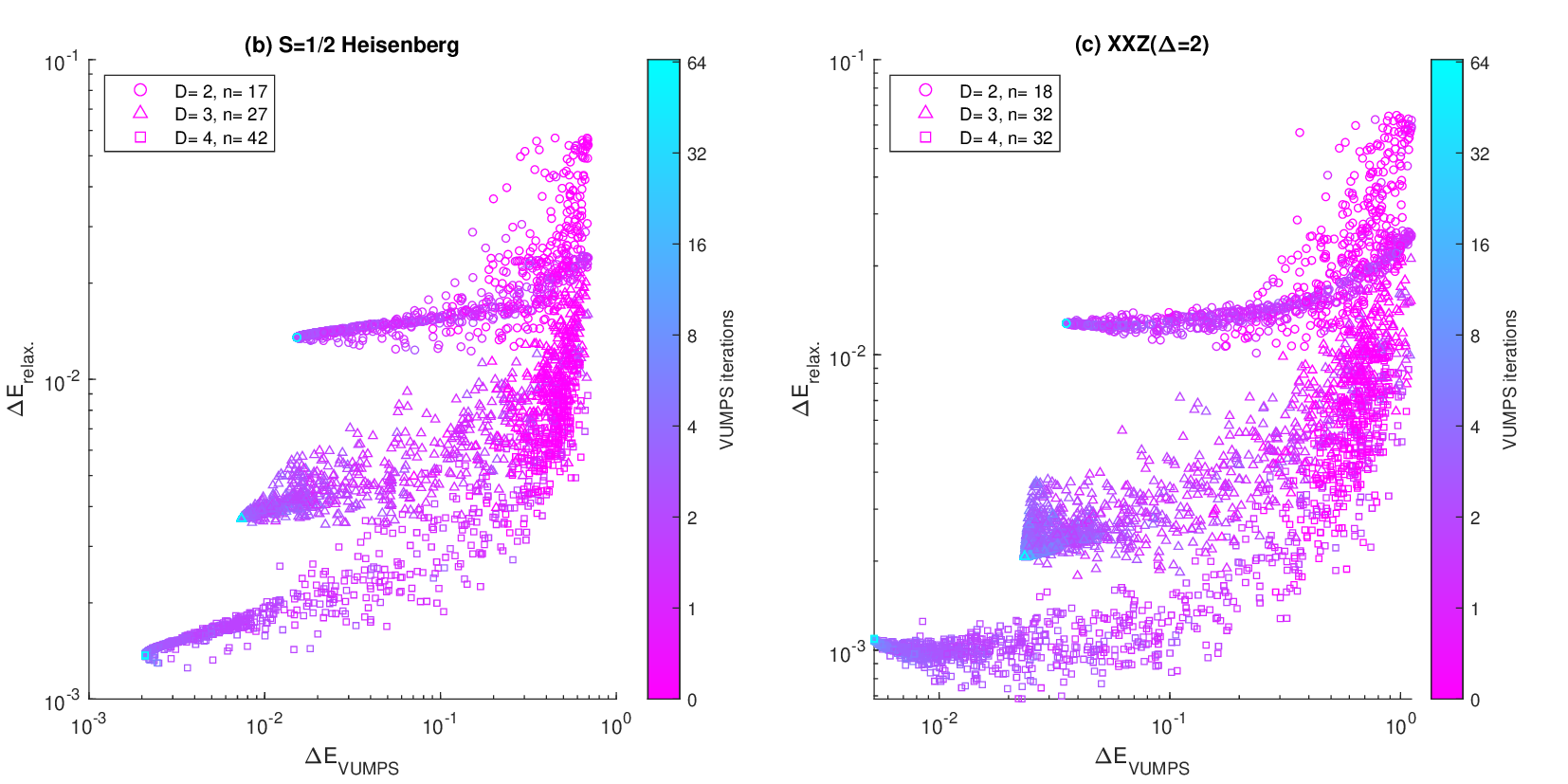}
	\caption{
		Precision of the lower bound obtained by solving  the MPS-based relaxation \cref{eq:locTInRelaxExpl} as a function of the precision of the variational  upper bound corresponding to the MPS that was used to construct the coarse-graining maps for the relaxation. For each $i\in\{0,1,2,4,8,\ldots,64\}$  300 random MPS tensors with bond dimension $D$ were generated and updated by $i$ iterations of the VUMPS algorithm. Each tensor was then used to obtain a lower bound using our relaxation scheme, and an upper bound by evaluating the energy density of the uniform MPS generated by the tensor.  
	}
	\label{fig:scatter}
\end{figure}

A remarkable observation was how well variationally optimized MPS performed when used as   coarse-graining maps for our relaxation, resulting in lower bounds of comparable precision. To further investigate this issue, we have numerically studied the relation between the quality of the upper and lower bounds obtained from a given MPS. 
To this end, we have generated random MPS tensors and applied the VUMPS algorithm to them for a number of iterations between $0$ and $64$; this way, we could sample MPS which provide variational upper bounds of increasing accuracy. 
The resulting data is shown in \cref{fig:scatter}.
For each tensor, the upper bound was obtained by evaluating the energy density of the corresponding MPS; its distance from the ground state energy,  $\Delta{E}_{\text{VUMPS}}$, is shown on the $x$-axis.
The lower bound for each tensor was obtained by solving the MPS-based relaxation,  \cref{eq:locTInRelaxExpl}, using the coarse-graining maps constructed from the given MPS. The corresponding error 
$\Delta{E}_{\text{relax.}}$ of the lower bound is shown on the $y$-axis.

\Cref{fig:scatter}  shows   a clear correlation between the precision of the upper and the lower bounds.  
It  also shows that while the variationally obtained MPS does not always give rise 
to the tightest lower bound attainable with that bond dimension, 
the difference between the best lower bound and that obtained from the MPS variationally optimized with VUMPS for a given bond dimension 
is small when compared to the improvement obtained when going to a larger bond dimension.

\subsection{Comparison with existing methods}

We   give a brief overview of related works, and compare their results with ours where relevant. 
The only results that can be compared directly are those of Ref.\ \cite{BarthelAndHubener} where  translation-invariant XXZ models on infinite lattices  in  1D and 2D   are treated. 
The errors of the bounds they obtained for $H^{1/2}_{\text{XXZ}}(\Delta)$    are $2\times 10^{-3}$ for $\Delta=1$; $3\times 10^{-3}$ for $\Delta=2$; and below $10^{-4}$ for $\Delta=0$ (i.e.\ the XX model, which is equivalent to a system of free fermions). We obtained $3\times 10^{-4}$,  $2\times 10^{-4}$, and $3\times 10^{-4}$ respectively. 
Several further works have implemented relaxation methods based on RDMT, or very similar approaches. Those, however, deal with periodic spin chains and therefore do not provide rigorous bounds on the infinite-system value. 
(We already mentioned above that the energy density on a ring of size $n$, $E_\text{PBC}(n)$,  is lower bounded by the LTI enery $E_{\text{LTI}}(n)$;
furthermore, one can   show that $E_{\text{LTI}}(n)\geq E_{\text{PBC}}(n) - c/n$ for a  constant $c$.) 
Ref.\ \cite{plenio} gives lower bounds for translation-invariant spin Hamiltonians: for the spin $1/2$ Heisenberg model on periodic chains of  sizes up to 50 they obtained energy densities  $10^{-2}$ below the true infinite-system value. 
In Ref.\ \cite{RDMTvariationalCorrelation} a method based on correlation matrices, equivalent to RDMT \footnote{A matrix of the form $\Gamma=\left(\begin{array}{cc}1&\bra{v}\\\ket{v}&\Gamma'\end{array}\right)$ is positive semidefinite iff ${\cal M}:=\Gamma'-\ket{v}\bra{v}$ is positive semidefinite \cite{matrix_analysis}. If $\Gamma$ is the system's moment matrix, with the first column denoting the identity operator, then the matrix ${\cal M}$ corresponds to the system's correlation matrix. Hence, by demanding the positive semidefiniteness of ${\cal M}$, the authors of \cite{RDMTvariationalCorrelation} are just implementing RDMT, as descibed in section \ref{sec:RDMT}.}, is applied to finite periodic spin chains. For the critical Ising model and   XXZ models with $\Delta \in [0,2]$ they obtained errors of the order $10^{-2}$ and $10^{-1}$, respectively.
Ref.\ \cite{WangRDMT2023} applies RDMT to bound the ground-state energy, as well as ground-state correlation functions in   the Heisenberg model with nearest-neighbor and next-nearest-neighbor interactions in periodic 1D and 2D lattices. They obtain an error of $4\times10^{-4}$ for the Heisenberg model on a periodic chain of length 100.

\subsection{Numerical implementation}
\label{sec:resultsNUM}
We conclude this section with some details regarding   our numerical implementations. Sample code is available at \cite{myCodeDemo}. We used predominantly general-purpose software packages to solve the SDPs involved. We used YALMIP \cite{YALMIP} for modeling the problems and solved primarily using MOSEK \cite{mosek}. The unfavorable memory scaling of interior-point methods (see e.g.\ \cite{scalableSDP}) prevented us from going to   bond dimensions higher than 7 in the MPS-based variant and 5 in the tree-tensor-network-based variant. To overcome this, we tried to use the splitting conic solver (SCS)   \cite{scs} which is a first-order method and requires less memory, but observed slow convergence and were not able to improve the results in any of the models, with the exception of the $S=1$ Heisenberg chain. In this model we used our own matrix-free implementation of SCS (where the constraint matrix is not explicitly constructed as a sparse matrix, but only implicitly applied, which provides significant savings due to the tensor-network structure of the constraints)   to produce the results for bond dimensions 6 through 9. In addition, SCS was used to solve   the larger instances of the LTI problem, \cref{eq:locTIn}, where it performed very well.

We have not explored further methods beyond the ones mentioned. 
 We believe the results we presented could be markedly  improved upon by using custom-tailored solvers. 
 The chain structure of the constraints in the relaxed problem \cref{eq:locTInRelaxExpl} (and more generally \cref{eq:primalGeneral}) suggests that splitting methods \cite{Parikh2013ProximalA} other than the one employed by SCS could lead to an advantage. 
 Combining those with matrix-free solvers  would further reduce the complexity and allow to perform computations for larger system sizes $n$ and bond dimesnions $D$.

Finally we note that the presence of local symmetries can be naturally accounted for within our method. 
In \cref{sec:symmetries} we explain how by using suitable coarse-graining maps, one ends up with  a relaxation involving symmetric states thus reducing the problem's complexity.  
 
\section{Conclusion}
\label{sec:conclusion}

In this paper, we have introduced a general method to obtain rigorous
lower bounds for a wide range of minimization problems in a quantum
many-body setting. As a prime application, we demonstrated how our method
can be applied to lower bound the ground state energy of
quantum spin systems. At the same time, the method lends itself to a wide
range of other optimization problems with a similar structure, such as the
certification of entanglement and non-locality in quantum information
theory. The method achieves this by providing a systematic and
customizable way to obtain outer approximations to the set of physical
quantum correlations; it is thus suggestive that it can also provide new
ideas on how to relax other numerical methods based on positivity constraints,
such as reduced-density-matrix theory or various bootstrap techniques.

The key idea of our method is to relax the convex set of few-body density
matrices or other correlations which are physical---that is, compatible
with a global quantum state---in two steps. In a first step, the global
compatibility constraint is decomposed into a hierarchy of constraints
between increasingly complex objects. A first relaxation can then be
obtained by truncating the hierarchy at a finite level $n$. Approaches of this
kind have been pursued previously in various settings; however, they
suffer from the exponential growth of the involved objects, which severely
limits the level of the hierarchy that one can reach.

The key novelty of our method lies in the second relaxation: We apply a
coarse-graining procedure to all objects in the hierarchy, in a way which
compresses them down to objects of a fixed (or otherwise tractable) size.
This compression must be chosen such as to keep the relevant degrees of
freedom, which can be accomplished, e.g., by employing
renormalization-based ideas.  At the same time, any such compression
scheme must act in a consistent way across all levels of the hierarchy, a
demand yet again met by renormalization-based coarse-graining maps.

By utilizing this compression, we can effectively reach very large levels
$n$ of the original hierarchy, and thus achieve results with a much
improved accuracy.  At the same time, the optimization problem remains a
semidefinite program (SDP), and can thus be efficiently solved.  In
addition, the possibility to smoothly adjust the coarse-graining maps to
best suit the problem at hand, and to optimize them as well in the course of the
optimization, makes our relaxation much more tunable than existing
approaches, which are based on discrete, and thus rigid, hierarchies. Moreover,
insights from renormalization procedures can be used as a guide in
choosing those maps.

We have worked out our method in detail for 1D quantum spin chains, and
demonstrated its power through the explicit numerical study of a range of
relevant models. Here, we have employed two coarse-graining schemes, both
of which are motivated by renormalization-based variational ansatzes: One
based on Matrix Product States (MPS), and one on Tree Tensor Networks
(TTN); in the latter case, we have additionally implemented an
optimization over the coarse-grainers. In both cases, we observed that the
variationally optimized wavefunction also provides a very good
coarse-graining map for the relaxation of the hierarchy, in line with the
intuition that the coarse-graining should keep low-energy degrees of
freedom, akin to renormalization schemes. We have tested both approaches
on a wide range of gapped as well as critical models. We found that both
schemes clearly outperform the exact truncated hierarchy, with the MPS
approach generally performing noticeably better. Specifically, using the
MPS-based relaxation, we were able to gain between one and two orders of
magnitude in precision in energy, on par with sizes of the truncated
hierarchy between $n\approx 40$ and $n\approx 120$ (while the
exact hierarchy was limited to $n=13$). Overall, we observed that the
resources required for our method scaled polynomially in the targeted
energy precision, as opposed to the exponential scaling of the exact
truncated hierarchy.

The general framework presented in this paper can be applied to a much
broader range of problems than those that we used as a demonstration.
First, the method opens avenues to obtain provable lower bounds on ground
state energies for two-dimensional systems, where good variational upper
bounds -e.g.: using tensor networks- are much harder to obtain, and where
the performance of those methods is less well understood. It can also be
applied to reduced-density-matrix theory and related approaches based on
correlation matrices, and thus to problems without an underlying locality
notion, such as in quantum chemistry. Finally, at its heart, our
relaxation presents a way to certify when a (very large or even infinite)
matrix that satisfies a set of linear and convex constraints on its
entries (such as having certain two-body marginals or few-point
correlations, or being a separable state) cannot be chosen to be positive;
as such, it is also applicable to problems such as entanglement or
non-locality detection, as well as to the so-called bootstrap approach to
many-body problems.

A key challenge in generalizing our method is the right choice of coarse-graining maps. In this regard, we have demonstrated that maps based on tensor networks, which can in most cases be interpreted as variational wavefunctions underlying different renormalization schemes \cite{Cirac_2009}, are a suitable option. It is reasonable to expect that tensor network ansatzes will work  in higher dimensional problems as well. Moreover, given the often excellent performance of tensor networks beyond quantum lattice problems, it is likely that suitably chosen tensor network ansatzes will also make good coarse-grainers in scenarios other than energy minimization. Remarkably, tensor networks optimized to derive variational upper bounds generally performed surprisingly well when used as  coarse-grainers in our relaxation scheme; a better understanding of this connection, as well as of the ways in which different tensor networks can be used as coarse-graining maps, remain as open problems for future work.

\begin{acknowledgements}
We thank Bram Vanhecke, Antoine Tilloy and Roger Mong for insightful suggestions, and the Erwin Schrödinger International Institute for Mathematics and Physics for its hospitality during the Thematic programme \emph{Tensor Networks: Mathematical Structures and Novel Algorithms}.
This research was funded in part by the Austrian Science Fund FWF
(Grant DOIs 
\href{https://doi.org/10.55776/P35509}{10.55776/P35509},
\href{https://doi.org/10.55776/P36305}{10.55776/P36305},
\href{https://doi.org/10.55776/F71}{10.55776/F71}),
and the European Union’s Horizon 2020 research and innovation programme through Grant No.\ 863476 (ERC-CoG SEQUAM). 
The computational results presented have been achieved   using the Vienna Scientific Cluster (VSC).
For open access purposes, the authors have applied a CC BY public copyright 
license to any author accepted manuscript version arising from this submission.

\end{acknowledgements}

\appendix

\section{Constructing  coarse-graining maps from tree tensor networks}
 \label{sec:CPTPmaps}
 Recall that in  \cref{sec:summary} we constructed  the coarse-graining from a  variational MPS ground state approximation. In the same spirit, here we  construct the coarse-graining maps $W_2^{(l)}$ composing each coarse-graining layer \cref{eq:treeCGlayers}  form the output of a tree-tensor-network (TTN) variational algorithm \cite{TTNimpl} which optimizes the three-legged tensors making up the tensor network layer by layer.
 The entire tensor network describes a pure state on a ring of size $2^L$, where $L$ is the number of layers. 
 Each layer $l$ is composed of tensor products of a single partial isometry  $T_l$ from two sites at  lower layer to one site in the layer above it. The output of the algorithm thus consists of $L$ partial isometries $T_l:\CC^{D_{l-1}}\otimes\CC^{D_{l-1}}\rightarrow\CC^{D_{l}}$
 where $ {D_{0}}=d$ is the dimension of the physical spins and for all the other layers  $D_l\leq D$, where $D$ is an input to the algorithm. At the last layer $D_L =1$ such that $T_L$ describes  a pure two-body state on $\CC^{D_{L-1}}\otimes\CC^{D_{L-1}}$.
 
 Next, recall that in order to satisfy \cref{eq:maps_cond_tree} we need the maps $W^{(l)}_2$ to be trace preserving.
 As the TTN algorithm outputs partial isometries $(T_l)_l$, we cannot simply use them to construct $W^{(l)}_2(\cdot)$ as $T_l(\cdot)T_l^\dagger$ as such maps do not preserve the trace. 
 To make trace preserving maps from the partial isometries $T_l$  we start with $T_1$ and  increase the output Hilbert space dimension by one
 $\CC^{D_{1}}\mapsto \CC^{D_{1}}\oplus \H_g \cong\CC^{D_{1}+1} $ where $\H_g$ is spanned by one vector $\ket{g}$. 
 Then for every vector $\ket{k}\in \ker(T_1)$ we define the Kraus operator $K_k=\ket{g}\bra{k}$. 
 The following map is then CPTP
 \begin{equation*}
 	W_2^{(1)}(X)=T_1XT_1^\dagger+\sum_{k=1}^{\dim \ker(T_1)} K_kXK_k^\dagger \; .
 \end{equation*}
 At the next layer we need to construct a map $W_2^{(2)}$ acting on states on 
 $(\CC^{D_{1}}\oplus \H_g)^{\otimes 2}$. We use $T_2$ to define the first Kraus operator of $W_2^{(2)}$ acting on the $\CC^{D_{1}}\otimes\CC^{D_{1}}$ subspace and mapping it into the $\CC^{D_{2}}$ subspace in  $\CC^{D_{2}}\oplus\H_{g^\prime}$.
 Then for every $\ket{k}$ in 
 $\ker(T_2)\oplus(\CC^{D_{1}}\otimes\H_g)\oplus(\H_g\otimes\CC^{D_{1}})\oplus(\H_g\otimes\H_g)$ 
 we construct a Kraus operator $\ket{g^\prime}\bra{k}$ mapping $\ket{k}$ to $\ket{g^\prime}$ in $\CC^{D_{2}}\oplus \H_{g^\prime}$.
 We iterate the last step using $T_l$ to make $W_2^{(l)}$ for $l=3,\ldots,L$.
 
%

 \section{The scaling of the local translation invariant energy in 2D}
 \label{sec:2D_details}
We solved the first few levels of the 2D LTI hierarchy (see \cref{fig:2DLTI}) for the Heisenberg and XX models on a  square lattice. 
In addition we computed the Anderson bounds \cite{Anderson} for square regions of sizes up to $5\times 5$ by exact diagonalization of the Hamiltonian term acting on each region.

The LTI condition in 2D can be formulated for regions of different shapes. The condition for square-shaped regions and triangle-shaped ones are depicted in \cref{fig:2DLTI}. In the case of square-shaped regions, optimizing over a state on a $4\times 4$ square is beyond the capability of the computer and software that we used (on a machine with 512GB of RAM we could solve systems of up to 13 spins using the SCS solver). In the case of triangular regions we could go up to a  triangle with side length 4.  

\begin{figure}[h]
	\centering
	\includegraphics[width=0.55\textwidth]{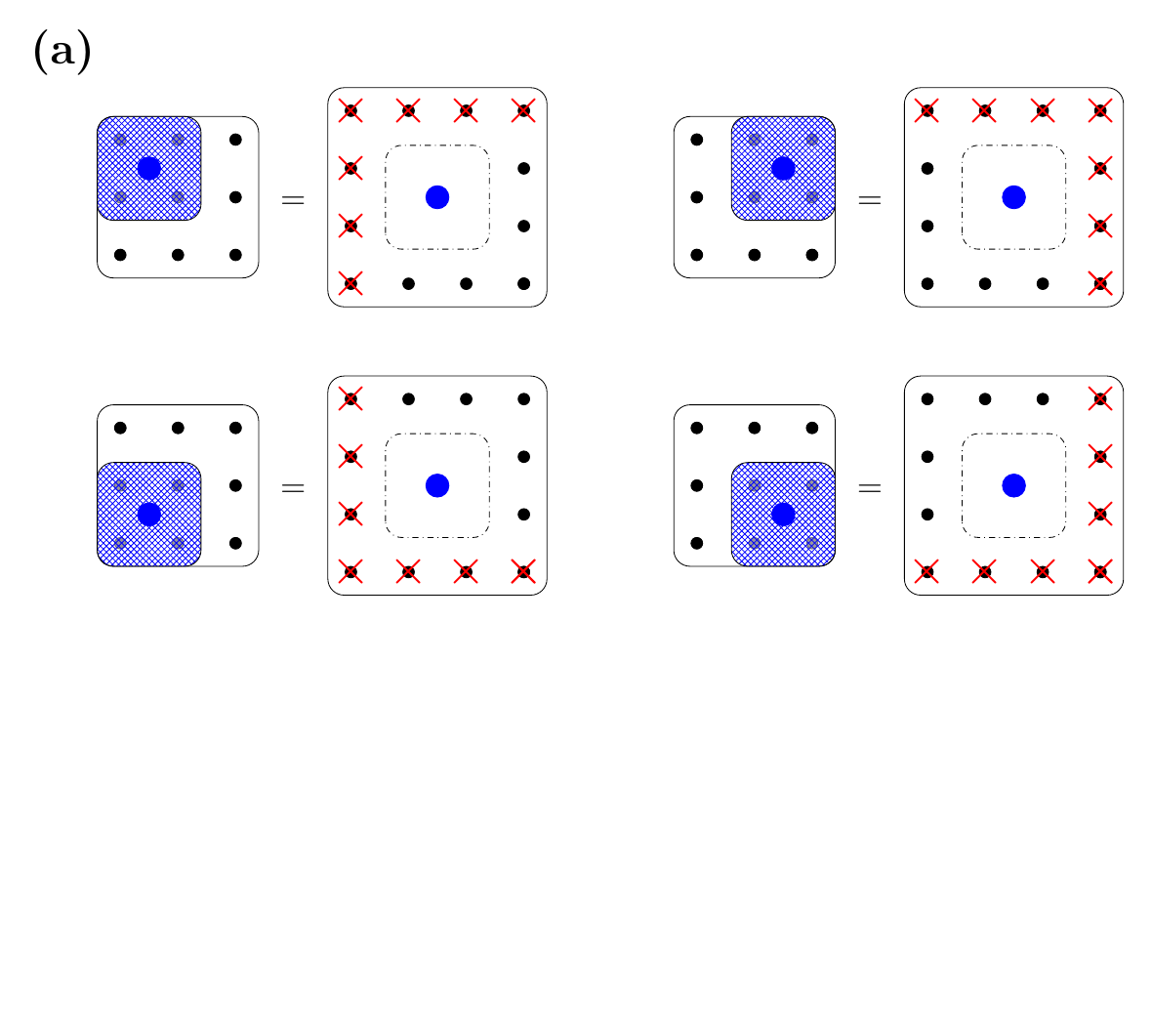}
	\hfill
	\includegraphics[width=0.35\textwidth]{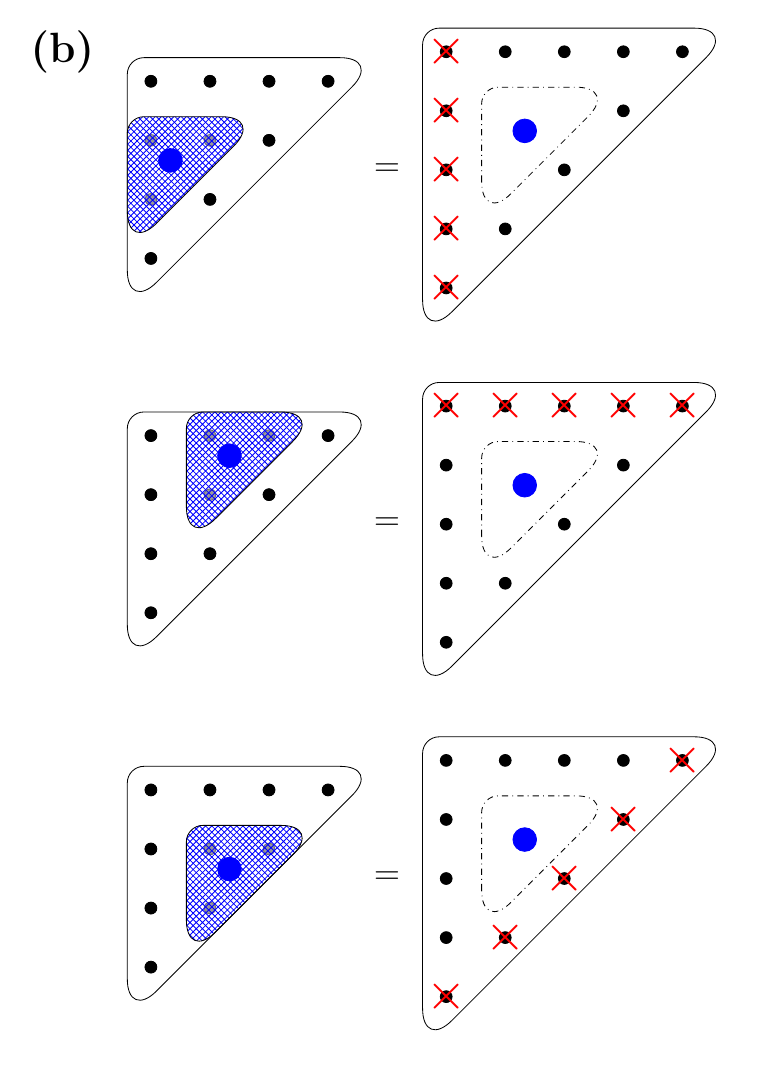}
	\caption{Relaxation of the constraints defining the 3rd and 4th levels of the 2D LTI hierarchies for square-shaped (\textbf{a}) and triangular (\textbf{b}) regions respectively.
		The previous levels of each hierarchy are depicted in \cref{fig:2DLTI}. 
		The notation is the same as in \cref{fig:2DLTI}, where spins are indicated by  black circles and partial traces by red {\textcolor{red}{$\times$}}s.
		In addition, on the right-hand side of each equality sign the spins absent from  the dash-dotted central region  have been   compressed into a $D$-dimensional system, which is  denoted by a big blue circle. On the left-hand side, the shaded region denotes a coarse-graining map that is applied to the spins within that region, mapping them to the $D$-dimensional system denoted by the blue circle.}
	\label{fig:2DLTIrelax}
\end{figure}

Next, by applying a single coarse-graining step we obtained relaxations of  the $4\times 4$ squares and the side-length-$5$ triangle LTI problems that we were still able to solve. The relaxations employed a single coarse-graining map that was applied to the central square or triangle of side length 2 within the $4\times 4$ square or side-length-$5$ triangle respectively. The relaxed constraints are shown in \cref{fig:2DLTIrelax}. The map we chose for the coarse-graining was a partial isometry consisting of the $D$ lowest energy eigenvectors of the Hamiltonian term on the patch of side length 2 (square or triangular respectively).

\begin{figure}[h]
	\centering
	\includegraphics[width=0.450\textwidth]{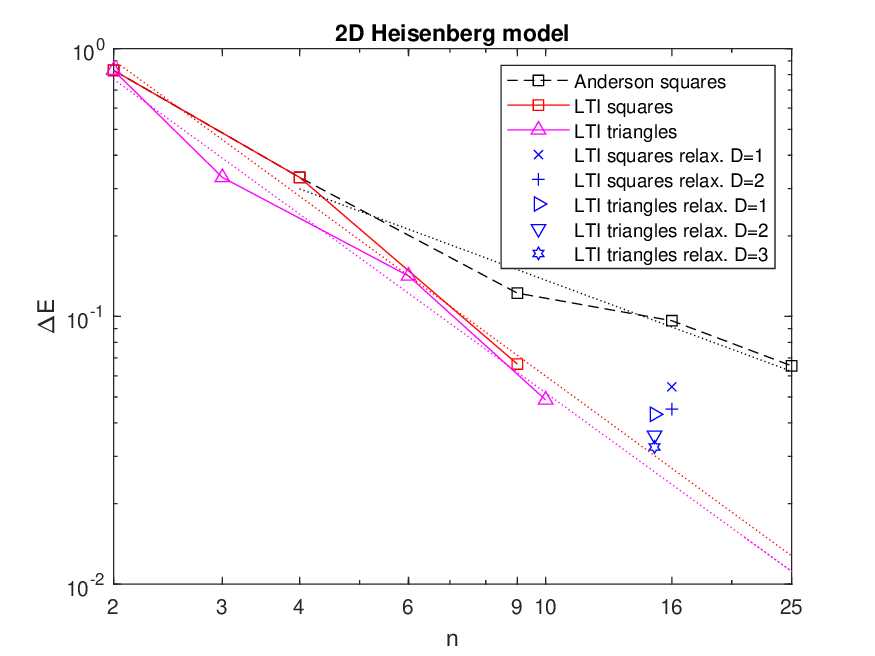}
	\includegraphics[width=0.450\textwidth]{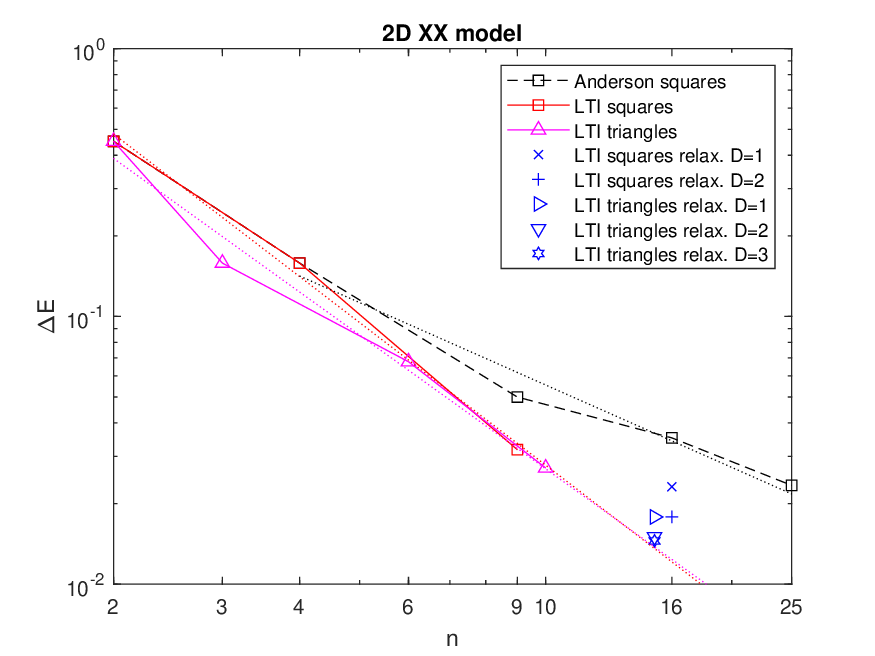}    
	\caption{2D LTI results. 
	The error $\Delta{E}$ of the lower bounds to the ground-state energies of the  Heisenberg and XX models on a 2D square lattice  	
	obtained by  solving 
	the  first two levels of the 2D LTI hierarchy for square-shaped  regions (\cref{fig:2DLTI} (\textbf{a})), the first three levels for triangle-shaped regions (\cref{fig:2DLTI} (\textbf{b})),  
	and the relaxations with coarse-graining dimension $1\leq D\leq 3$ of the next level for both squares and triangles   (\cref{fig:2DLTIrelax} (\textbf{a}) and (\textbf{b}) respectively).
	In addition Anderson bounds for square-shaped regions of sizes up to $5\times5$ are shown. 
	The error $\Delta{E}$ is plotted against the number of spins $n$ in the largest region appearing  in the problem (before compression) on a log-log scale, and was computed with respect to energies obtained from   quantum Monte Carlo computations  (Refs.\ \cite{qmc2DHeisenberg,qmc2DXX}).
	For each of the series: LTI squares, LTI triangles, and Anderson squares a linear extrapolation is plotted by  a dotted line.
}
	\label{fig:2DLTIresults}
\end{figure}

The results are shown in \cref{fig:2DLTIresults}. The figure shows the error of the lower bounds with respect to energies computed by extrapolating quantum  Monte-Carlo results (Ref.\ \cite{qmc2DHeisenberg} for the Heisenberg model and \cite{qmc2DXX} for the XX model) $\Delta{E}=E_{\text{QMC}}-E_{\text{relax.}}$   as a function of the number of spins $n$, on a log-log scale. The relaxation energies $E_{\text{relax.}}$ were  obtained using each of the above mentioned methods: Anderson bound for square regions, LTI bound for square or triangular regions, and the relaxations of the latter using one coarse graining step with coarse-graining dimensions $D=1,2$ and in the triangles case also $D=3$. 
In addition, we plotted the linear extrapolation of the results  for the Anderson and LTI bounds.

	There are several things to note about the results in \cref{fig:2DLTIresults}. First of all we see the markedly steeper slope of the LTI bounds as compared with the Anderson bounds. 
	(The fitted values for the Heisenberg model are:
	$-0.86$ for the Anderson bounds, $-1.69$ for the LTI bounds for squares,  and $-1.68$ for LTI triangles; for the  XX model the values are: $-1.02,-1.77$, and $-1.66$ respectively.)
	A similar  difference between the  LTI and Anderson slopes was observed in all the 1D models (not shown in \cref{fig:results}).
	Second, The slopes of the LTI squares and LTI triangles bounds are very close in both models. It seems that the LTI conditions for the two different shapes have similar power in terms of  ``constraints per spin".
	When comparing the LTI slopes of the 2D models with those observed in 1D we notice that the convergence is faster in 1D (where the scaling was $n^{-2}$).
	Finally, as expected, for increasing coarse-graining dimension $D$ the value of the lower bound from the relaxation defined in \cref{fig:2DLTIrelax} converges towards the extrapolation of the LTI line. (The compression would be lossless for $D=8$ in the case of triangles and $D=16$ in the case of squares.)

These results demonstrate that   our method can be used in 2D to push beyond   what is possible using the LTI hierarchy. The bounds  we obtained with this rudimentary demonstration are already very close to the best lower bounds we are aware of for those systems (given in Ref.\ \cite{BarthelAndHubener}).  
From the linear extrapolation in \cref{fig:2DLTIresults} we can read off which region  sizes have to be reached in a relaxation scheme based on the 2D LTI hierarchy in order to achieve  a desired accuracy.  
The SDPs we solved to produce these results are very simple and are easy  to set up using SDP modeling software packages (50-100 lines of code). Sample code is available at \cite{myCodeDemo}. 
Larger coarse graining dimensions could be handled by implementing matrix-free SCS for this problem.

It also demonstrates the challenges of applying our approach to relax the 2D-LTI hierarchy: to preserve the LTI symmetry of the problem when formulating the next level of the hierarchy we must increase the size of the patch of side length $l$ by $2l+1$ spins in the case of squares, and by $l+1$ spins in the case of triangles. 
In addition, even after coarse-graining we end up with a system consisting of all the spins on the boundary of the patch, which quickly becomes too large.
Finally, it is not obvious how to implement the   composition properties required from the coarse-graining maps if we would like to continue with a further coarse-graining step which is compatible with the LTI symmetry (maps based on PEPS would work but they result in a coarse-graining dimension which is too large).

We already mentioned in the main text in \cref{sec:2D} that those difficulties can be overcome by using a tree coarse-graining procedure, or by first formulating the problem within the RDMT framework and applying the procedure described in \cref{sec:RDMT}.
Another way would be to give  up the explicit LTI symmetry in our relaxation and instead relax a chain of constraints in which one increases the   triangular region   by at most two spins at each step, such that an additional side-length-2 triangle can fit in it. 
The region already covered can be then coarse-grained with an MPS using a snake pattern in order to cover triangles of increasing  size.
In this scheme the system size still grows from one step to the next, but it requires only $l+1$ spins to be kept to relax the 2D LTI problem  for a side-length-$l$ triangle. We can use the extrapolation of the LTI bounds in \cref{fig:2DLTIresults} to estimate the precision we could hope to achieve by implementing this scheme. For example, in order to improve the current best result for the Heisenberg model (Ref.\ \cite{BarthelAndHubener}) by an order of magnitude we would need to solve  for a triangular region of side length $l\gtrsim 11$, which is still a tractable SDP. We leave this for future work. 


\section{Accounting for symmetries}
\label{sec:symmetries}
If the Hamiltonian  has a symmetry we can simplify the ground state energy problem by restricting the search to symmetric states. Such states can be represented with  fewer variables than general states that do not observe the symmetry. 
In \cref{sec:summary} we already used translation  symmetry when formulating the relaxation of the translation-invariant problem by expressing everything in terms of one state of each size. 

Hamiltonians of interest often possess further symmetries. Consider a Hamiltonian which is invariant under the action of a group $G$ simultaneously on all spins:\ $[U_g, H] = 0$, where $U_g:=(\otimes_{i } u_g)$ and $u:G\mapsto \L(\H)$ is a unitary group representation of $G$ on the local spin Hilbert space $\H$.
We will now describe how one can incorporate such symmetries in our relaxation scheme.
 
\begin{figure}[H]
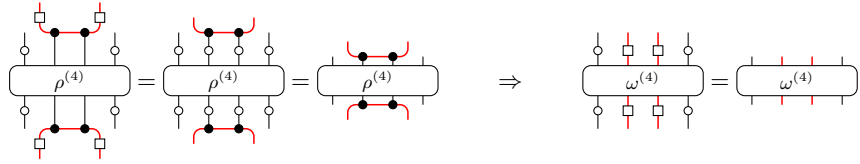

	\begin{equation*}
	\tikzPic{cgSigma4SYMMtwine } = \tikzPic{cgSigma4SYMM } = \tikzPic{cgSigma4 } \hspace{20pt} \Rightarrow \hspace{20pt} 
	\tikzPic{tilde4SYMM } = \tikzPic{tilde4}  
\end{equation*}

\caption{Optimization over symmetric states can be relaxed in terms of symmetric coarse-grained variables. 
The black circles in the diagram represent symmetric MPS tensors which intertwine between  the group action on the physical spins, $u_g$ (white circles), and the virtual representations on the bond degrees of freedom, $v_g$ and $\overline{v}_g$ (white squares). }
\label{fig:intertwineMPS}
\end{figure}

Consider the MPS-based relaxation  scheme in the 1D setting, \cref{eq:locTInRelaxExpl}. In this case we can construct coarse-graining maps from symmetric MPS. Symmetric MPS have the property that they intertwine between the representation $u_g$ on the physical Hilbert space and a representation on the virtual degrees of freedom ($v_g\otimes \overline{v}_g$) \cite{stringOrder}. Now consider for example the symmetry constraint on a $4$-body state $\rho_{\{1,2,3,4\}}$. When we compress this variable using a coarse-graining map constructed out of a symmetric MPS we can use the intertwining property of the MPS tensor to obtain a symmetry condition for the compressed variable $\omega_{\{1,o^l,o^r,4\}}$ as shown in \cref{fig:intertwineMPS}. This means that we can formulate our relaxation in term of symmetric states which can have significantly less free parameters than general states. 
One could similarly use symmetric tree tensor networks \cite{TTNsymm} when formulating \cref{eq:TTNRelaxExpl}.

 \end{document}

%% file: tikzPreFabForNewSection.tex
\tikzset{
	psi/.pic = {
		\tikzmath{ \up=0.4;  \step=0.5;   \n=36; \s=0.22;} 
		\coordinate (R) at (0,0);
		\coordinate (L) at ($ (R) - \n*(\step,0)$);
		\foreach \x in {0,...,\n}
		\draw[-] ($(L)+ (\x*\step,\up)$) -- ($(L)+ (\x*\step,-\up)$) ;
		\draw[fill=white,rounded corners=3pt] ($ (L) - (\s,\s) $) rectangle ($ (R) + (\s,\s) $) ;
		\node at ($ 0.5*(R)+0.5*(L) $) {\tikzpictext};
	}
}

\tikzset{
	psiS/.pic = {
		\tikzmath{ \up=0.4;  \step=0.5;   \n=12; \s=0.22;} 
		\coordinate (R) at (0,0);
		\coordinate (L) at ($ (R) - \n*(\step,0)$);
		\foreach \x in {0,...,\n}
		\draw[-] ($(L)+ (\x*\step,\up)$) -- ($(L)+ (\x*\step,-\up)$) ;
		\draw[fill=white,rounded corners=3pt] ($ (L) - (\s,\s) $) rectangle ($ (R) + (\s,\s) $) ;
		\node at ($ 0.5*(R)+0.5*(L) $) {\tikzpictext};
	}
}

\tikzset{
	iso2/.pic = {
		\tikzmath{  \step=0.5;   \n=3; \s=0.25;} 
	\coordinate (R) at (0,0);
	\coordinate (L) at ($ (R) - \n*(\step,0)$);
	\foreach \x in {1,...,2}
	\draw[-] ($(L)+ (\x*\step,0)$) -- ($(L)+ (\x*\step,-\s)$) ;
	\draw[line width=0.5mm] ($(L)+ (\step+\s,0)$) -- ($(L)+ (\step+\s,0+2*\s)$) ;
	\draw[line width=0.5mm] ($(L)+ (\step+2.5*\s,2*\s)$) -- ($(L)+ (\step+2.5*\s,4*\s)$) ;
	\draw[-] ($(L)+ (3*\step,2*\s)$) -- ($(L)+ (3*\step,-\s)$) ;
	\draw[fill=white,rounded corners=3pt] ($ (L)+(\step-\s,0) $) rectangle ($ (R) + (-\step+\s,0)+ (0,\s) $) ;
	\draw[fill=white,rounded corners=3pt] ($ (L)+(\step-\s,2*\s)+(\s,0) $) rectangle ($ (R) + (0,2*\s)+(\s,\s)$) ;
}
}

\tikzset{
	iso3/.pic = {
	\tikzmath{ \up=0.4;  \step=0.5;   \n=4; \s=0.25;} 
\coordinate (R) at (0,0);
\coordinate (L) at ($ (R) - \n*(\step,0)$);
\foreach \x in {1,...,3}
\draw[-] ($(L)+ (\x*\step,0)$) -- ($(L)+ (\x*\step,-\s)$) ;
\draw[line width=0.5mm] ($(L)+ (\step+2*\s,0)$) -- ($(L)+ (\step+2*\s,0+2*\s)$) ;
\draw[fill=white,rounded corners=3pt] ($ (L)+(\step-\s,0) $) rectangle ($ (R) + (-\step+\s,0)+ (0,\s) $) ;
}
}

\tikzset{
	mpsCirc/.pic = {
		\draw[rounded corners=3pt, line width=0.4mm, red] (-0.5,0)--(0.5,0);
		\draw (0,0)--(0,-0.5);
		\draw[fill] (0,0)  circle (.5ex);
		\node at (0.2,0.2) {\tikzpictext};
	}
}

\tikzset{
	rho/.pic = {
		\tikzmath{\step = 1; \up=0.4;  \step=0.5;   \n=1; \s=0.22;} 
		\coordinate (R) at (0,0);
		\coordinate (L) at ($ (R) - \n*(\step,0)$);
		\foreach \x in {0,...,\n}
		\draw[-] ($(L)+ (\x*\step,\up)$) -- ($(L)+ (\x*\step,-\up)$) ;
		\draw[fill=white,rounded corners=3pt] ($ (L) - (\s,\s) $) rectangle ($ (R) + (\s,\s) $) ;
		\node at ($ 0.5*(R)+0.5*(L) $) {\tikzpictext};		 
	}
}

\tikzset{
	sigma3Tilde/.pic = {
		\tikzmath{ \up=0.4;  \step=0.5;   \n=2; \s=0.22;} 
		\coordinate (R) at (0,0);
		\coordinate (L) at ($ (R) - \n*(\step,0)$);
		\foreach \x in {0,\n}
		\draw[-] ($(L)+ (\x*\step,\up)$) -- ($(L)+ (\x*\step,-\up)$) ;
		\draw[-,thick,red] ($(L)+ (\step,\up)$) -- ($(L)+ (\step,-\up)$) ;
		\draw[fill=white,rounded corners=3pt] ($ (L) - (\s,\s) $) rectangle ($ (R) + (\s,\s) $) ;
		\node at ($ 0.5*(R)+0.5*(L) $) {\tikzpictext};
	}
}

\tikzset{
	sigma3IsoR/.pic = {
 			\tikzmath{ \up=0.4;  \step=0.5;   \n=2; \s=0.25;} 
			\coordinate (R) at (0,0);
			\coordinate (L) at ($ (R) - \n*(\step,0)$);
			\foreach \x in {0,...,\n}
			\draw[-] ($(L)+ (\x*\step,\up)$) -- ($(L)+ (\x*\step,-\up)$) ;
			\draw[fill=white,rounded corners=3pt] ($ (L) - (\s,\s) $) rectangle ($ (R) + (\s,\s) $) ;
			\node at ($ 0.5*(R)+0.5*(L) $) {\tikzpictext};
			\draw[line width=0.5mm] ($(L)+ (\step+\s,\up)$) -- ($(L)+ (\step+\s,\up+2*\s)$) ;
			\draw[line width=0.5mm] ($(L)+ (\step+\s,-\up)$) -- ($(L)+ (\step+\s,-\up-2*\s)$) ;
			\draw[fill=white,rounded corners=3pt] ($ (L)+(\step-\s,\up) $) rectangle ($ (R) + (\s,\up)+ (0,\s) $) ;
			\draw[fill=white,rounded corners=3pt] ($ (L)+(\step-\s,-\up-\s) $) rectangle ($ (R) + (\s,-\up) $) ;
	}
}

\tikzset{
	sigma3IsoL/.pic = {
	\tikzmath{ \up=0.4;  \step=0.5;   \n=2; \s=0.25;} 
	\coordinate (R) at (0,0);
	\coordinate (L) at ($ (R) - \n*(\step,0)$);
	\foreach \x in {0,...,\n}
	\draw[-] ($(L)+ (\x*\step,\up)$) -- ($(L)+ (\x*\step,-\up)$) ;
	\draw[fill=white,rounded corners=3pt] ($ (L) - (\s,\s) $) rectangle ($ (R) + (\s,\s) $) ;
	\node at ($ 0.5*(R)+0.5*(L) $) {\tikzpictext};
	\draw[line width=0.5mm] ($(L)+ ( +\s,\up)$) -- ($(L)+ (+\s,\up+2*\s)$) ;
	\draw[line width=0.5mm] ($(L)+ ( +\s,-\up)$) -- ($(L)+ (+\s,-\up-2*\s)$) ;
	\draw[fill=white,rounded corners=3pt] ($ (L)+(-\s,\up) $) rectangle ($ (R) + (-\step+\s,\up)+ (0,\s) $) ;
	\draw[fill=white,rounded corners=3pt] ($ (L)+(-\s,-\up-\s) $) rectangle ($ (R) + (-\step+\s,-\up) $) ;
	}
}

\tikzset{
	sigma4IsoR/.pic = {
		\tikzmath{ \up=0.4;  \step=0.5;   \n=3; \s=0.25;} 
		\coordinate (R) at (0,0);
		\coordinate (L) at ($ (R) - \n*(\step,0)$);
		\foreach \x in {0,...,\n}
		\draw[-] ($(L)+ (\x*\step,\up)$) -- ($(L)+ (\x*\step,-\up)$) ;
		\draw[fill=white,rounded corners=3pt] ($ (L) - (\s,\s) $) rectangle ($ (R) + (\s,\s) $) ;
		\node at ($ 0.5*(R)+0.5*(L) $) {\tikzpictext};
		\draw[line width=0.5mm] ($(L)+ (\step+2*\s,\up)$) -- ($(L)+ (\step+2*\s,\up+2*\s)$) ;
		\draw[line width=0.5mm] ($(L)+ (\step+2*\s,-\up)$) -- ($(L)+ (\step+2*\s,-\up-2*\s)$) ;
		\draw[fill=white,rounded corners=3pt] ($ (L)+(\step-\s,\up) $) rectangle ($ (R) + (\s,\up)+ (0,\s) $) ;
		\draw[fill=white,rounded corners=3pt] ($ (L)+(\step-\s,-\up-\s) $) rectangle ($ (R) + (\s,-\up) $) ;
	}
}

\tikzset{
	sigma4IsoL/.pic = {
		\tikzmath{ \up=0.4;  \step=0.5;   \n=3; \s=0.25;} 
		\coordinate (R) at (0,0);
		\coordinate (L) at ($ (R) - \n*(\step,0)$);
		\foreach \x in {0,...,\n}
		\draw[-] ($(L)+ (\x*\step,\up)$) -- ($(L)+ (\x*\step,-\up)$) ;
		\draw[fill=white,rounded corners=3pt] ($ (L) - (\s,\s) $) rectangle ($ (R) + (\s,\s) $) ;
		\node at ($ 0.5*(R)+0.5*(L) $) {\tikzpictext};
		\draw[line width=0.5mm] ($(L)+ ( +2*\s,\up)$) -- ($(L)+ (+2*\s,\up+2*\s)$) ;
		\draw[line width=0.5mm] ($(L)+ ( +2*\s,-\up)$) -- ($(L)+ (+2*\s,-\up-2*\s)$) ;
		\draw[fill=white,rounded corners=3pt] ($ (L)+(-\s,\up) $) rectangle ($ (R) + (-\step+\s,\up)+ (0,\s) $) ;
		\draw[fill=white,rounded corners=3pt] ($ (L)+(-\s,-\up-\s) $) rectangle ($ (R) + (-\step+\s,-\up) $) ;
	}
}

\tikzset{
	sigma4IsoC/.pic = {
	\tikzmath{ \up=0.4;  \step=0.5;   \n=3; \s=0.25;} 
	\coordinate (R) at (0,0);
	\coordinate (L) at ($ (R) - \n*(\step,0)$);
	\foreach \x in {0,...,3}
	\draw[-] ($(L)+ (\x*\step,\up)$) -- ($(L)+ (\x*\step,-\up)$) ;
	\draw[fill=white,rounded corners=3pt] ($ (L) - (\s,\s) $) rectangle ($ (R) + (\s,\s) $) ;
	\node at ($ 0.5*(R)+0.5*(L) $) {\tikzpictext};
	\draw[line width=0.5mm] ($(L)+ (\step+\s,\up)$) -- ($(L)+ (\step+\s,\up+2*\s)$) ;
	\draw[line width=0.5mm] ($(L)+ (\step+\s,-\up)$) -- ($(L)+ (\step+\s,-\up-2*\s)$) ;
	\draw[fill=white,rounded corners=3pt] ($ (L)+(\step-\s,\up) $) rectangle ($ (R) + (-\step+\s,\up)+ (0,\s) $) ;
	\draw[fill=white,rounded corners=3pt] ($ (L)+(\step-\s,-\up-\s) $) rectangle ($ (R) + (-\step+\s,-\up) $) ;
	}
}

\tikzset{
	sigma4IsoCTrR/.pic = {
	\tikzmath{ \up=0.4;  \step=0.5;   \n=3; \s=0.25;} 
	\coordinate (R) at (0,0);
	\coordinate (L) at ($ (R) - \n*(\step,0)$);
	\draw[rounded corners=3pt,white] ($(L)+(0,\up)$) -- ($(L)+(-\step,\up)$) --  ($(L)+(-\step,-\up)$) -- ($(L)+(0,-\up)$) --cycle;
	\foreach \x in {0,...,2}
	\draw[-] ($(L)+ (\x*\step,\up)$) -- ($(L)+ (\x*\step,-\up)$) ;
	\draw[rounded corners=3pt] ($(R)+(0,\up)$) -- ($(R)+(\step,\up)$) --  ($(R)+(\step,-\up)$) -- ($(R)+(0,-\up)$) -- cycle ;
	\draw[fill=white,rounded corners=3pt] ($ (L) - (\s,\s) $) rectangle ($ (R) + (\s,\s) $) ;
	\node at ($ 0.5*(R)+0.5*(L) $) {\tikzpictext};
	\draw[line width=0.5mm] ($(L)+ (\step+\s,\up)$) -- ($(L)+ (\step+\s,\up+2*\s)$) ;
	\draw[line width=0.5mm] ($(L)+ (\step+\s,-\up)$) -- ($(L)+ (\step+\s,-\up-2*\s)$) ;
	\draw[fill=white,rounded corners=3pt] ($ (L)+(\step-\s,\up) $) rectangle ($ (R) + (-\step+\s,\up)+ (0,\s) $) ;
	\draw[fill=white,rounded corners=3pt] ($ (L)+(\step-\s,-\up-\s) $) rectangle ($ (R) + (-\step+\s,-\up) $) ;
	}
}

\tikzset{
	sigma4IsoCTrL/.pic = {
		\tikzmath{ \up=0.4;  \step=0.5;   \n=3; \s=0.25;} 
	\coordinate (R) at (0,0);
	\coordinate (L) at ($ (R) - \n*(\step,0)$);
	\foreach \x in {1,...,\n}
	\draw[-] ($(L)+ (\x*\step,\up)$) -- ($(L)+ (\x*\step,-\up)$) ;
	\draw[rounded corners=3pt] ($(L)+(0,\up)$) -- ($(L)+(-\step,\up)$) --  ($(L)+(-\step,-\up)$) -- ($(L)+(0,-\up)$) -- cycle;
	\draw[fill=white,rounded corners=3pt] ($ (L) - (\s,\s) $) rectangle ($ (R) + (\s,\s) $) ;
	\node at ($ 0.5*(R)+0.5*(L) $) {\tikzpictext};
	\draw[line width=0.5mm] ($(L)+ (\step+\s,\up)$) -- ($(L)+ (\step+\s,\up+2*\s)$) ;
	\draw[line width=0.5mm] ($(L)+ (\step+\s,-\up)$) -- ($(L)+ (\step+\s,-\up-2*\s)$) ;
	\draw[fill=white,rounded corners=3pt] ($ (L)+(\step-\s,\up) $) rectangle ($ (R) + (-\step+\s,\up)+ (0,\s) $) ;
	\draw[fill=white,rounded corners=3pt] ($ (L)+(\step-\s,-\up-\s) $) rectangle ($ (R) + (-\step+\s,-\up) $) ;
	}
}

\tikzset{
	sigma5IsoCTrR/.pic = {
		\tikzmath{ \up=0.4;  \step=0.5;   \n=4; \s=0.25;} 
		\coordinate (R) at (0,0);
		\coordinate (L) at ($ (R) - \n*(\step,0)$);
		\draw[rounded corners=3pt,white] ($(L)+(0,\up)$) -- ($(L)+(-\step,\up)$) --  ($(L)+(-\step,-\up)$) -- ($(L)+(0,-\up)$) --cycle;
		\foreach \x in {0,...,3}
		\draw[-] ($(L)+ (\x*\step,\up)$) -- ($(L)+ (\x*\step,-\up)$) ;
		\draw[rounded corners=3pt] ($(R)+(0,\up)$) -- ($(R)+(\step,\up)$) --  ($(R)+(\step,-\up)$) -- ($(R)+(0,-\up)$) -- cycle ;
		\draw[fill=white,rounded corners=3pt] ($ (L) - (\s,\s) $) rectangle ($ (R) + (\s,\s) $) ;
		\node at ($ 0.5*(R)+0.5*(L) $) {\tikzpictext};
		\draw[line width=0.5mm] ($(L)+ (\step+2*\s,\up)$) -- ($(L)+ (\step+2*\s,\up+2*\s)$) ;
		\draw[line width=0.5mm] ($(L)+ (\step+2*\s,-\up)$) -- ($(L)+ (\step+2*\s,-\up-2*\s)$) ;
		\draw[fill=white,rounded corners=3pt] ($ (L)+(\step-\s,\up) $) rectangle ($ (R) + (-\step+\s,\up)+ (0,\s) $) ;
		\draw[fill=white,rounded corners=3pt] ($ (L)+(\step-\s,-\up-\s) $) rectangle ($ (R) + (-\step+\s,-\up) $) ;
	}
}

\tikzset{
	sigma5IsoCTrL/.pic = {
		\tikzmath{ \up=0.4;  \step=0.5;   \n=4; \s=0.25;} 
		\coordinate (R) at (0,0);
		\coordinate (L) at ($ (R) - \n*(\step,0)$);
		\foreach \x in {1,...,\n}
		\draw[-] ($(L)+ (\x*\step,\up)$) -- ($(L)+ (\x*\step,-\up)$) ;
		\draw[rounded corners=3pt] ($(L)+(0,\up)$) -- ($(L)+(-\step,\up)$) --  ($(L)+(-\step,-\up)$) -- ($(L)+(0,-\up)$) -- cycle;
		\draw[fill=white,rounded corners=3pt] ($ (L) - (\s,\s) $) rectangle ($ (R) + (\s,\s) $) ;
		\node at ($ 0.5*(R)+0.5*(L) $) {\tikzpictext};
		\draw[line width=0.5mm] ($(L)+ (\step+2*\s,\up)$) -- ($(L)+ (\step+2*\s,\up+2*\s)$) ;
		\draw[line width=0.5mm] ($(L)+ (\step+2*\s,-\up)$) -- ($(L)+ (\step+2*\s,-\up-2*\s)$) ;
		\draw[fill=white,rounded corners=3pt] ($ (L)+(\step-\s,\up) $) rectangle ($ (R) + (-\step+\s,\up)+ (0,\s) $) ;
		\draw[fill=white,rounded corners=3pt] ($ (L)+(\step-\s,-\up-\s) $) rectangle ($ (R) + (-\step+\s,-\up) $) ;
	}
}

\tikzset{
	sigmaTilde1/.pic = {
		\tikzmath{ \up=0.4;  \step=0.35;   \n=3; \s=0.25;} 
	\coordinate (R) at (0,0);
	\coordinate (L) at ($ (R) - \n*(\step,0)$);
	\foreach \x in {0,3}
	\draw[-] ($(L)+ (\x*\step,\up)$) -- ($(L)+ (\x*\step,-\up)$) ;
	\draw[line width=0.5mm] ($(L)+ (1.5*\step,-\up)$) -- ($(L)+ (1.5*\step,\up)$) ;
	\draw[fill=white,rounded corners=3pt] ($ (L) - (\s,\s) $) rectangle ($ (R) + (\s,\s) $) ;
	\node at ($ 0.5*(R)+0.5*(L) $) {\tikzpictext};
		}
}

\tikzset{
	sigmaTilde1TrR/.pic = {
	\tikzmath{ \up=0.4;  \step=0.35;   \n=3; \s=0.25;} 
	\coordinate (R) at (0,0);
	\coordinate (L) at ($ (R) - \n*(\step,0)$);
	\draw[rounded corners=3pt,white] ($(L)+(0,\up)$) -- ($(L)+(-\step,\up)$) --  ($(L)+(-\step,-\up)$) -- ($(L)+(0,-\up)$) --cycle;		 
		\foreach \x in {0}
		\draw[-] ($(L)+ (\x*\step,\up)$) -- ($(L)+ (\x*\step,-\up)$) ;
	\draw[rounded corners=3pt] ($(R)+(0,\up)$) -- ($(R)+(\step,\up)$) --  ($(R)+(\step,-\up)$) -- ($(R)+(0,-\up)$) -- cycle ;
	\draw[line width=0.5mm] ($(L)+ (1.5*\step,-\up)$) -- ($(L)+ (1.5*\step,\up)$) ;
	\draw[fill=white,rounded corners=3pt] ($ (L) - (\s,\s) $) rectangle ($ (R) + (\s,\s) $) ;
	\node at ($ 0.5*(R)+0.5*(L) $) {\tikzpictext};
}
}

\tikzset{
	sigmaTilde1TrL/.pic = {
\tikzmath{ \up=0.4;  \step=0.35;   \n=3; \s=0.25;} 
\coordinate (R) at (0,0);
\coordinate (L) at ($ (R) - \n*(\step,0)$);
\draw[rounded corners=3pt] ($(L)+(0,\up)$) -- ($(L)+(-\step,\up)$) --  ($(L)+(-\step,-\up)$) -- ($(L)+(0,-\up)$) -- cycle;
\draw[-] ($(L)+ (3*\step,\up)$) -- ($(L)+ (3*\step,-\up)$) ;
\draw[line width=0.5mm] ($(L)+ (1.5*\step,-\up)$) -- ($(L)+ (1.5*\step,\up)$) ;
\draw[fill=white,rounded corners=3pt] ($ (L) - (\s,\s) $) rectangle ($ (R) + (\s,\s) $) ;
\node at ($ 0.5*(R)+0.5*(L) $) {\tikzpictext};
}
}

\tikzset{
	sigma3/.pic = {
		\tikzmath{ \up=0.4;  \step=0.5;   \n=2; \s=0.22;} 
		\coordinate (R) at (0,0);
		\coordinate (L) at ($ (R) - \n*(\step,0)$);
		\foreach \x in {0,...,\n}
		\draw[-] ($(L)+ (\x*\step,\up)$) -- ($(L)+ (\x*\step,-\up)$) ;
		\draw[fill=white,rounded corners=3pt] ($ (L) - (\s,\s) $) rectangle ($ (R) + (\s,\s) $) ;
		\node at ($ 0.5*(R)+0.5*(L) $) {$\rhoM{3}$};
	}
}

\tikzset{
	sigma4/.pic = {
		\tikzmath{ \up=0.4;  \step=0.5;   \n=3; \s=0.22;} 
		\coordinate (R) at (0,0);
		\coordinate (L) at ($ (R) - \n*(\step,0)$);
		\foreach \x in {0,...,\n}
		\draw[-] ($(L)+ (\x*\step,\up)$) -- ($(L)+ (\x*\step,-\up)$) ;
		\draw[fill=white,rounded corners=3pt] ($ (L) - (\s,\s) $) rectangle ($ (R) + (\s,\s) $) ;
		\node at ($ 0.5*(R)+0.5*(L) $) {$\rhoM{4}$};
	}
}

\tikzset{
	sigma5/.pic = {
		\tikzmath{ \up=0.4;  \step=0.5;   \n=4; \s=0.22;} 
		\coordinate (R) at (0,0);
		\coordinate (L) at ($ (R) - \n*(\step,0)$);
		\foreach \x in {0,...,\n}
		\draw[-] ($(L)+ (\x*\step,\up)$) -- ($(L)+ (\x*\step,-\up)$) ;
		\draw[fill=white,rounded corners=3pt] ($ (L) - (\s,\s) $) rectangle ($ (R) + (\s,\s) $) ;
		\node at ($ 0.5*(R)+0.5*(L) $) {\tikzpictext};
	}
}

\tikzset{
	sigma6/.pic = {
		\tikzmath{ \up=0.4;  \step=0.5;   \n=5; \s=0.25;} 
		\coordinate (R) at (0,0);
		\coordinate (L) at ($ (R) - \n*(\step,0)$);
		\foreach \x in {0,...,\n}
		\draw[-] ($(L)+ (\x*\step,\up)$) -- ($(L)+ (\x*\step,-\up)$) ;
		\draw[fill=white,rounded corners=3pt] ($ (L) - (\s,\s) $) rectangle ($ (R) + (\s,\s) $) ;
		\node at ($ 0.5*(R)+0.5*(L) $) {\tikzpictext};
	}
}

\tikzset{
	TLSigma3/.pic = {
	 	\tikzmath{\scale = 1; \up=0.4;  \step=0.5;   \n=2; \s=0.25;} 
		 \coordinate (R) at (0,0);
		 \coordinate (L) at ($ (R) - \n*(\step,0)$);
		 \foreach \x in {1,...,\n}
		 \draw[-] ($(L)+ (\x*\step,\up)$) -- ($(L)+ (\x*\step,-\up)$) ;
		 \draw[rounded corners=3pt] ($(L)+(0,\up)$) -- ($(L)+(-\step,\up)$) --  ($(L)+(-\step,-\up)$) -- ($(L)+(0,-\up)$) -- cycle ;
		 \draw[fill=white,rounded corners=3pt] ($ (L) - (\s,\s) $) rectangle ($ (R) + (\s,\s) $) ;
		 \node at ($ 0.5*(R)+0.5*(L) $) {$\rhoM{3}$};
	}
}

\tikzset{
	TRSigma3/.pic = {
		\tikzmath{\scale = 1; \up=0.4;  \step=0.5;   \n=2; \s=0.25;} 
		\coordinate (R) at (0,0);
		\coordinate (L) at ($ (R) - \n*(\step,0)$);
		\draw[rounded corners=3pt,white] ($(L)+(0,\up)$) -- ($(L)+(-\step,\up)$) --  ($(L)+(-\step,-\up)$) -- ($(L)+(0,-\up)$) -- cycle ;
		\foreach \x in {0,...,1}
		\draw[-] ($(L)+ (\x*\step,\up)$) -- ($(L)+ (\x*\step,-\up)$) ;
		\draw[rounded corners=3pt] ($(R)+(0,\up)$) -- ($(R)+(\step,\up)$) --  ($(R)+(\step,-\up)$) -- ($(R)+(0,-\up)$) -- cycle;
		\draw[fill=white,rounded corners=3pt] ($ (L) - (\s,\s) $) rectangle ($ (R) + (\s,\s) $) ;
		\node at ($ 0.5*(R)+0.5*(L) $) {$\rhoM{3}$};
	}
}

\tikzset{
	TLSigma4/.pic = {
			\tikzmath{ \up=0.4;  \step=0.5;   \n=3; \s=0.25;} 
		\coordinate (R) at (0,0);
		\coordinate (L) at ($ (R) - \n*(\step,0)$);
		\foreach \x in {1,...,\n}
		\draw[-] ($(L)+ (\x*\step,\up)$) -- ($(L)+ (\x*\step,-\up)$) ;
		\draw[rounded corners=3pt] ($(L)+(0,\up)$) -- ($(L)+(-\step,\up)$) --  ($(L)+(-\step,-\up)$) -- ($(L)+(0,-\up)$) -- cycle ;
		\draw[fill=white,rounded corners=3pt] ($ (L) - (\s,\s) $) rectangle ($ (R) + (\s,\s) $) ;
		\node at ($ 0.5*(R)+0.5*(L) $) {$\rhoM{4}$};
	}
}

\tikzset{
	TRSigma4/.pic = {
			\tikzmath{ \up=0.4;  \step=0.5;   \n=3; \s=0.25;} 
		\coordinate (R) at (0,0);
		\coordinate (L) at ($ (R) - \n*(\step,0)$);
		\draw[rounded corners=3pt,white] ($(L)+(0,\up)$) -- ($(L)+(-\step,\up)$) --  ($(L)+(-\step,-\up)$) -- ($(L)+(0,-\up)$) -- cycle ;
		\foreach \x in {0,...,2}
		\draw[-] ($(L)+ (\x*\step,\up)$) -- ($(L)+ (\x*\step,-\up)$) ;
		\draw[rounded corners=3pt] ($(R)+(0,\up)$) -- ($(R)+(\step,\up)$) --  ($(R)+(\step,-\up)$) -- ($(R)+(0,-\up)$) -- cycle ;
		\draw[fill=white,rounded corners=3pt] ($ (L) - (\s,\s) $) rectangle ($ (R) + (\s,\s) $) ;
		\node at ($ 0.5*(R)+0.5*(L) $) {$\rhoM{4}$};
	}
}

\tikzset{
	cgLSigma3/.pic = {
	\tikzmath{ \up=0.4;  \step=0.5;   \n=2;\s=0.25;} 
	\coordinate (R) at (0,0);
	\coordinate (L) at ($ (R) - \n*(\step,0)$);
	\foreach \x in {0,...,\n}
	\draw[-] ($(L)+ (\x*\step,\up)$) -- ($(L)+ (\x*\step,-\up)$) ;
	\draw[fill=white,rounded corners=3pt] ($ (L) - (\s,\s) $) rectangle ($ (R) + (\s,\s) $) ;
	\node at ($ 0.5*(R)+0.5*(L) $) {$\rhoM{3}$};
	\draw[rounded corners=3pt, line width=0.2mm, red]  ($(L)+ (0,\up)$) -- ($(L)+ (\step,\up)$);
	\draw[rounded corners=3pt, line width=0.2mm, red] ($(L)+(0,\up)$) -- ($(L)+(0,\up)+(-0.5*\step,0)$) -- ($(L)+(0,\up)+(-0.5*\step,0.5*\step)$);
	\draw[rounded corners=3pt, line width=0.2mm, red] ($(R)+(-\step,\up)$) -- ($(R)+(-\step,\up)+(0.5*\step,0)$) -- ($(R)+(-\step,\up)+(0.5*\step,0.5*\step)$);
	\foreach \x in {0,...,1}
	\draw[fill] ($(L)+ (\x*\step,\up)$)  circle (.5ex);
	\draw[rounded corners=3pt, line width=0.2mm, red]  ($(L)+ (0,-\up)$) -- ($(L)+ (\step,-\up)$);
	\draw[rounded corners=3pt, line width=0.2mm, red] ($(L)+(0,-\up)$) -- ($(L)+(0,-\up)+(-0.5*\step,0)$) -- ($(L)+(0,-\up)+(-0.5*\step,-0.5*\step)$);
	\draw[rounded corners=3pt, line width=0.2mm, red] ($(R)+(-\step,-\up)$) -- ($(R)+(-\step,-\up)+(0.5*\step,0)$) -- ($(R)+(-\step,-\up)+(0.5*\step,-0.5*\step)$);
	\foreach \x in {0,...,1}
	\draw[fill] ($(L)+ (\x*\step,-\up)$)  circle (.5ex);
	}
}

\tikzset{
	cgRSigma3/.pic = {
			\tikzmath{ \up=0.4;  \step=0.5;   \n=2;\s=0.25;} 
		\coordinate (R) at (0,0);
		\coordinate (L) at ($ (R) - \n*(\step,0)$);
		\foreach \x in {0,...,\n}
		\draw[-] ($(L)+ (\x*\step,\up)$) -- ($(L)+ (\x*\step,-\up)$) ;
		\draw[fill=white,rounded corners=3pt] ($ (L) - (\s,\s) $) rectangle ($ (R) + (\s,\s) $) ;
		\node at ($ 0.5*(R)+0.5*(L) $) {$\rhoM{3}$};
		\draw[rounded corners=3pt, line width=0.2mm, red]  ($(L)+ (\step,\up)$) -- ($(L)+ (2*\step,\up)$);
		\draw[rounded corners=3pt, line width=0.2mm, red] ($(L)+(\step,\up)$) -- ($(L)+(\step,\up)+(-0.5*\step,0)$) -- ($(L)+ (\step,\up)+(-0.5*\step,0.5*\step)$);
		\draw[rounded corners=3pt, line width=0.2mm, red] ($(R)+(0,\up)$) -- ($(R)+(0,\up)+(0.5*\step,0)$) -- ($(R)+(0,\up)+(0.5*\step,0.5*\step)$);
		\foreach \x in {1,...,2}
		\draw[fill] ($(L)+ (\x*\step,\up)$)  circle (.5ex);
		\draw[rounded corners=3pt, line width=0.2mm, red]  ($(L)+ (\step,-\up)$) -- ($(L)+ (2*\step,-\up)$);
		\draw[rounded corners=3pt, line width=0.2mm, red] ($(L)+(\step,-\up)$) -- ($(L)+(\step,-\up)+(-0.5*\step,0)$) -- ($(L)+(\step,-\up)+(-0.5*\step,-0.5*\step)$);
		\draw[rounded corners=3pt, line width=0.2mm, red] ($(R)+(0,-\up)$) -- ($(R)+(0,-\up)+(0.5*\step,0)$) -- ($(R)+(0,-\up)+(0.5*\step,-0.5*\step)$);
		\foreach \x in {1,...,2}
		\draw[fill] ($(L)+ (\x*\step,-\up)$)  circle (.5ex);
	}
}

\tikzset{
	TLcgSigma4/.pic = {
		\tikzmath{ \up=0.4;  \step=0.5;   \n=3;\s=0.25;} 
	\coordinate (R) at (0,0);
	\coordinate (L) at ($ (R) - \n*(\step,0)$);
	\foreach \x in {1,...,\n}
	\draw[-] ($(L)+ (\x*\step,\up)$) -- ($(L)+ (\x*\step,-\up)$) ;
	\draw[rounded corners=3pt] ($(L)+(0,\up)$) -- ($(L)+(-\step,\up)$) --  ($(L)+(-\step,-\up)$) -- ($(L)+(0,-\up)$) -- cycle ;
	\draw[fill=white,rounded corners=3pt] ($ (L) - (\s,\s) $) rectangle ($ (R) + (\s,\s) $) ;
	\node at ($ 0.5*(R)+0.5*(L) $) {$\rhoM{4}$};
	\draw[rounded corners=3pt, line width=0.2mm, red]  ($(L)+ (\step,\up)$) -- ($(L)+ (2*\step,\up)$);
	\draw[rounded corners=3pt, line width=0.2mm, red] ($(L)+(\step,\up)$) -- ($(L)+(\step,\up)+(-0.5*\step,0)$) -- ($(L)+ (\step,\up)+(-0.5*\step,0.5*\step)$);
	\draw[rounded corners=3pt, line width=0.2mm, red] ($(R)+(-\step,\up)$) -- ($(R)+(-\step,\up)+(0.5*\step,0)$) -- ($(R)+(-\step,\up)+(0.5*\step,0.5*\step)$);
	\foreach \x in {1,...,2}
	\draw[fill] ($(L)+ (\x*\step,\up)$)  circle (.5ex);
	\draw[rounded corners=3pt, line width=0.2mm, red]  ($(L)+ (\step,-\up)$) -- ($(L)+ (2*\step,-\up)$);
	\draw[rounded corners=3pt, line width=0.2mm, red]($(L)+(\step,-\up)$) -- ($(L)+(\step,-\up)+(-0.5*\step,0)$) -- ($(L)+(\step,-\up)+(-0.5*\step,-0.5*\step)$);
	\draw[rounded corners=3pt, line width=0.2mm, red] ($(R)+(-\step,-\up)$) -- ($(R)+(-\step,-\up)+(0.5*\step,0)$) -- ($(R)+(-\step,-\up)+(0.5*\step,-0.5*\step)$);
	\foreach \x in {1,...,2}
	\draw[fill] ($(L)+ (\x*\step,-\up)$)  circle (.5ex);
	}
}

\tikzset{
	TRcgSigma4/.pic = {
	 	\tikzmath{ \up=0.4;  \step=0.5;   \n=3;\s=0.25;} 
	 \coordinate (R) at (0,0);
	 \coordinate (L) at ($ (R) - \n*(\step,0)$);
	 \draw[rounded corners=3pt,white] ($(L)+(0,\up)$) -- ($(L)+(-\step,\up)$) --  ($(L)+(-\step,-\up)$) -- ($(L)+(0,-\up)$) -- cycle ;
	 \foreach \x in {0,...,2}
	 \draw[-] ($(L)+ (\x*\step,\up)$) -- ($(L)+ (\x*\step,-\up)$) ;
	 \draw[rounded corners=3pt] ($(R)+(0,\up)$) -- ($(R)+(\step,\up)$) --  ($(R)+(\step,-\up)$) -- ($(R)+(0,-\up)$) -- cycle ;
	 \draw[fill=white,rounded corners=3pt] ($ (L) - (\s,\s) $) rectangle ($ (R) + (\s,\s) $) ;
	 \node at ($ 0.5*(R)+0.5*(L) $) {$\rhoM{4}$};
	 \draw[rounded corners=3pt, line width=0.2mm, red]  ($(L)+ (\step,\up)$) -- ($(L)+ (2*\step,\up)$);
	 \draw[rounded corners=3pt, line width=0.2mm, red] ($(L)+(\step,\up)$) -- ($(L)+(\step,\up)+(-0.5*\step,0)$) -- ($(L)+ (\step,\up)+(-0.5*\step,0.5*\step)$);
	 \draw[rounded corners=3pt, line width=0.2mm, red] ($(R)+(-\step,\up)$) -- ($(R)+(-\step,\up)+(0.5*\step,0)$) -- ($(R)+(-\step,\up)+(0.5*\step,0.5*\step)$);
	 \foreach \x in {1,...,2}
	 \draw[fill] ($(L)+ (\x*\step,\up)$)  circle (.5ex);
	 \draw[rounded corners=3pt, line width=0.2mm, red]   ($(L)+ (\step,-\up)$) -- ($(L)+ (2*\step,-\up)$);
	 \draw[rounded corners=3pt, line width=0.2mm, red]  ($(L)+(\step,-\up)$) -- ($(L)+(\step,-\up)+(-0.5*\step,0)$) -- ($(L)+(\step,-\up)+(-0.5*\step,-0.5*\step)$);
	 \draw[rounded corners=3pt, line width=0.2mm, red]  ($(R)+(-\step,-\up)$) -- ($(R)+(-\step,-\up)+(0.5*\step,0)$) -- ($(R)+(-\step,-\up)+(0.5*\step,-0.5*\step)$);
	 \foreach \x in {1,...,2}
	 \draw[fill] ($(L)+ (\x*\step,-\up)$)  circle (.5ex);
	}
}

\tikzset{
	cgSigma4/.pic = {
		\tikzmath{ \up=0.4;  \step=0.5;   \n=3;\s=0.25;} 
		\coordinate (R) at (0,0);
		\coordinate (L) at ($ (R) - \n*(\step,0)$);
		\foreach \x in {0,...,3}
		\draw[-] ($(L)+ (\x*\step,\up)$) -- ($(L)+ (\x*\step,-\up)$) ;
		\draw[fill=white,rounded corners=3pt] ($ (L) - (\s,\s) $) rectangle ($ (R) + (\s,\s) $) ;
		\node at ($ 0.5*(R)+0.5*(L) $) {$\rhoM{4}$};
		\draw[rounded corners=3pt, line width=0.2mm, red]  ($(L)+ (\step,\up)$) -- ($(L)+ (2*\step,\up)$);
		\draw[rounded corners=3pt, line width=0.2mm, red] ($(L)+(\step,\up)$) -- ($(L)+(\step,\up)+(-0.5*\step,0)$) -- ($(L)+ (\step,\up)+(-0.5*\step,0.5*\step)$);
		\draw[rounded corners=3pt, line width=0.2mm, red] ($(R)+(-\step,\up)$) -- ($(R)+(-\step,\up)+(0.5*\step,0)$) -- ($(R)+(-\step,\up)+(0.5*\step,0.5*\step)$);
		\foreach \x in {1,...,2}
		\draw[fill] ($(L)+ (\x*\step,\up)$)  circle (.5ex);
		\draw[rounded corners=3pt, line width=0.2mm, red]   ($(L)+ (\step,-\up)$) -- ($(L)+ (2*\step,-\up)$);
		\draw[rounded corners=3pt, line width=0.2mm, red]  ($(L)+(\step,-\up)$) -- ($(L)+(\step,-\up)+(-0.5*\step,0)$) -- ($(L)+(\step,-\up)+(-0.5*\step,-0.5*\step)$);
		\draw[rounded corners=3pt, line width=0.2mm, red]  ($(R)+(-\step,-\up)$) -- ($(R)+(-\step,-\up)+(0.5*\step,0)$) -- ($(R)+(-\step,-\up)+(0.5*\step,-0.5*\step)$);
		\foreach \x in {1,...,2}
		\draw[fill] ($(L)+ (\x*\step,-\up)$)  circle (.5ex);
	}
}

\tikzset{
	Sigma4SYMM/.pic = {
		\tikzmath{ \up=0.4;  \step=0.5;   \n=3;\s=0.25;} 
		\coordinate (R) at (0,0);
		\coordinate (L) at ($ (R) - \n*(\step,0)$);
		\foreach \x in {0,...,3}
		\draw[-] ($(L)+ (\x*\step,2*\up)$) -- ($(L)+ (\x*\step,-2*\up)$) ;
		\draw[fill=white,rounded corners=3pt] ($ (L) - (\s,\s) $) rectangle ($ (R) + (\s,\s) $) ;
		\node at ($ 0.5*(R)+0.5*(L) $) {$\rhoM{4}$};
		\foreach \x in {0,...,3}
		\draw[fill=white] ($(L)+ (\x*\step,1.25*\up)$)  circle (.5ex);
		\foreach \x in {0,...,3}
		\draw[fill=white] ($(L)+ (\x*\step,-1.25*\up)$)  circle (.5ex);
		
	}
}

\tikzset{
	cgSigma4SYMM/.pic = {
		\tikzmath{ \up=0.4;  \step=0.5;   \n=3;\s=0.25;} 
		\coordinate (R) at (0,0);
		\coordinate (L) at ($ (R) - \n*(\step,0)$);
		 \foreach \x in {0,...,3}
		\draw[-] ($(L)+ (\x*\step,2*\up)$) -- ($(L)+ (\x*\step,-2*\up)$) ;
		\draw[fill=white,rounded corners=3pt] ($ (L) - (\s,\s) $) rectangle ($ (R) + (\s,\s) $) ;
		\node at ($ 0.5*(R)+0.5*(L) $) {$\rhoM{4}$};
		\foreach \x in {0,...,3}
		\draw[fill=white] ($(L)+ (\x*\step,1.25*\up)$)  circle (.5ex);
		\foreach \x in {0,...,3}
		\draw[fill=white] ($(L)+ (\x*\step,-1.25*\up)$)  circle (.5ex);
		\draw[rounded corners=3pt, line width=0.2mm, red]  ($(L)+ (\step,2*\up)$) -- ($(L)+ (2*\step,2*\up)$);
		\draw[rounded corners=3pt, line width=0.2mm, red] ($(L)+(\step,2*\up)$) -- ($(L)+(\step,2*\up)+(-0.5*\step,0)$) -- ($(L)+ (\step,2*\up)+(-0.5*\step,0.5*\step)$);
		\draw[rounded corners=3pt, line width=0.2mm, red] ($(R)+(-\step,2*\up)$) -- ($(R)+(-\step,2*\up)+(0.5*\step,0)$) -- ($(R)+(-\step,2*\up)+(0.5*\step,0.5*\step)$);
		\foreach \x in {1,...,2}
		\draw[fill] ($(L)+ (\x*\step,2*\up)$)  circle (.5ex);
		\draw[rounded corners=3pt, line width=0.2mm, red]   ($(L)+ (\step,-2*\up)$) -- ($(L)+ (2*\step,-2*\up)$);
		\draw[rounded corners=3pt, line width=0.2mm, red]  ($(L)+(\step,-2*\up)$) -- ($(L)+(\step,-2*\up)+(-0.5*\step,0)$) -- ($(L)+(\step,-2*\up)+(-0.5*\step,-0.5*\step)$);
		\draw[rounded corners=3pt, line width=0.2mm, red]  ($(R)+(-\step,-2*\up)$) -- ($(R)+(-\step,-2*\up)+(0.5*\step,0)$) -- ($(R)+(-\step,-2*\up)+(0.5*\step,-0.5*\step)$);
		\foreach \x in {1,...,2}
		\draw[fill] ($(L)+ (\x*\step,-2*\up)$)  circle (.5ex);
	}
}

\tikzset{
	cgSigma4SYMMtwine/.pic = {
		\tikzmath{ \up=0.4;  \step=0.5;   \n=3;\s=0.25;\q=0.08;} 
		\coordinate (R) at (0,0);
		\coordinate (L) at ($ (R) - \n*(\step,0)$);
		\foreach \x in {0,...,3}
		\draw[-] ($(L)+ (\x*\step,2*\up)$) -- ($(L)+ (\x*\step,-2*\up)$) ;
		\draw[fill=white,rounded corners=3pt] ($ (L) - (\s,\s) $) rectangle ($ (R) + (\s,\s) $) ;
		\node at ($ 0.5*(R)+0.5*(L) $) {$\rhoM{4}$};
		\draw[rounded corners=3pt, line width=0.2mm, red]  ($(L)+ (\step,2*\up)$) -- ($(L)+ (2*\step,2*\up)$);
		\draw[rounded corners=3pt, line width=0.2mm, red] ($(L)+(\step,2*\up)$) -- ($(L)+(\step,2*\up)+(-0.5*\step,0)$) -- ($(L)+ (\step,2*\up)+(-0.5*\step,1*\step)$);
		\draw[rounded corners=3pt, line width=0.2mm, red] ($(R)+(-\step,2*\up)$) -- ($(R)+(-\step,2*\up)+(0.5*\step,0)$) -- ($(R)+(-\step,2*\up)+(0.5*\step,1*\step)$);
		\foreach \x in {1,...,2}
		\draw[fill] ($(L)+ (\x*\step,2*\up)$)  circle (.5ex);
		\draw[rounded corners=3pt, line width=0.2mm, red]   ($(L)+ (\step,-2*\up)$) -- ($(L)+ (2*\step,-2*\up)$);
		\draw[rounded corners=3pt, line width=0.2mm, red]  ($(L)+(\step,-2*\up)$) -- ($(L)+(\step,-2*\up)+(-0.5*\step,0)$) -- ($(L)+(\step,-2*\up)+(-0.5*\step,-1*\step)$);
		\draw[rounded corners=3pt, line width=0.2mm, red]  ($(R)+(-\step,-2*\up)$) -- ($(R)+(-\step,-2*\up)+(0.5*\step,0)$) -- ($(R)+(-\step,-2*\up)+(0.5*\step,-1*\step)$);
		\foreach \x in {1,...,2}
		\draw[fill] ($(L)+ (\x*\step,-2*\up)$)  circle (.5ex);
		\foreach \x in {0,3}
		\draw[fill=white] ($(L)+ (\x*\step,1.25*\up)$)  circle (.5ex);
		\foreach \x in {0,3}
		\draw[fill=white] ($(L)+ (\x*\step,-1.25*\up)$)  circle (.5ex);
		\draw[fill=white] ($(L)+ (1*\step,2*\up)+(-0.5*\step,0.5*\step) -(\q,\q)$) rectangle ($(L)+ (1*\step,2*\up)+(-0.5*\step,0.5*\step) +(\q,\q)$);
		\draw[fill=white] ($(L)+ (2*\step,2*\up)+(+0.5*\step,0.5*\step)-(\q,\q)$)  rectangle ($(L)+ (2*\step,2*\up)+(+0.5*\step,0.5*\step)+(\q,\q)$);
		\draw[fill=white] ($(L)+ (1*\step,-2*\up)+(-0.5*\step,-0.5*\step)-(\q,\q)$)  rectangle ($(L)+ (1*\step,-2*\up)+(-0.5*\step,-0.5*\step)+(\q,\q)$);
		\draw[fill=white] ($(L)+ (2*\step,-2*\up)+(+0.5*\step,-0.5*\step)-(\q,\q)$)  rectangle ($(L)+ (2*\step,-2*\up)+(+0.5*\step,-0.5*\step)+(\q,\q)$) ;
	}
}

\tikzset{
	tilde4/.pic = {
		\tikzmath{ \up=0.4;  \step=0.5;   \n=3;\s=0.25;} 
		\coordinate (R) at (0,0);
		\coordinate (L) at ($ (R) - \n*(\step,0)$);
		\foreach \x in {0,3}
		\draw[-] ($(L)+ (\x*\step,\up)$) -- ($(L)+ (\x*\step,-\up)$) ;
		\foreach \x in {1,2}
		\draw[rounded corners=2pt, line width=0.2mm, red]  ($(L)+ (\x*\step,\up)$) -- ($(L)+ (\x*\step,-\up)$) ;
		\draw[fill=white,rounded corners=3pt] ($ (L) - (\s,\s) $) rectangle ($ (R) + (\s,\s) $) ;
		\node at ($ 0.5*(R)+0.5*(L) $) {$\omegaM{4}$};
	}
}

\tikzset{
	tilde4SYMM/.pic = {
		\tikzmath{ \up=0.4;  \step=0.5;   \n=3;\s=0.25; \q=0.08;} 
		\coordinate (R) at (0,0);
		\coordinate (L) at ($ (R) - \n*(\step,0)$);
		\foreach \x in {0,3}
		\draw[-] ($(L)+ (\x*\step,2*\up)$) -- ($(L)+ (\x*\step,-2*\up)$) ;
		\foreach \x in {1,2}
		\draw[rounded corners=2pt, line width=0.2mm, red]  ($(L)+ (\x*\step,2*\up)$) -- ($(L)+ (\x*\step,-2*\up)$) ;
		\draw[fill=white,rounded corners=3pt] ($ (L) - (\s,\s) $) rectangle ($ (R) + (\s,\s) $) ;
		\node at ($ 0.5*(R)+0.5*(L) $) {$\omegaM{4}$};
		\foreach \x in {0,3}
		\draw[fill=white] ($(L)+ (\x*\step,1.25*\up)$)  circle (.5ex);
		\foreach \x in {0,3}
		\draw[fill=white] ($(L)+ (\x*\step,-1.25*\up)$)  circle (.5ex);
		\draw[fill=white] ($(L)+ (1*\step,1.25*\up) -(\q,\q)$) rectangle ($(L)+ (1*\step,1.25*\up)+(\q,\q)$);
		\draw[fill=white] ($(L)+ (2*\step,1.25*\up)-(\q,\q)$)  rectangle ($(L)+ (2*\step,1.25*\up)+(\q,\q)$);
		\draw[fill=white] ($(L)+ (1*\step,-1.25*\up)-(\q,\q)$)  rectangle ($(L)+ (1*\step,-1.25*\up)+(\q,\q)$);
		\draw[fill=white] ($(L)+ (2*\step,-1.25*\up)-(\q,\q)$)  rectangle ($(L)+ (2*\step,-1.25*\up)+(\q,\q)$) ;
	}
}

\tikzset{
	TLtilde4/.pic = {
			\tikzmath{ \up=0.4;  \step=0.5;   \n=3;\s=0.25;} 
		\coordinate (R) at (0,0);
		\coordinate (L) at ($ (R) - \n*(\step,0)$);
		\draw[-] ($(L)+ (3*\step,\up)$) -- ($(L)+ (3*\step,-\up)$) ;
		\foreach \x in {1,2}
		\draw[rounded corners=2pt, line width=0.2mm, red]  ($(L)+ (\x*\step,\up)$) -- ($(L)+ (\x*\step,-\up)$) ;
		\draw[rounded corners=3pt] ($(L)+(0,\up)$) -- ($(L)+(-\step,\up)$) --  ($(L)+(-\step,-\up)$) -- ($(L)+(0,-\up)$) -- cycle ;
		\draw[fill=white,rounded corners=3pt] ($ (L) - (\s,\s) $) rectangle ($ (R) + (\s,\s) $) ;
		\node at ($ 0.5*(R)+0.5*(L) $) {$\omegaM{4}$};
	}
}

\tikzset{
	TRtilde4/.pic = {
		 	\tikzmath{ \up=0.4;  \step=0.5;   \n=3;\s=0.25;} 
		 \coordinate (R) at (0,0);
		 \coordinate (L) at ($ (R) - \n*(\step,0)$);
		 \draw[rounded corners=3pt, white] ($(L)+(0,\up)$) -- ($(L)+(-\step,\up)$) --  ($(L)+(-\step,-\up)$) -- ($(L)+(0,-\up)$) -- cycle ;
		 \foreach \x in {0}
		 \draw[-] ($(L)+ (\x*\step,\up)$) -- ($(L)+ (\x*\step,-\up)$) ;
		 \foreach \x in {1,2}
		 \draw[rounded corners=2pt, line width=0.2mm, red] ($(L)+ (\x*\step,\up)$) -- ($(L)+ (\x*\step,-\up)$) ;
		 \draw[rounded corners=3pt] ($(R)+(0,\up)$) -- ($(R)+(\step,\up)$) --  ($(R)+(\step,-\up)$) -- ($(R)+(0,-\up)$) -- cycle ;
		 \draw[fill=white,rounded corners=3pt] ($ (L) - (\s,\s) $) rectangle ($ (R) + (\s,\s) $) ;
		 \node at ($ 0.5*(R)+0.5*(L) $) {$\omegaM{4}$};
	}
}

\tikzset{
	TLSigma5/.pic = {
			\tikzmath{ \up=0.4;  \step=0.5;   \n=4; \s=0.25;} 
		\coordinate (R) at (0,0);
		\coordinate (L) at ($ (R) - \n*(\step,0)$);
		\foreach \x in {1,...,\n}
		\draw[-] ($(L)+ (\x*\step,\up)$) -- ($(L)+ (\x*\step,-\up)$) ;
		\draw[rounded corners=3pt] ($(L)+(0,\up)$) -- ($(L)+(-\step,\up)$) --  ($(L)+(-\step,-\up)$) -- ($(L)+(0,-\up)$) -- cycle ;
		\draw[fill=white,rounded corners=3pt] ($ (L) - (\s,\s) $) rectangle ($ (R) + (\s,\s) $) ;
		\node at ($ 0.5*(R)+0.5*(L) $) {$\rhoM{5}$};	
	}
}

\tikzset{
	cgLSigma4/.pic = {
		\tikzmath{ \up=0.4;  \step=0.5;   \n=3;\s=0.25;} 
	\coordinate (R) at (0,0);
	\coordinate (L) at ($ (R) - \n*(\step,0)$);
	\foreach \x in {0,...,\n}
	\draw[-] ($(L)+ (\x*\step,\up)$) -- ($(L)+ (\x*\step,-\up)$) ;
	\draw[fill=white,rounded corners=3pt] ($ (L) - (\s,\s) $) rectangle ($ (R) + (\s,\s) $) ;
	\node at ($ 0.5*(R)+0.5*(L) $) {$\rhoM{4}$};
	\draw[rounded corners=3pt, line width=0.2mm, red]  ($(L)+ (0,\up)$) -- ($(L)+ (2*\step,\up)$);
	\draw[rounded corners=3pt, line width=0.2mm, red] ($(L)+(0,\up)$) -- ($(L)+(0,\up)+(-0.5*\step,0)$) -- ($(L)+(0,\up)+(-0.5*\step,0.5*\step)$);
	\draw[rounded corners=3pt, line width=0.2mm, red]($(R)+(-\step,\up)$) -- ($(R)+(-\step,\up)+(0.5*\step,0)$) -- ($(R)+(-\step,\up)+(0.5*\step,0.5*\step)$);
	\foreach \x in {0,...,2}
	\draw[fill] ($(L)+ (\x*\step,\up)$)  circle (.5ex);
	\draw[rounded corners=3pt, line width=0.2mm, red]  ($(L)+ (0,-\up)$) -- ($(L)+ (2*\step,-\up)$);
	\draw[rounded corners=3pt, line width=0.2mm, red] ($(L)+(0,-\up)$) -- ($(L)+(0,-\up)+(-0.5*\step,0)$) -- ($(L)+(0,-\up)+(-0.5*\step,-0.5*\step)$);
	\draw[rounded corners=3pt, line width=0.2mm, red] ($(R)+(-\step,-\up)$) -- ($(R)+(-\step,-\up)+(0.5*\step,0)$) -- ($(R)+(-\step,-\up)+(0.5*\step,-0.5*\step)$);
	\foreach \x in {0,...,2}
	\draw[fill] ($(L)+ (\x*\step,-\up)$)  circle (.5ex);
	}
}

\tikzset{
	TLcgSigma5/.pic = {
			\tikzmath{ \up=0.4;  \step=0.5;   \n=4;\s=0.25;} 
		\coordinate (R) at (0,0);
		\coordinate (L) at ($ (R) - \n*(\step,0)$);
		\foreach \x in {1,...,\n}
		\draw[-] ($(L)+ (\x*\step,\up)$) -- ($(L)+ (\x*\step,-\up)$) ;
		\draw[rounded corners=3pt] ($(L)+(0,\up)$) -- ($(L)+(-\step,\up)$) --  ($(L)+(-\step,-\up)$) -- ($(L)+(0,-\up)$) -- cycle ;
		\draw[fill=white,rounded corners=3pt] ($ (L) - (\s,\s) $) rectangle ($ (R) + (\s,\s) $) ;
		\node at ($ 0.5*(R)+0.5*(L) $) {$\rhoM{5}$};
		\draw[rounded corners=3pt, line width=0.2mm, red]  ($(L)+ (\step,\up)$) -- ($(L)+ (3*\step,\up)$);
		\draw[rounded corners=3pt, line width=0.2mm, red] ($(L)+(\step,\up)$) -- ($(L)+(\step,\up)+(-0.5*\step,0)$) -- ($(L)+ (\step,\up)+(-0.5*\step,0.5*\step)$);
		\draw[rounded corners=3pt, line width=0.2mm, red] ($(R)+(-\step,\up)$) -- ($(R)+(-\step,\up)+(0.5*\step,0)$) -- ($(R)+(-\step,\up)+(0.5*\step,0.5*\step)$);
		\foreach \x in {1,...,3}
		\draw[fill] ($(L)+ (\x*\step,\up)$)  circle (.5ex);
		\draw[rounded corners=3pt, line width=0.2mm, red]  ($(L)+ (\step,-\up)$) -- ($(L)+ (3*\step,-\up)$);
		\draw[rounded corners=3pt, line width=0.2mm, red] ($(L)+(\step,-\up)$) -- ($(L)+(\step,-\up)+(-0.5*\step,0)$) -- ($(L)+(\step,-\up)+(-0.5*\step,-0.5*\step)$);
		\draw[rounded corners=3pt, line width=0.2mm, red] ($(R)+(-\step,-\up)$) -- ($(R)+(-\step,-\up)+(0.5*\step,0)$) -- ($(R)+(-\step,-\up)+(0.5*\step,-0.5*\step)$);
		\foreach \x in {1,...,3}
		\draw[fill] ($(L)+ (\x*\step,-\up)$)  circle (.5ex);
	}
}

\tikzset{
	 TLtilde5/.pic = {
	 			\tikzmath{ \up=0.4;  \step=0.5;   \n=3;\s=0.25;} 
	 		\coordinate (R) at (0,0);
	 		\coordinate (L) at ($ (R) - \n*(\step,0)$);
	 		\draw[-] ($(L)+ (3*\step,\up)$) -- ($(L)+ (3*\step,-\up)$) ;
	 		\foreach \x in {1,2}
	 		\draw[rounded corners=2pt, line width=0.2mm, red]  ($(L)+ (\x*\step,\up)$) -- ($(L)+ (\x*\step,-\up)$) ;
	 		\draw[rounded corners=3pt] ($(L)+(0,\up)$) -- ($(L)+(-\step,\up)$) --  ($(L)+(-\step,-\up)$) -- ($(L)+(0,-\up)$) -- cycle ;
	 		\draw[fill=white,rounded corners=3pt] ($ (L) - (\s,\s) $) rectangle ($ (R) + (\s,\s) $) ;
	 		\node at ($ 0.5*(R)+0.5*(L) $) {$\omegaM{5}$};	
	 }
 }

\tikzset{
	TRSigma5/.pic = {
		\tikzmath{ \up=0.4;  \step=0.5;   \n=4; \s=0.25;} 
		\coordinate (R) at (0,0);
		\coordinate (L) at ($ (R) - \n*(\step,0)$);
		\draw[rounded corners=3pt,white] ($(L)+(0,\up)$) -- ($(L)+(-\step,\up)$) --  ($(L)+(-\step,-\up)$) -- ($(L)+(0,-\up)$) -- cycle ;
		\foreach \x in {0,...,3}
		\draw[-] ($(L)+ (\x*\step,\up)$) -- ($(L)+ (\x*\step,-\up)$) ;
		\draw[rounded corners=3pt] ($(R)+(0,\up)$) -- ($(R)+(\step,\up)$) --  ($(R)+(\step,-\up)$) -- ($(R)+(0,-\up)$) -- cycle ;
		\draw[fill=white,rounded corners=3pt] ($ (L) - (\s,\s) $) rectangle ($ (R) + (\s,\s) $) ;
		\node at ($ 0.5*(R)+0.5*(L) $) {$\rhoM{5}$};
	}
}

\tikzset{
	cgLTilde4/.pic = {
		\tikzmath{ \up=0.4;  \step=0.5;   \n=3;\s=0.25;} 
		\coordinate (R) at (0,0);
		\coordinate (L) at ($ (R) - \n*(\step,0)$);
		\foreach \x in {0,...,\n}
		\draw[-] ($(L)+ (\x*\step,\up)$) -- ($(L)+ (\x*\step,-\up)$) ;	
		\foreach \x in {1,2}	
		\draw[rounded corners=2pt, line width=0.2mm, red] ($(L)+ (\x*\step,\up)$) -- ($(L)+ (\x*\step,-\up)$) ;
		\draw[fill=white,rounded corners=3pt] ($ (L) - (\s,\s) $) rectangle ($ (R) + (\s,\s) $) ;
		\node at ($ 0.5*(R)+0.5*(L) $) {$\omegaM{4}$};
		\draw[-] ($(L)+ (0,\up)$) -- ($(L)+ (0,\up+0.25*\step)$);
		\draw[rounded corners=2pt, line width=0.2mm, red] ($(L)+ (\step,\up)$) -- ($(L)+ (\step,\up+0.25*\step)$) -- ($(L)+(-0.5*\step,\up+0.25*\step)$) -- ($(L)+ (-0.5*\step,\up+0.5*\step)$) ;
		\draw[fill] ($(L)+ (0,\up+0.25*\step)$)  circle (.5ex);
		\draw[-] ($(L)+ (0,-\up)$) -- ($(L)+ (0,-\up-0.25*\step)$);
		\draw[rounded corners=2pt, line width=0.2mm, red] ($(L)+ (\step,-\up)$) -- ($(L)+ (\step,-\up-0.25*\step)$) -- ($(L)+(-0.5*\step,-\up-.25*\step)$) -- ($(L)+ (-0.5*\step,-\up-0.5*\step)$) ;
		\draw[fill] ($(L)+ (0,-\up-0.25*\step)$)  circle (.5ex);
	}
}

\tikzset{
	cgRSigma4/.pic = {
			\tikzmath{ \up=0.4;  \step=0.5;   \n=3;\s=0.25;} 
		\coordinate (R) at (0,0);
		\coordinate (L) at ($ (R) - \n*(\step,0)$);
		\foreach \x in {0,...,\n}
		\draw[-] ($(L)+ (\x*\step,\up)$) -- ($(L)+ (\x*\step,-\up)$) ;
		\draw[fill=white,rounded corners=3pt] ($ (L) - (\s,\s) $) rectangle ($ (R) + (\s,\s) $) ;
		\node at ($ 0.5*(R)+0.5*(L) $) {$\rhoM{4}$};
		\draw[rounded corners=3pt, line width=0.2mm, red]  ($(L)+ (\step,\up)$) -- ($(L)+ (3*\step,\up)$);
		\draw[rounded corners=3pt, line width=0.2mm, red] ($(L)+(\step,\up)$) -- ($(L)+(\step,\up)+(-0.5*\step,0)$) -- ($(L)+(\step,\up)+(-0.5*\step,0.5*\step)$);
		\draw[rounded corners=3pt, line width=0.2mm, red] ($(R)+(0,\up)$) -- ($(R)+(0,\up)+(0.5*\step,0)$) -- ($(R)+(0,\up)+(0.5*\step,0.5*\step)$);
		\foreach \x in {1,...,3}
		\draw[fill] ($(L)+ (\x*\step,\up)$)  circle (.5ex);
		\draw[rounded corners=3pt, line width=0.2mm, red]  ($(L)+ (\step,-\up)$) -- ($(L)+ (3*\step,-\up)$);
		\draw[rounded corners=3pt, line width=0.2mm, red] ($(L)+(\step,-\up)$) -- ($(L)+(\step,-\up)+(-0.5*\step,0)$) -- ($(L)+(\step,-\up)+(-0.5*\step,-0.5*\step)$);
		\draw[rounded corners=3pt, line width=0.2mm, red] ($(R)+(0,-\up)$) -- ($(R)+(0,-\up)+(0.5*\step,0)$) -- ($(R)+(0,-\up)+(0.5*\step,-0.5*\step)$);
		\foreach \x in {1,...,3}
		\draw[fill] ($(L)+ (\x*\step,-\up)$)  circle (.5ex);
	}
}

\tikzset{
	TRcgSigma5/.pic = {
			\tikzmath{ \up=0.4;  \step=0.5;   \n=4;\s=0.25;} 
		\coordinate (R) at (0,0);
		\coordinate (L) at ($ (R) - \n*(\step,0)$);
		\draw[rounded corners=3pt,white] ($(L)+(0,\up)$) -- ($(L)+(-\step,\up)$) --  ($(L)+(-\step,-\up)$) -- ($(L)+(0,-\up)$) -- cycle ;
		\foreach \x in {0,...,3}
		\draw[-] ($(L)+ (\x*\step,\up)$) -- ($(L)+ (\x*\step,-\up)$) ;
		\draw[rounded corners=3pt] ($(R)+(0,\up)$) -- ($(R)+(\step,\up)$) --  ($(R)+(\step,-\up)$) -- ($(R)+(0,-\up)$) -- cycle ;
		\draw[fill=white,rounded corners=3pt] ($ (L) - (\s,\s) $) rectangle ($ (R) + (\s,\s) $) ;
		\node at ($ 0.5*(R)+0.5*(L) $) {$\rhoM{5}$};
		\draw[rounded corners=3pt, line width=0.2mm, red]  ($(L)+ (\step,\up)$) -- ($(L)+ (3*\step,\up)$);
		\draw[rounded corners=3pt, line width=0.2mm, red] ($(L)+(\step,\up)$) -- ($(L)+(\step,\up)+(-0.5*\step,0)$) -- ($(L)+ (\step,\up)+(-0.5*\step,0.5*\step)$);
		\draw[rounded corners=3pt, line width=0.2mm, red] ($(R)+(-\step,\up)$) -- ($(R)+(-\step,\up)+(0.5*\step,0)$) -- ($(R)+(-\step,\up)+(0.5*\step,0.5*\step)$);
		\foreach \x in {1,...,3}
		\draw[fill] ($(L)+ (\x*\step,\up)$)  circle (.5ex);
		\draw[rounded corners=3pt, line width=0.2mm, red]  ($(L)+ (\step,-\up)$) -- ($(L)+ (3*\step,-\up)$);
		\draw[rounded corners=3pt, line width=0.2mm, red] ($(L)+(\step,-\up)$) -- ($(L)+(\step,-\up)+(-0.5*\step,0)$) -- ($(L)+(\step,-\up)+(-0.5*\step,-0.5*\step)$);
		\draw[rounded corners=3pt, line width=0.2mm, red] ($(R)+(-\step,-\up)$) -- ($(R)+(-\step,-\up)+(0.5*\step,0)$) -- ($(R)+(-\step,-\up)+(0.5*\step,-0.5*\step)$);
		\foreach \x in {1,...,3}
		\draw[fill] ($(L)+ (\x*\step,-\up)$)  circle (.5ex);
	}
}

\tikzset{
	cgRTilde4/.pic = {
			\tikzmath{ \up=0.4;  \step=0.5;   \n=3;\s=0.25;} 
		\coordinate (R) at (0,0);
		\coordinate (L) at ($ (R) - \n*(\step,0)$);
		\foreach \x in {0,...,\n}
		\draw[-] ($(L)+ (\x*\step,\up)$) -- ($(L)+ (\x*\step,-\up)$) ;
		\foreach \x in {1,2}
		\draw[rounded corners=2pt, line width=0.2mm, red] ($(L)+ (\x*\step,\up)$) -- ($(L)+ (\x*\step,-\up)$) ;
		\draw[fill=white,rounded corners=3pt] ($ (L) - (\s,\s) $) rectangle ($ (R) + (\s,\s) $) ;
		\node at ($ 0.5*(R)+0.5*(L) $) {$\omegaM{4}$};
		\draw[-] ($(R)+ (0,\up)$) -- ($(R)+ (0,\up+0.25*\step)$);
		\draw[rounded corners=2pt, line width=0.2mm, red] ($(R)+ (-\step,\up)$) --  ($(R)+ (-\step,\up+0.25*\step)$) -- ($(R)+(0.5*\step,\up+0.25*\step)$) -- ($(R)+ (0.5*\step,\up+0.5*\step)$) ;
		\draw[fill] ($(R)+ (0,\up+0.25*\step)$)  circle (.5ex);
		\draw[-] ($(R)+ (0,-\up)$) -- ($(R)+ (0,-\up-0.25*\step)$);
		\draw[rounded corners=2pt, line width=0.2mm, red] ($(R)+ (-\step,-\up)$) --  ($(R)+ (-\step,-\up-0.25*\step)$) -- 	($(R)+(0.5*\step,-\up-0.25*\step)$) -- ($(R)+ (0.5*\step,-\up-0.5*\step)$) ;
		\draw[fill] ($(R)+ (0,-\up-0.25*\step)$)  circle (.5ex);
	}
}

\tikzset{
	TRtilde5/.pic = {
				\tikzmath{ \up=0.4;  \step=0.5;   \n=3;\s=0.25;} 
			\coordinate (R) at (0,0);
			\coordinate (L) at ($ (R) - \n*(\step,0)$);
			\draw[rounded corners=3pt, white] ($(L)+(0,\up)$) -- ($(L)+(-\step,\up)$) --  ($(L)+(-\step,-\up)$) -- ($(L)+(0,-\up)$) -- cycle ;
			\foreach \x in {0}
			\draw[-] ($(L)+ (\x*\step,\up)$) -- ($(L)+ (\x*\step,-\up)$) ;
			\foreach \x in {1,2}
			\draw[rounded corners=2pt, line width=0.2mm, red] ($(L)+ (\x*\step,\up)$) -- ($(L)+ (\x*\step,-\up)$) ;
			\draw[rounded corners=3pt] ($(R)+(0,\up)$) -- ($(R)+(\step,\up)$) --  ($(R)+(\step,-\up)$) -- ($(R)+(0,-\up)$) -- cycle ;
			\draw[fill=white,rounded corners=3pt] ($ (L) - (\s,\s) $) rectangle ($ (R) + (\s,\s) $) ;
			\node at ($ 0.5*(R)+0.5*(L) $) {$\omegaM{5}$};
	}
}

\tikzset{
state3/.pic = {
	\tikzmath{ \up=0.4;  \step=0.5;   \n=2; \s=0.25;} 
	\coordinate (R) at (0,0);
	\coordinate (L) at ($ (R) - \n*(\step,0)$);
	\foreach \x in {0,...,\n}
	\draw[-] ($(L)+ (\x*\step,\up)$) -- ($(L)+ (\x*\step,-\up)$) ;
	\draw[fill=white,rounded corners=3pt] ($ (L) - (\s,\s) $) rectangle ($ (R) + (\s,\s) $) ;
	\node at ($ 0.5*(R)+0.5*(L) $) {$\rhoM{3}$};
		}
}

\tikzset{
	double arrow/.style args={#1 colored by #2 and #3}{
		->,line cap=round,line width=#1,#2, 
		postaction={draw,->,line cap=round,#3,line width=(#1)/3,
			shorten <=(#1)/3,shorten >=2*(#1)/3}, 
	}
}